\documentclass[pra,twocolumn,superscriptaddress,longbibliography]{revtex4-2}

\usepackage{siunitx}
\usepackage{braket}
\usepackage{amsmath}
\usepackage{graphicx}
\usepackage{chngcntr}
\usepackage{tabularray}
\usepackage{enumitem}

\usepackage[colorlinks=true, allcolors=blue]{hyperref}

\let\originalleft\left
\let\originalright\right
\renewcommand{\left}{\mathopen{}\mathclose\bgroup\originalleft}
\renewcommand{\right}{\aftergroup\egroup\originalright}

\begin{document}
\title{Fast quantum interferometry at the nanometer and attosecond scales with energy-entangled photons}

\author{Colin P. Lualdi}\thanks{C.P.L. (clualdi2@illinois.edu) and S.J.J. contributed equally to this work.}
\affiliation{Department of Physics, The Grainger College of Engineering, University of Illinois Urbana-Champaign; Urbana, IL, USA}
\affiliation{Illinois Quantum Information Science and Technology Center, The Grainger College of Engineering, University of Illinois Urbana-Champaign; Urbana, IL, USA}

\author{Spencer J. Johnson}
\altaffiliation{Present address: Jet Propulsion Laboratory (JPL), California Institute of Technology; Pasadena, CA, USA}
\affiliation{Department of Physics, The Grainger College of Engineering, University of Illinois Urbana-Champaign; Urbana, IL, USA}
\affiliation{Illinois Quantum Information Science and Technology Center, The Grainger College of Engineering, University of Illinois Urbana-Champaign; Urbana, IL, USA}

\author{Michael Vayninger}
\altaffiliation{Present address: Department of Physics, University of Chicago; Chicago, IL, USA}
\affiliation{Department of Physics, The Grainger College of Engineering, University of Illinois Urbana-Champaign; Urbana, IL, USA}
\affiliation{Illinois Quantum Information Science and Technology Center, The Grainger College of Engineering, University of Illinois Urbana-Champaign; Urbana, IL, USA}

\author{Kristina A. Meier}
\altaffiliation{Present address: Intelligence and Space Research Division, Los Alamos National Laboratory; Los Alamos, NM, USA}
\affiliation{Department of Physics, The Grainger College of Engineering, University of Illinois Urbana-Champaign; Urbana, IL, USA}
\affiliation{Illinois Quantum Information Science and Technology Center, The Grainger College of Engineering, University of Illinois Urbana-Champaign; Urbana, IL, USA}

\author{Swetapadma Sahoo}
\affiliation{Illinois Quantum Information Science and Technology Center, The Grainger College of Engineering, University of Illinois Urbana-Champaign; Urbana, IL, USA}
\affiliation{Department of Electrical and Computer Engineering, The Grainger College of Engineering, University of Illinois Urbana-Champaign; Urbana, IL, USA}
\affiliation{Holonyak Micro and Nanotechnology Lab, The Grainger College of Engineering, University of Illinois Urbana-Champaign; Urbana, IL, USA}

\author{Simeon I. Bogdanov}
\affiliation{Illinois Quantum Information Science and Technology Center, The Grainger College of Engineering, University of Illinois Urbana-Champaign; Urbana, IL, USA}
\affiliation{Department of Electrical and Computer Engineering, The Grainger College of Engineering, University of Illinois Urbana-Champaign; Urbana, IL, USA}
\affiliation{Holonyak Micro and Nanotechnology Lab, The Grainger College of Engineering, University of Illinois Urbana-Champaign; Urbana, IL, USA}

\author{Paul G. Kwiat}
\affiliation{Department of Physics, The Grainger College of Engineering, University of Illinois Urbana-Champaign; Urbana, IL, USA}
\affiliation{Illinois Quantum Information Science and Technology Center, The Grainger College of Engineering, University of Illinois Urbana-Champaign; Urbana, IL, USA}
\affiliation{Department of Electrical and Computer Engineering, The Grainger College of Engineering, University of Illinois Urbana-Champaign; Urbana, IL, USA}

\begin{abstract}
In classical optical interferometry, loss and background complicate achieving fast nanometer-resolution measurements with illumination at low light levels. Conversely, quantum two-photon interference is unaffected by loss and background, but nanometer-scale resolution is physically difficult to realize. As a solution, we enhance two-photon   interference with highly non-degenerate energy entanglement featuring photon frequencies separated by 177 THz. We observe measurement resolution at the nanometer (attosecond) scale with only $O(10^4)$ photon pairs, despite the presence of background and loss. Our non-destructive thickness measurement of a metallic thin film agrees with atomic force microscopy, which often achieves better resolution via destructive means. With contactless, non-destructive measurements in seconds or faster, our instrument enables metrological studies in optically challenging contexts where background, loss, or photosensitivity are factors.
\end{abstract}

\maketitle

\section{Introduction}

Optical interferometry is an effective technique for high-resolution measurement and imaging; diverse applications include gravitational-wave detection \cite{ligo2016}, long-baseline astronomy \cite{monnier2003}, and optical coherence tomography \cite{huang1991}. Most of these applications utilize classical interference, in which an electromagnetic wave travels in a superposition of two paths and either constructively or destructively interferes with itself on a balanced beamsplitter depending on the relative phase between the two paths. While this technique can easily detect relative delays at the nanometer (or, equivalently, attosecond) scale, it is ill-suited for contexts with imbalanced path loss and optical background, which reduce the interference visibility, and hence the attainable resolution. In contrast, quantum interference involves two photons incident on the two inputs of a balanced beamsplitter; when the photons are indistinguishable, including in their time of arrival, their bosonic nature induces them to always exit the beamsplitter in the same port \cite{hong1987}. The visibility of two-photon interference is inherently robust against imbalanced path loss and optical background, motivating its use in quantum optical coherence tomography \cite{abouraddy2002, nasr2003}, quantum microscopy \cite{ndagano2022}, clock synchronization \cite{bahder2004, xie2021}, and other metrological applications. However, the measurement utility can be limited as achieving high resolution typically requires long measurements or ultra-broadband photons. For the former, nanometer-scale resolution has been demonstrated with hours-long measurements \cite{lyons2018}, and for the latter, resolutions range from the submicron \cite{nasr2008submicron,okano2015} to nanometer \cite{singh2023} scales, depending on the measurement technique. 

Introducing energy entanglement between the two interfering photons reveals an alternative path towards improved resolution with quantum interference \cite{chen2019}. Conventional two-photon interference features a dip in the probability of two photons exiting the beamsplitter in separate ports, as a function of the relative temporal delay between the two paths incident on the beamsplitter. The attainable measurement resolution is determined by the dip width, inversely related to the photons’ bandwidth. When the photons are energy-entangled, the dip is modulated sinusoidally with a period varying inversely with the difference (or ``beat note'') between the frequencies of the entangled photons \cite{ou1988, rarity1990}, resulting in interference fringes of similar form as those obtained via classical interference. The measurement information acquired per photon pair is therefore greatly increased. Indeed, this measurement scheme saturates the quantum Cram\'er–Rao bound, which sets the maximum attainable measurement precision, given a quantum probe state \cite{chen2019, helstrom1969, fujiwara1995}.

In this work, we fully realize the potential of this technique with highly non-degenerate, narrowband entangled photons, performing measurements at the nanometer and attosecond scales in seconds, even in the presence of substantial imbalanced path loss and optical background. While the features of loss and background insensitivity do not require entanglement, but only two-photon interference, the fact that such features persist in the presence of the energy entanglement needed to achieve high resolution makes this methodology superior to classical interference techniques.  

\section{Results}

\subsection{Theoretical description}

Consider the energy-entangled two-photon state
\begin{equation}
	\ket{\psi} = \frac{1}{\sqrt{2}}\left(\ket{\omega_1}_a\ket{\omega_2}_b+\ket{\omega_2}_a\ket{\omega_1}_b\right),
	\label{eq:1}
\end{equation}
where $\omega_i$ denotes the photon’s angular frequency, proportional to its energy, and $a$, $b$ denote the two inputs of a balanced beamsplitter. When these two otherwise identical photons impinge on a beamsplitter, the probability the photons exit in separate beamsplitter outputs and result in a coincidence detection between detectors placed at each output is given by 
\begin{equation}
	P_C = \frac{1}{2}\left[1-\cos((\Delta\omega)\tau)e^{-2\sigma^2\tau^2}\right],
	\label{eq:2}
\end{equation}
where $\Delta\omega \equiv \omega_1 - \omega_2$ is the angular frequency detuning of the entangled photons, $\tau$ is the relative temporal delay between the beamsplitter input paths, and $\sigma$ is the photons’ angular frequency half bandwidth (see Supplementary Materials). In conventional two-photon interference, $\omega_1 = \omega_2$ such that the cosine factor reduces to unity and Eq.~\ref{eq:2} reduces to the functional form of the so-called Hong-Ou-Mandel dip. As in \cite{lyons2018}, we set $\tau$ to maximize $dP_C/d\tau$ such that the interferometer yields the maximum Fisher information when measuring the change in $P_C$ induced by a small change in the relative delay $\delta\tau$. Via the quantum Cram\'er–Rao bound, the quantum Fisher information provides a lower bound on the standard deviation $\sigma_\tau$ of an estimation of $\delta\tau$, which defines the interferometer resolution. In the ideal case, we have
\begin{equation}
	\sigma_\tau \geq \frac{1}{\sqrt{\mathstrut N}\sqrt{\mathstrut Q}}=\frac{1}{\sqrt{\mathstrut N}\sqrt{\mathstrut (\Delta\omega)^2 + 4\sigma^2}},
	\label{eq:3}
\end{equation}
with $N$ the number of measurements and $Q$ the quantum Fisher information. As noted above, increasing $N$ or $\sigma$ has been demonstrated to yield a smaller $\sigma_\tau$ \cite{lyons2018,nasr2008submicron,okano2015,singh2023}, but this introduces practical challenges. Instead, with non-degenerate energy entanglement, a large detuning $\Delta\omega$ can achieve a comparable $\sigma_\tau$, but with narrow-bandwidth photons and far fewer measurements.

\subsection{Experimental setup}

\begin{figure*}[htb!]
	\begin{center}
		\includegraphics[width=\textwidth]{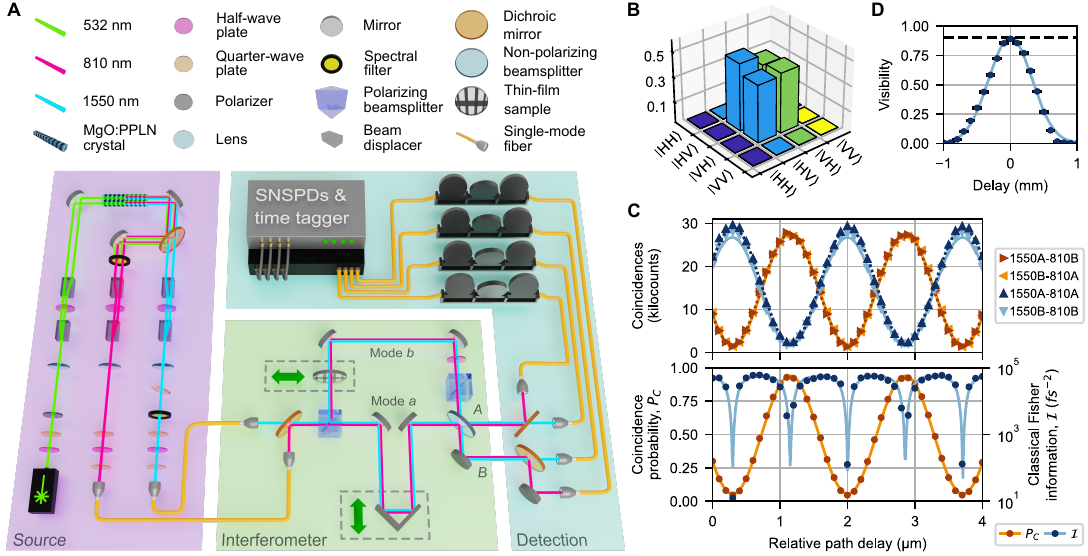}
	\end{center}
	\caption{\textbf{Highly non-degenerate energy-entangled two-photon interferometer.} \textbf{(A)} The experimental apparatus. In the source module (purple), beam displacers separate a diagonally polarized \mbox{532-nm} continuous-wave laser beam into horizontally and vertically polarized components. Each component pumps its own magnesium-oxide-doped periodically poled lithium niobate (MgO:PPLN) crystal to generate non-degenerate photon pairs at 810 and 1550~nm via SPDC. One of the two crystals is rotated $90^\circ$ around the optical propagation axis such that photon pairs from the two crystals are orthogonally polarized. A dichroic mirror separates the two down-converted wavelengths into individual spatial modes, where beam displacers spatially recombine the two polarization components. The resulting polarization-entangled photons are collected into single-mode fibers and transferred to the interferometer module (green), where polarization entanglement is converted into energy entanglement. Energy-entangled photon pairs then interfere on a 50:50 beamsplitter; a tunable path delay controls the relative temporal delay between modes $a$ and $b$. The detection module (blue) sorts the exiting photons by wavelength for detection by superconducting nanowire single-photon detectors. \textbf{(B)} Density matrix of the non-degenerate polarization-entangled state incident on the PBS in the interferometer module, prior to conversion into energy entanglement. \textbf{(C)} Top panel: Energy-entangled two-photon interference fringes as a function of relative path delay for each coincidence and anti-coincidence channel, with sinusoidal fits (\mbox{4-\si{\micro\meter}} snippet of a \mbox{20-\si{\micro\meter}} scan shown; the fits utilize the full scan). Accidentals are negligible with a mean coincidence to accidental ratio of 1,100(200) per measurement for the four channels combined. Bottom panel: The normalized coincidence probability fringe with a sinusoidal fit (solid curve). The corresponding classical Fisher information $\mathcal{I}$ is also shown; the solid curve is the theoretical prediction based on the sinusoidal fringe fits. \textbf{(D)} Fringe visibility as a function of relative path delay. The exponential fit yields a maximum visibility of $90.5(1.0)\%$ (dashed line).}
	\label{fig:fig1}
\end{figure*}

Our apparatus consists of source, interferometer, and detection modules (Fig.~\hyperref[fig:fig1]{\ref*{fig:fig1}A}). To generate the required energy-entangled state (Eq.~\ref{eq:1}), we first generate non-degenerate polarization-entangled photon pairs \cite{ramelow2009}. Our source utilizes a revised version of the beam-displacer geometry described in \cite{evans2010} to coherently drive two highly non-degenerate spontaneous parametric down-conversion (SPDC) processes, producing the state
\begin{equation}
	\begin{split}
	\ket{\phi_{SPDC}} =& \frac{1}{\sqrt{2}}\big(\ket{H}_{1550}\ket{H}_{810}\\
	&\qquad+e^{i\phi}\ket{V}_{1550}\ket{V}_{810}\big),
	\end{split}
	\label{eq:4}
\end{equation} 
where $H$, $V$ are the horizontal and vertical polarization states, the subscripts denote the \mbox{1550-nm} and \mbox{810-nm} photon wavelengths, and $\phi$ is an arbitrary relative phase. Type-0 phase matching enables a high source brightness: $>$$10^5$ detected pairs per second per milliwatt. 

Photons at each wavelength are separated into individual spatial modes and coupled into their own single-mode fiber for transfer between the source and interferometer modules. The photons are then collimated into free space and multiplexed (via a dichroic mirror) into a common spatial mode incident on the interferometer input. A polarization state tomography at the interferometer input yields a state with a high purity, concurrence, and entangled singlet fraction of $90.3(2)\%$, $89.6(2)\%$, and $94.8(1)\%$, respectively (Fig.~\hyperref[fig:fig1]{\ref*{fig:fig1}B}).

A pair of half-wave and quarter-wave plates at the input of each source collection fiber are set such that the waveplate-fiber system performs the operation $I_{1550}X_{810}$ on $\ket{\phi_{SPDC}}$, making the two wavelengths orthogonally polarized. At the interferometer input, the bit-flipped state passes through a polarizing beamsplitter (PBS) that transmits $H$ photons into spatial mode $a$ and reflects $V$ into $b$. Since the state’s two wavelengths correspond to orthogonal polarizations, it transforms into a superposition of a \mbox{1550-nm} (\mbox{810-nm}) photon in mode $a$ ($b$) and vice versa. After using a half-wave plate to rotate the polarization of the mode $b$ photons from $V$ to $H$, we obtain the energy-entangled state Eq.~\ref{eq:1}. With \mbox{1550-nm} and \mbox{810-nm} photons, our detuning $\Delta\omega \equiv \omega_1 - \omega_2 = 2\pi\,177~\text{THz}$ is nearly an order of magnitude larger than the previous best result, $2\pi\,30.1~\text{THz}$ \cite{torre2023}.

The energy-entangled photons then impinge on a balanced beamsplitter and undergo interference. The relative temporal delay between the beamsplitter input paths ($\tau$) is adjusted by tuning the relative optical path lengths for modes $a$ and $b$ via an optical trombone in path $a$. The trombone position is controlled by both a piezoelectric nano-positioning stage and a servo actuator, enabling nanometer resolution with centimeters of travel. 

Upon interference, the photons exit the beamsplitter in either port $A$ or $B$. At each output port a dichroic mirror separates \mbox{1550-nm} and \mbox{810-nm} photons for coupling into single-mode fiber leading to superconducting nanowire single-photon detectors. Four pairs of coincident detections (1550$A$-810$B$, 1550$B$-810$A$, 1550$A$-810$A$, and 1550$B$-810$B$) are monitored via a time tagger, with the first two corresponding to ``coincidence'' events (coincident detections in the opposite ports) and the last two ``anti-coincidence'' events (coincident detections in the same port). 

With these four detection pairs we directly measure the normalized coincidence probability $P_C$ (Eq.~\ref{eq:2}) as $N_C/(N_C+N_A)$ where $N_C$ and $N_A$ are the total number of coincidence and anti-coincidence events, respectively. Directly detecting all interfering photons simplifies previous methods, which relied on extensive characterization of system losses \cite{chen2019} or probabilistic detector trees \cite{torre2023} to extract $P_C$ from a given measurement of $N_C$.

Figure~\hyperref[fig:fig1]{\ref*{fig:fig1}C} shows the interference fringes observed in our experiment. As the relative path delay between modes $a$ and $b$ is scanned, $P_C$ oscillates sinusoidally with a fitted period of 1705.9(2)~nm, close to the \mbox{1701.87(1)-nm} period expected from the photon center wavelengths of 810.504(1)~nm (measured) and 1547.484(5)~nm (inferred via energy conservation). The given errors are based on fitting errors; the slight discrepancy between the fringe and spectral measurements is likely due to systematic errors, e.g., we observe a small drift in the interferometer relative phase (\mbox{$\sim$1} degree per minute, see fig.~\ref{fig:figs6}), which would be sufficient to account for the inferred wavelength mismatch. The fitted fringe visibility of $88.9(2)\%$ is close to the \mbox{$\sim$$87.4\%$} expected given our entangled state purity, PBS extinction ratio, and beamsplitter splitting ratio. Finally, while the fitted visibilities of the four coincident detection fringes range between $87.7(4)\%$ and $89.2(5)\%$, the individual-detector fringes have visibilities below $1\%$, indicating that two-photon, not single-photon, interference dominates.

From the four coincident detection fringe data and fits, we extract the corresponding classical Fisher information and estimate the maximum attainable measurement resolution. With \mbox{1-mW} source pumping power and a \mbox{1-s} integration time, we observe a mean of 59,000(1,000) total coincident detections per measurement. The corresponding resolution is 1.26~nm (1.26~nm/$c$ = 4.2~as, where $c$ is the speed of light), an $88\%$ saturation of the Cram\'er–Rao bound.

Figure~\hyperref[fig:fig1]{\ref*{fig:fig1}D} illustrates a key advantage of introducing energy entanglement to two-photon interference. To achieve nanometer-scale resolution, conventional two-photon interference would require ultrabroadband photons (\mbox{$\sim$177 THz}), corresponding to a narrow dip width of \mbox{$\sim$317 nm} full width at half maximum (FWHM), making the dip difficult to locate. Additionally, optical systems that support such a large spectral spread are difficult to realize, as is such a broadband SPDC source \cite{nasr2008submicron, okano2015, singh2023, nasr2008ultrabroadband}. In contrast, our experiment utilizes narrowband photons (e.g., $\sigma_{810} = 0.495(2)$~nm FWHM) such that the modulated interference dip envelope is much wider: 0.76(1)~mm FWHM. Non-zero interference visibility over such a large window greatly simplifies initially calibrating the relative path lengths, and can enable a very large dynamic range, e.g., by counting fringes. 

\subsection{Entanglement-enhanced quantum metrology}

Our interferometer can achieve nanometer (attosecond) resolution with only $O(10^4)$ detected photon pairs; with a detected pair rate of at least 150,000 per second (enabled by tuning the source pumping power) we can attain nanometer resolution in a timescale of seconds (or less).

\begin{figure}[htb!]
	\begin{center}
		\includegraphics[scale=0.931]{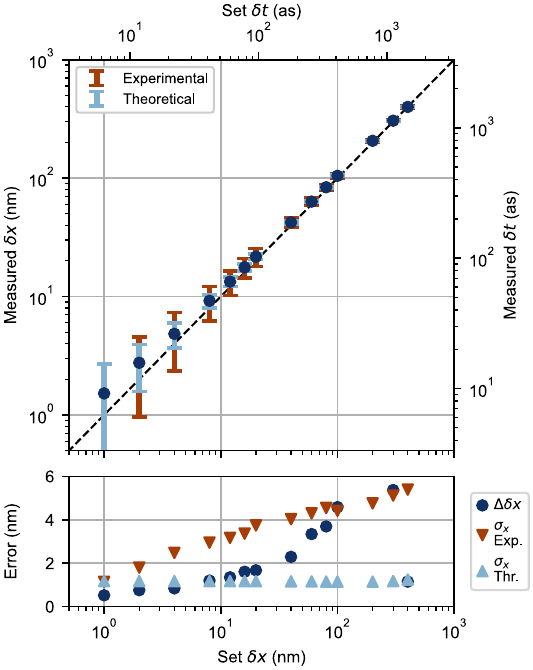}
	\end{center}
	\caption{\textbf{Demonstrating nanometer-scale resolution.} Experimental data illustrating measurements at the nanometer (attosecond) scale. Displacements were introduced by a nano-positioning stage displacing a retroreflector; a displacement of $\delta x$ in the optical path length is therefore realized by translating the nano-positioning stage by $\delta x/2$. The bottom panel shows the ``error'' in terms of both accuracy ($\Delta\delta x$) and precision ($\sigma_x$).}
	\label{fig:fig2}
\end{figure}

To validate the resolution estimated from the fringes in Fig.~\hyperref[fig:fig1]{\ref*{fig:fig1}C}, we displace the trombone retroreflector by a set displacement $\delta x/2$, measure the resulting change in the coincidence probability $P_C$ (with a 1-second integration time), and extract the corresponding $\delta x$. The top panel of Fig. \ref{fig:fig2} shows the mean measured $\delta x$ and its standard deviation $\sigma_x$ (for 100 trials) for multiple values of $\delta x$. The dynamic range is set by our standard measurement scheme, which is restricted to displacements on one rising or falling fringe, i.e., half of the interference period. As the Fisher information is high near the fringe center (where  $P_C \approx 0.5$), we typically operate within this region, which spans a few hundred nanometers. The bottom panel shows the corresponding measurement error $\left(\Delta\delta x \equiv \left|\delta x_{measured} - \delta x_{set}\right| \right)$ as well as the experimental and theoretical $\sigma_x$. We observe nanometer-scale accuracy and precision. We attribute the increase in error and decrease in resolution as $\delta x$ increases to interferometer drift; all 14 set displacements were measured sequentially from a common zero-displacement point.  

\begin{figure}[htb!]
	\begin{center}
		\includegraphics[scale=0.931]{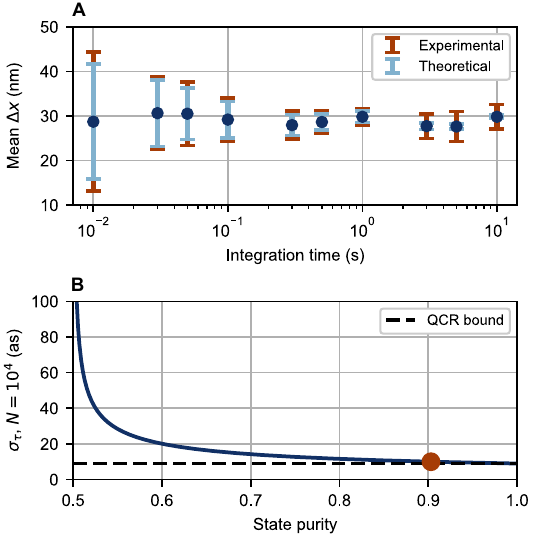}
	\end{center}
	\caption{\textbf{Effects of integration time and state purity on measurement resolution.} \textbf{(A)} Experimental data illustrating trade-offs between resolution and measurement time. The interferometer displacement $\Delta x$ is the path-length change introduced by moving from the initial sample position (uncoated region) to the final position (coated region). The ``interferometer displacement'' is extracted by assuming that the fringe period (in terms of path length) is unchanged by the indices of refraction of the sapphire wafer and nickel coating. Using a sample with a fixed displacement instead of the optical trombone eliminates additional uncertainty from the trombone positioner. Each trial for each data point involves two $P_C$ measurements (both with identical integration times), one for each sample position. While the upper bound on the resolution is in principle set by the phase drift of our passively stabilized interferometer, one could perform multiple measurements with a short integration time (where drift is negligible) and attain higher resolution via the standard error of the mean. \textbf{(B)} The maximum theoretical measurement resolution $\sigma_\tau$ as a function of state purity for $N = 10^4$, with our result highlighted.}
	\label{fig:fig3}
\end{figure}

Our measurement resolution is dominated by the frequency detuning $\Delta\omega$ and the number of detected photon pairs $N$ (Eq.~\ref{eq:3}). Our high detected photon pair rate therefore enables tuning the tradeoff between the measurement resolution and the integration time (Fig.~\hyperref[fig:fig3]{\ref*{fig:fig3}A}), which shows the average measured interferometer displacement $\left(\Delta x\right)$ and $\sigma_x$ for a fixed set displacement (100 trials each). To introduce the displacement, we insert a sapphire wafer in one path of the interferometer and translate it between two transverse positions with respect to the optical beam; one position corresponds to an uncoated region and the other to a nickel-coated region with a nominal thickness of 5~nm. With a baseline detection rate of 128,000(3,000) and 68,000(2,000) pairs per second for the uncoated and coated regions, respectively, we achieve an interferometer resolution of $\sigma_x = 1.8$~nm with a \mbox{1-second} integration time. As expected, a shorter \mbox{0.1-s} integration time yields a slightly reduced resolution of 4.9~nm. Longer integration times introduce measurement error from interferometer drift, as our apparatus currently has no active stabilization. For example, increasing the integration time to 10~seconds yields a resolution of 2.7~nm instead of the 0.4~nm predicted by theory. However, since nanometer-scale resolution is still achievable with shorter integration times, measurements may be made much faster relative to the interferometer drift such that active stabilization is not strictly necessary. Overall, our measurements approach the fundamental resolution limit given our probe state, as dictated by the quantum Cram\'er–Rao bound, despite an imperfect entangled state purity of $90.3(2)\%$ (Fig.~\hyperref[fig:fig3]{\ref*{fig:fig3}B}).

\begin{figure*}[htb!]
	\begin{center}
		\includegraphics[scale=0.931]{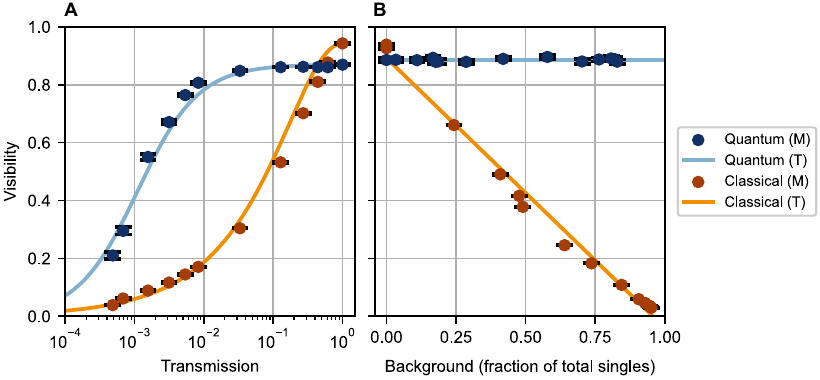}
	\end{center}
	\caption{\textbf{Robustness against imbalanced path loss and optical background.} \textbf{(A)} Measured (M) interference visibility as a function of transmission in mode $b$ of the interferometer for both quantum and classical interference. Solid lines are theoretical (T) predictions. The reduction in two-photon visibility beyond 10~dB of loss is attributed to noise events. Contributions to noise include both accidentals (coincident detections between uncorrelated photons) and a small $\ket{\omega_1}_a\ket{\omega_2}_a$ term in our state (Eq.~\ref{eq:1}) arising from the finite extinction ratio of our PBS and the imperfect application of $I_{1550}X_{810}$ between the source and interferometer modules. Present in mode $a$, these sources of noise become relevant only when the loss in mode $b$ reduces the relative rate of genuine to noise coincident detections. \textbf{(B)} Observed interference visibility as a function of optical background, expressed as a fraction of total individual detector clicks (``singles''). Solid lines are fits to theoretical models. Even when nearly $100\%$ of the detector clicks are background events, the quantum visibility is unaffected whereas the classical visibility is severely degraded.}
	\label{fig:fig4}
\end{figure*}

We also verify the robustness of energy-entangled two-photon interference against imbalanced path loss and optical background. Since quantum interference is a two-photon process, imbalanced path loss acts globally on the two-photon state to reduce the detection rate of two-photon states, with the interference visibility unaffected. Conversely, in classical interference imbalanced loss degrades the photon’s superposition state, diminishing both the photon rate and interference visibility. We introduce tunable optical loss to mode $b$ of our interferometer by inserting a PBS after the half-wave plate; the PBS transmission is adjusted by rotating the half-wave plate. Measuring the interference visibility for increasing loss (up to \mbox{$\sim$33 dB}), we observe that the two-photon interference visibility is largely unaffected up to \mbox{$\sim$10 dB} of loss and only slightly affected up to \mbox{$\sim$20 dB} of loss. In contrast, the classical interference visibility (using \mbox{1550-nm} photons) decreases immediately and precipitously, dropping to approximately half its starting value with only \mbox{$\sim$10 dB} of loss (Fig.~\hyperref[fig:fig4]{\ref*{fig:fig4}A}). The reduction in quantum visibility for $>$10~dB of loss is attributed to the presence of noise photons resulting from experimental imperfections; our noise-adjusted theoretical model closely tracks the experimental data. 

Optical background can cause coincident detections uncorrelated with actual energy-entangled photon pairs. The resulting accidentals background reduces fringe visibility for the same reason classical interference visibility is degraded by detector background; in both cases the reduced visibility leads to a decreased measurement resolution. However, unlike classical interference, two-photon interference is measured in coincidence; thus, any background photons would need to appear in the same coincidence window as the photons undergoing interference to register as an accidental event. By utilizing low-jitter detectors and electronics (\mbox{75-ps} mean total FWHM jitter per coincident detection channel) and a tight coincidence window radius ($\pm$50~ps), we suppress the per-second background rate by 100~dB while maintaining a high coincidence rate. Since our photons are narrowband, an even greater suppression may be achieved with tight spectral filtering at the detectors; such filtering cannot be leveraged to the same extent for conventional two-photon interference with broadband photons. We experimentally introduce background by shining a halogen lamp onto our apparatus at varying intensities. Even with background photons approaching $100\%$ of all individual detector clicks, we observe unchanged two-photon interference visibility (Fig.~\hyperref[fig:fig4]{\ref*{fig:fig4}B}). In contrast, repeating the same experiment with classical interference (using \mbox{1550-nm} photons) results in a $97\%$ visibility reduction.

\begin{figure*}[htb!]
	\begin{center}
		\includegraphics[scale=0.931]{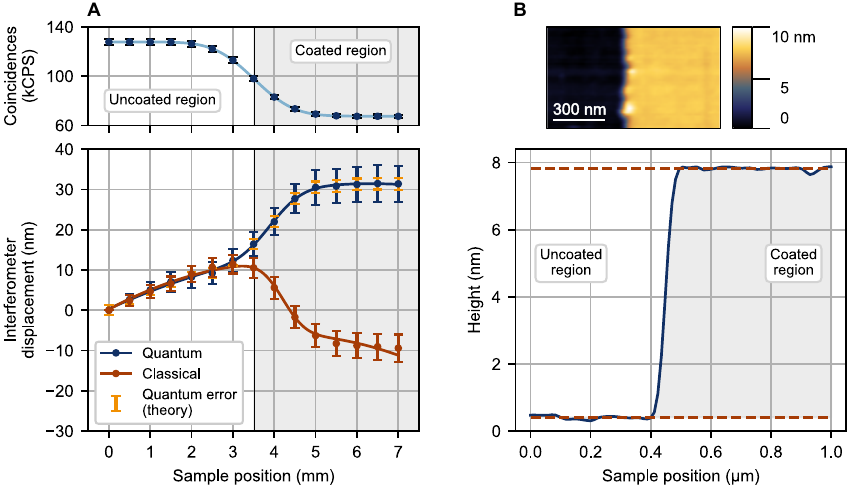}
	\end{center}
	\caption{\textbf{Measuring the nanometer-scale thickness of a lossy metallic thin film.} \textbf{(A)} Top panel: The coincident detection rate (for quantum interference) as a function of sample position, showing a clear delineation between the uncoated and coated regions. The fit (solid curve) indicates a quantum probe transmission of $52.8(1)\%$. Bottom panel: The interferometer displacement as a function of sample position, for both quantum and classical interference; the film thickness is related to the interferometer displacement via the refractive index of the film material. The blue and red error bars show the standard deviation for 100 trials and indicate the measurement resolution for the \mbox{1-second} per-point integration time. The larger error bars for the coated region are due to both a loss-induced reduction in $N$ and drift in the interferometer phase as the sample is scanned, as evidenced by the theoretical quantum error bars (yellow), which assume no drift. Solid curves are fits to a theoretical model (see Supplementary Materials). \textbf{(B)} Top panel: A false-color AFM image of the sample, showing uncoated (left) and coated (right) regions. Bottom panel: Profile view of top-panel image, showing the average from each horizontal scan line. The red dashed lines indicate the average height for the first and last 340~nm of the bottom and top regions; their mean and standard deviation values yield a step height of 7.4(1)~nm. Unlike the interferometric measurements, our AFM measurement was destructive: to address poor edge conditions that rendered non-destructive measurement inaccurate, we had to mechanically remove a small region of the film (using the AFM tip) to expose the substrate before measuring the height difference.}
	\label{fig:fig5}
\end{figure*}

We demonstrate the metrological capabilities of our system by measuring the thickness of a thin metallic film (Ni) on a \mbox{3-inch} sapphire wafer. We measure in transmission: the wafer is placed in the $b$ mode of the interferometer and scanned transversely such that the optical beam (with a mean effective $1/e^2$ diameter of 1.21(4)~mm) moves horizontally across the wafer, from an uncoated region to a coated region. By monitoring the coincidence probability $P_C$ as a function of the sample position, we can infer the optical delay introduced by the thin film. Using a calibrated effective refractive index acquired by measuring a calibration sample with a well-defined thickness (see Supplementary Materials) and the known interference fringe period, we convert phase to film thickness. The average (100 trials) measured displacement as a function of sample position is shown in Fig.~\hyperref[fig:fig5]{\ref*{fig:fig5}A}, for both quantum and classical interference. The data fits well to a model describing a Gaussian optical probe scanning over an infinitely sharp step in height on a substrate with a linear wedge and quadratic curvature (see Supplementary Materials). From the fit we extract a film thickness of 7(1)~nm for the quantum measurement, in excellent agreement with the 7.4(1)~nm obtained via atomic force microscopy (AFM), as shown in Fig.~\hyperref[fig:fig5]{\ref*{fig:fig5}B}. A scanning-stylus profilometry measurement yielded a similar thickness of 5.9(1)~nm. In contrast, three-dimensional (3D) optical profilometry could only provide a semi-quantitative thickness estimate of 12(1)~nm because of measurement noise. The classical interference measurement is even worse, returning a film thickness of $-$9(1)~nm; the deposited film instead appears as an etched substrate. We attribute this inaccurate result to degraded fringe quality due to the lossy film; observed degradations include decreased interference visibility ($95.7(1)\% \rightarrow 82.7(2)\%$). In contrast, the quantum fringes are largely unchanged (e.g., the quantum visibility is negligibly affected: $88.5(3)\% \rightarrow 88.2(4)\%$).

\section{Discussion}

Leveraging a bright source of highly non-degenerate energy entanglement, we realize fast, loss-tolerant nanometer-scale metrology at the single-photon level. With only $O(10^4)$ photons required for nanometer-scale (attosecond-scale) resolution, our performance surpasses the state-of-the-art in conventional (non-ultrabroadband) quantum interferometry, which requires $O(10^{11})$ photons collected over hours to achieve comparable resolution \cite{lyons2018}. Our 177-THz frequency detuning also improves upon the tens of THz utilized in the state of the art, enabling a resolution improvement of two orders of magnitude \cite{chen2019,torre2023}. Our combination of high measurement resolution and robustness against imbalanced loss enabled an accurate thickness measurement of a lossy metallic thin film, an exercise that classical interferometry failed. Although classical interferometry may be made to work with lossy samples in certain situations by calibrating the visibility as a function of sample location, such calibration represents an unnecessary additional step with its own complications. Our method stands out for its contactless (unlike scanning-stylus profilometry), non-destructive nature (unlike AFM in some circumstances), single-photon level of illumination (unlike 3D optical profilometry), and a large scanning range of millimeters rather than micrometers. The AFM measurement was destructive in our case because of the need to remove film material to measure the relative height between the film and substrate for our sample. 

Straightforward engineering improvements can enable improved lateral resolution via probe focusing and two-dimensional imaging via raster scanning, both of which have been demonstrated for energy-entangled interferometry \cite{torre2023}, along with measuring in reflection, as demonstrated in quantum optical coherence tomography experiments \cite{nasr2003}. Overall, our system enables nanometer-scale metrological studies, such as probing lossy, light-sensitive biological samples, for which existing classical techniques are ill-suited. 

Additionally, as discussed in the Supplementary Materials, our interferometer can be operated in two additional modes. The first uses non-entangled, non-degenerate photon pairs to observe beating between \mbox{1550-nm} and \mbox{810-nm} single-photon interference fringes. The second involves replacing $\ket{\psi}$ (Eq.~\ref{eq:1}) with $\ket{\psi'}\propto\left(\ket{\omega_1}_a\ket{\omega_2}_a+\ket{\omega_2}_b\ket{\omega_1}_b\right)$, a frequency-dependent variant of the two-photon N00N state \cite{lee2002}. With this state, the sum (rather than the difference) of the two frequencies determines the interference period (532~nm in our case). These modes enable additional metrological possibilities, such as probing samples opaque at 532~nm but transparent at 810 and 1550~nm.

\section{Materials and Methods}

\subsection{Theory}

\subsubsection{Quantum Fisher information}

For a given measurement of some parameter $\tau$ using quantum probe state $\ket{\psi\left(\tau\right)}$, the resolution of the measurement $\sigma_\tau$ is given by the quantum Cram\'er–Rao bound,
\begin{equation}
	\sigma_\tau \geq \sigma_{\tau,QCR} = \frac{1}{\sqrt{N}}\frac{1}{\sqrt{Q}},
\end{equation}
where $N$ is the total number of trials in the measurement, and 
\begin{equation}
	Q=4\left(\left\langle\frac{\partial\psi(\tau)}{\partial\tau}\middle|\frac{\partial\psi(\tau)}{\partial\tau}\right\rangle-\left|\left\langle\psi(\tau)\middle|\frac{\partial\psi(\tau)}{\partial\tau}\right\rangle\right|^2 \right),
	\label{eq:6}
\end{equation}
is the quantum Fisher information \cite{helstrom1969,fujiwara1995}. Evaluating Eq.~\ref{eq:6} with our probe state (see Supplementary Materials), we obtain
\begin{equation}
	Q=(\Delta\omega)^2 + 4\sigma^2,
	\label{eq:7}
\end{equation}
which leads to Eq.~\ref{eq:3}.

\subsubsection{Classical Fisher information}

When performing measurements, it is often convenient to measure the classical Fisher information $\mathcal{I}$ rather than the quantum Fisher information $Q$. This is because the classical Fisher information can be directly calculated from the coincidence and anti-coincidence probabilities associated with our interference fringes and provides additional information about imperfections in the measurement scheme. The classical Fisher information for a discrete probability distribution $P(X;\tau)$ is given by
\begin{equation}
	\begin{split}
	\mathcal{I} &= E\left[\left(\frac{\partial\log(P(X;\tau))}{\partial\tau}\right)^2\right] \\
	&= \sum_x\left(\frac{\frac{\partial P(x;\tau)}{\partial\tau}}{P(x;\tau)}\right)^2 P(x;\tau)\\
	&= \sum_x\frac{\left(\frac{\partial P(x;\tau)}{\partial\tau}\right)^2}{P(x;\tau)},
	\end{split}
	\label{eq:8}
\end{equation} 
where $E[f]$ is the expectation value of $f$. Evaluating Eq.~\ref{eq:8} for the case of ideal energy-entangled two-photon interference (see Supplementary Materials), we obtain the single-event Fisher information:
\begin{equation}
	\mathcal{I} = \frac{\left((\Delta\omega)\sin((\Delta\omega)\tau)+4\sigma^2\tau\cos((\Delta\omega)\tau)\right)^2}{e^{4\sigma^2\tau^2}-\cos^2((\Delta\omega)\tau)}.
\end{equation} 
In the limit of $\tau \rightarrow 0$, we recover
\begin{equation}
	\lim_{\tau \rightarrow 0}\mathcal{I} = (\Delta\omega)^2 + 4\sigma^2,
\end{equation}
identical to the quantum Fisher information $Q$ (Eq.~\ref{eq:7}). Experimentally, $\mathcal{I}$ is calculated via Eq.~\ref{eq:8}, where the probabilities $P_{AA}$, $P_{AB}$, $P_{BA}$, and $P_{BB}$ correspond to the four measured coincidence and anti-coincidence probability fringes. We note that, for calculational convenience, we assume in our normalization of each probability fringe that the fringe is centered around $P = 1/4$, which is consistent with our observed fringes. 

\subsubsection{Experimental saturation of the Cram\'er–Rao bound}

While the quantum Cram\'er–Rao bound describes the fundamental resolution limit a measurement scheme can achieve for a given probe state, in practice, experimental imperfections result in the observed resolution being below the bound. In our system, the primary contribution to resolution degradation is imperfect entanglement purity, which leads to reduced interferometer visibility and therefore resolution. A mixed energy-entangled (EE) state with purity $(1+\epsilon^2)/2$ produces interference fringes
\begin{equation}
	P_C(\tau)=\frac{1}{2}\left[1-\epsilon\cos((\Delta\omega)\tau)e^{-2\sigma^2\tau^2}\right].
\end{equation}
A full derivation is given in the Supplementary Materials. These fringes have visibility $\epsilon$, allowing this to also illustrate the effect of imperfect interference. By error propagation,
\begin{equation}
	\text{Var}[P_C] = \left(\frac{\partial P_C}{\partial\tau}\right)^2\sigma_\tau^2.
\end{equation}
Solving for the single-measurement error $\sigma_\tau$ and evaluating the right-hand side of the equation (see Supplementary Materials) yields
\begin{equation}
	\sigma_\tau = \frac{1}{\epsilon}\frac{\sqrt{e^{4\sigma^2\tau^2}-\epsilon^2\cos^2((\Delta\omega)\tau)}}{(\Delta\omega)\sin((\Delta\omega)\tau)+4\sigma^2\tau\cos((\Delta\omega)\tau)}.
	\label{eq:13}
\end{equation}
In the limit of $\tau \rightarrow 0$, $\epsilon\rightarrow 1$, Eq.~\ref{eq:13} becomes Eq.~\ref{eq:3} with $N = 1$. We note that for $\epsilon < 1$, the optimal Fisher information resides at $\tau = (\pi/2)/\Delta\omega$ rather than at $\tau = 0$.

\subsubsection{Experimental displacement extraction}

To extract the corresponding displacement arising from a given interferometric measurement, we utilize maximum-likelihood estimation and a set of reference fringes. These reference fringes are measured by sweeping the interferometer through \SI{4}{\micro\meter} of path-length difference (\SI{2}{\micro\meter} of trombone displacement) and recording the four coincidence and anti-coincidence fringes. The measured fringes are normalized and fit to sinusoidal curves
\begin{equation}
	P(x) = a + b\cos\left(\frac{2\pi}{c}x-d\right).
\end{equation}
We ignore the exponential envelope, which is wide compared to the period of an interference fringe. From this fitting function, we can extract the total power $a + b$, the fringe visibility $|b/a|$, the wavelength $c$, and the phase offset $d$. Unless otherwise noted, this fitting provides all the visibility values reported in this work, with the uncertainty obtained by propagating the fit errors for $a$ and $b$. 

Performing a fit for each combination of detector pairs, we produce a set of curves: $P_{AA}(x)$, $P_{AB}(x)$, $P_{BA}(x)$, and $P_{BB}(x)$. The use of four individual reference curves --- as opposed to a single fringe based on the coincidence probability $P_C$ --- allows us to account for potential visibility and phase differences associated with experimental imperfections in the interferometer, improving extraction accuracy. 

For a given displacement measurement, we measure four pairs of coincident detections: $N_{AA}$,  $N_{AB}$, $N_{BA}$, and $N_{BB}$. From here, we construct the log-likelihood function
\begin{widetext}
\begin{equation}
	\begin{split}
	\ell(x,N_{AA},N_{AB},N_{BA}, N_{BB}) &= - \left[\frac{\left(P_{AA}(x)-\frac{N_{AA}}{N_{tot}}\right)^2}{P_{AA}(x)}+\cdots+\frac{\left(P_{BB}(x)-\frac{N_{BB}}{N_{tot}}\right)^2}{P_{BB}(x)}\right]\\
	N_{tot} &= N_{AA}+N_{AB}+N_{BA}+N_{BB}.
	\end{split}
\end{equation}
\end{widetext}
This form assumes that each measurement is sampled from a Gaussian distribution, which is a good approximation for Poissonian statistics in the high-count regime. Finally, the displacement $x^*$ is recovered by using Python to perform the numerical optimization
\begin{equation}
	x^* = \text{ArgMax}_x\ell(x,N_{AA},N_{AB},N_{BA}, N_{BB}).
\end{equation}
Theoretical error bars are produced via Eq.~\ref{eq:8}, with the four reference curves used to calculate the single-event Fisher information $\mathcal{I}$. From here, the theoretical resolution is
\begin{equation}
	\sigma_{x,theory} \equiv \frac{1}{\sqrt{N_{tot}}}\frac{1}{\sqrt{\mathcal{I}}}
\end{equation}
This calculation neglects the non-zero bandwidth of the photons due to the relative size of the frequency detuning employed.

\subsubsection{Modeling the effect of loss}

For the energy-entangled state employed in this study, loss in one arm of the interferometer (corresponding to a transmission $\eta$ in, e.g., mode $b$; see Supplementary Materials) is frequency dependent: 
\begin{widetext}
	\begin{equation}
		\frac{1}{\sqrt{2}}\left(\ket{\omega_1}_a\ket{\omega_2}_b+\ket{\omega_2}_a\ket{\omega_1}_b\right) \rightarrow \frac{1}{\sqrt{\eta_{\omega_1}+\eta_{\omega_2}}}\left(\sqrt{\eta_{\omega_2}}\ket{\omega_1}_a\ket{\omega_2}_b+\sqrt{\eta_{\omega_1}}\ket{\omega_2}_a\ket{\omega_1}_b\right).
	\end{equation}
\end{widetext}
So long as $\eta_{\omega_1} = \eta_{\omega_2} = \eta > 0$, the loss manifests as a global factor on the quantum state that reduces the coincident detection rate (see Supplementary Materials); the visibility of the interference fringe is unaffected.

However, if the interferometer is subject to loss-independent optical noise, the visibility is affected. Let $C_0 = C_{LD} + C_{LI}$ be the total number of coincident detections when $\eta=1$, with contributions from both loss-dependent detections $C_{LD}$, which includes most accidentals,  and loss-independent noise detections $C_{LI}$. The loss-dependent detections correspond to fringes with visibility $V_{LD}$, and the loss-independent noise detections correspond to noise fringes with visibility $V_{LI}$. We assume $V_{LI}=0$ since the noise photons are independent of the interference process. We can rewrite the coincidence fringe $N_c$ as
\begin{equation}
	N_c = C_{LD}\left(\frac{1}{2}+\frac{1}{2}V_{LD}\cos(\phi)\right)+\frac{C_{LI}}{2},
\end{equation}
where $\phi$ is the interferometer phase. The net fringe visibility $V_0$ is
\begin{equation}
	\begin{split}
		V_0 &= \frac{\text{max}(N_c)-\text{min}(N_c)}{\text{max}(N_c)+\text{min}(N_c)}\\
		&= \frac{C_{LD}}{C_{LD}+C_{LI}}V_{LD}\\
		&= \frac{C_{LD}}{C_0}V_{LD}.
	\end{split}
	\label{eq:20}
\end{equation}
When $\eta < 1$, only the number of non-noise detections is reduced: $C_{LD} \rightarrow \left(C_{LD}\right)\eta$. Eq.~\ref{eq:20} therefore becomes 
\begin{equation}
	V\left(\eta\right) = \frac{(C_{LD})\eta}{(C_{LD})\eta + C_{LI}}V_{LD}.
	\label{eq:21}
\end{equation}
We can rewrite Eq.~\ref{eq:21} in terms of the experimentally measured quantities $V_0$, $C_0$, $\eta$, and $C_{LI}$ by substituting in $(C_{LD})V_{LD} = (C_0)V_0$ (Eq.~\ref{eq:20}) and $C_{LD} = C_0 = C_{LI}$:
\begin{equation}
	V\left(\eta\right) = \frac{(C_0)\eta}{(C_0 - C_{LI})\eta + C_{LI}}V_0.
\end{equation}
Fringe scans performed without imbalanced path loss yield $V_0$ and $C_0$. We directly measured $\eta$ and $C_{LI}$ when characterizing our apparatus (see Supplementary Materials).

In contrast, loss in one arm of a classical interferometer will always affect the state
\begin{equation}
	\frac{1}{\sqrt{2}}\left(\ket{1}_a + \ket{1}_b\right) \rightarrow \frac{1}{\sqrt{1+\eta}}\left(\ket{1}_a + \sqrt{\eta}\ket{1}_b\right),
\end{equation}
assuming the loss occurs in mode $b$. This unbalanced state produces interference fringes with reduced visibility
\begin{equation}
	V(\eta)=\frac{2\sqrt{\eta}}{1+\eta}V_0.
\end{equation}

\subsubsection{Modeling the effect of background}

In a two-photon interferometer, optical background will introduce noise in the form of increased accidentals (beyond those resulting from SPDC and imperfect optics), assuming the background light is continuously distributed and uncorrelated in time. Let $\left\langle S_i \right\rangle$ be the mean total single-detector events when our background source is set to the $i^\text{th}$ brightness setting, and  $\Delta T$ be the temporal width of the detector coincidence window. The total number of accidental coincident detections $A_i$ is then
\begin{equation}
	A_i \approx \left\langle S_i \right\rangle^2 \Delta T.
\end{equation} 
If $A_i$ is the total number of accidental coincident detections corresponding to the $i^\text{th}$ brightness setting of the background source, the total measured coincidence fringe is 
\begin{equation}
	N_{c,i} = \frac{1}{2}\left[C_0(1+V_0\cos(\phi))+(A_i-A_0)\right],
\end{equation}
where $C_0$, $A_0$, and $V_0$ are the total number of coincident detections (with accidentals), accidentals, and interferometer visibility with no background, respectively, and $\phi$ is the interferometer phase. The corresponding interference visibility is then
\begin{equation}
	\begin{split}
		V_i &= \frac{\text{max}(N_{c,i})-\text{min}(N_{c,i})}{\text{max}(N_{c,i})+\text{min}(N_{c,i})}\\
		&= \frac{C_0}{C_0+(A_i - A_0)}V_{0}.
	\end{split}
	\label{eq:27}
\end{equation}
In a classical interferometer, the background will act linearly on the detected photon rate $N_i$,
\begin{equation}
	N_i = \frac{1}{2}\left[\left\langle S_0 \right\rangle(1\pm V_0\cos(\phi))+(\left\langle S_i \right\rangle - \left\langle S_0\right\rangle)\right],
\end{equation}
where $\left\langle S_0 \right\rangle$ is the mean total single-detector events in the presence of no background (i.e., the $0^\text{th}$ brightness setting). The corresponding interference visibility is
\begin{equation}
	V_i = \frac{\left\langle S_0 \right\rangle}{\left\langle S_i \right\rangle}V_0.
	\label{eq:29}
\end{equation}
Eqs. \ref{eq:27} and \ref{eq:29} may be rewritten in terms of the background, quantified as the fraction of total singles; see Supplementary Materials.

\subsection{Entanglement source}

Our entanglement source features a double beam displacer configuration to avoid potential effects from birefringent focusing. Each wavelength involved in the SPDC process (532~nm, 810~nm, and 1550~nm) has its own beam displacer assembly featuring a half-wave plate placed between two identical calcite beam displacers. For the pump, the first beam displacer (BD1) laterally displaces the $H$ component from the $V$ component. The optic axis of the half-wave plate is rotated by $45^\circ$ with respect to $H$ such that the $H$ and $V$ components are swapped prior to transmission through the second beam displacer (BD2). BD2 is rotated $180^\circ$ about the optical axis relative to BD1 such that the $H$ (originally $V$) component is displaced laterally in the opposite direction of the displacement introduced by BD1. The calcite crystals are cut such that the final separation between the $H$ and $V$ components is 4.8~mm, distributed symmetrically about the initial beam (see Supplementary Materials). The same process occurs in reverse for the 810~nm and 1550~nm wavelengths, where the spatially separated beams are recombined into a single spatial mode for each wavelength. 

The two MgO:PPLN SPDC crystals are 20~mm long with a \SI{7.4}{\micro\meter} poling period. These crystals are designed to down-convert ordinary-polarized \mbox{532-nm} photons into a pair of collinear, ordinary-polarized photons at 810~nm and 1550~nm via \mbox{Type-0} SPDC ($\text{o} \rightarrow \text{o} + \text{o}$). With \mbox{1-mm} square facets, they are mounted in a custom housing such that their centers are separated by 4.8~mm (see Supplementary Materials), matching the beam displacers. An oven maintains the crystals’ temperature at \mbox{$\sim$130 $^\circ\text{C}$}. 

The \mbox{4-mm} diameter beam from the \mbox{532-nm} continuous-wave pump laser is focused via a \mbox{400-mm} plano-convex lens for a \mbox{$\sim$65-\si{\micro\meter}} waist at the SPDC crystal. The down-converted photons are recollimated with plano-convex lenses with focal lengths of 300~mm and 125~mm for the \mbox{810-nm} and \mbox{1550-nm} photons, respectively. The focal lengths of the fiber-coupling aspheric lenses (15.29~mm for 810~nm, 18.4~mm for 1550~nm) correspond to beam waists of \mbox{$\sim$82 \si{\micro\meter}} and \mbox{$\sim$85 \si{\micro\meter}}, respectively, at the down-conversion crystals. After down-conversion, a long-pass filter assembly removes residual \mbox{532-nm} pump photons in the \mbox{810-nm} arm, and a \mbox{12-nm} bandwidth bandpass filter does the same in the \mbox{1550-nm} arm.

After the \mbox{810-nm} and \mbox{1550-nm} recombination beam displacers, the generated photons are in the maximally entangled state Eq.~\ref{eq:4}. The phase $\phi$ is determined by the phase of the beam-displacer interferometer, and is inconsequential for the main interferometer experiment, corresponding to a static phase offset in the measured interference fringes. However, to produce a more ideal entangled state, we use a tiltable quarter-wave plate in the \mbox{1550-nm} arm to minimize $\phi$. To further adjust the state, we use a trio of waveplates --- two quarter-wave plates and a half-wave plate --- in each arm to rotate the polarization of the photons. These waveplates serve a dual purpose: to perform the rotations necessary for a quantum state tomography of the source and to provide correction for the unitary transformations applied by the collection fibers (see Supplementary Materials).

The Supplementary Materials contains details regarding spectral, brightness, and heralding efficiency measurements for our source, as well as descriptions of our procedures for optimizing and characterizing the generated entanglement. 

\subsection{Dual-wavelength interferometer}

The PBSs and the 50:50 (nominally) non-polarizing beamsplitter are custom cube beamsplitters coated for 810 and 1550~nm. For stability, these were epoxied directly to stainless steel \mbox{one-inch}-diameter pedestal pillar posts. A detailed characterization of the beamsplitters used and their impact on the interference visibility is presented in the Supplementary Materials. Aside from the achromatic half-wave plate, the remainder of the optics within the interferometer are gold-coated mirrors, a cost-effective option for obtaining relatively high reflectivity for both 1550 and 810~nm. 

To reduce phase drift, the interferometer module is fully enclosed by plastic panels to minimize the effects of air flow in addition to being built atop a standard floating optics table. We also utilize commercially available thermally compensated stainless steel mounts for all kinematic tip-tilt mounts within the interferometer and the optics leading up to its input. These mounts are designed to have minimal thermally induced deflections; a temperature sensor within the enclosure measured an ambient temperature range of \mbox{$\sim$0.4 $^\circ\text{C}$} during a typical \mbox{24-hour} period. Elsewhere, stainless steel is utilized instead of aluminum where possible because of steel’s lower coefficient of thermal expansion, e.g., the kinematic mounts are mounted on one-inch-diameter steel pedestal pillar posts clamped directly onto the optical table. A characterization of the interferometer drift is given in the Supplementary Materials.

The single-axis piezoelectric nano-positioning stage integrated into the optical trombone system (path $a$) is specified for \mbox{30-\si{\micro\meter}} travel with a unidirectional repeatability of $\pm2$~nm.

\subsection{Spatial and temporal mode overlap} 

We performed knife-edge scans to ensure that the \mbox{1550-nm} and \mbox{810-nm} spatial modes overlapped in the interferometer (see Supplementary Materials) so that both wavelengths probe the same target. 

However, while the observed interference effect requires the detection of two photons, these photons do not need to be temporally overlapped, i.e., they need not arrive together. This is in stark contrast to conventional two-photon interference, in which --- without a quantum eraser for temporal which-path information \cite{pittman1996} --- the interference visibility scales with temporal mismatch as
\begin{equation}
	V = e^{-2\sigma^2\tau^2}.
\end{equation}
In energy-entangled two-photon interference, the two photons technically interfere directly, but that effect is suppressed by many orders of magnitude because of the frequency difference between the photons; this is the exponential envelope in Eq.~\ref{eq:2}. The dominant effect is instead equivalent to two individual single-photon interferometers, each with a single photon interfering with itself. These interferometers are entangled with one another, which creates the observed beat note. The two interference processes only rely on the photon of each wavelength being coherent with itself; interference will be observed as long as the path-length difference of the interferometer is within a coherence length for each individual photon, determined by their individual bandwidths.

\subsection{End-to-end system calibration and performance}

To prepare our apparatus for interferometric measurements, the source, interferometer, and detection modules undergo a six-step calibration protocol (see Supplementary Materials) to optimize the entangled probe state and configure the interferometer for maximum measurement sensitivity. Typical system performance as a function of pumping power (after calibration) is reported in the Supplementary Materials.

\subsection{Sample fabrication}

We fabricated both the test and calibration samples using identical methodology. We used double-side polished C-plane sapphire wafers (\mbox{76.2-mm} diameter, \mbox{0.5-mm} thickness) as the substrate (see Supplementary Materials for wafer specifications). We prepared wafers with a basic solvent clean and applied parallel strips of Kapton tape (\mbox{12.7-mm} wide) across the wafer in one direction with a typical separation of \mbox{$\sim$12.7 mm}. We applied force to the tape edges to maximize adhesion.

We then coated the prepared samples via electron-beam physical vapor deposition with a nickel target ($99.995\%$ purity). The deposition time for the nominal requested film thickness was determined automatically based on pre-calibrated deposition rate data. After coating, we removed the Kapton tape, leaving strips of uncoated and coated regions with well-defined edges (see Supplementary Materials).

We cleaved a small, coated piece from the test sample for inspection under an atomic force microscope to verify film smoothness and uniformity. We measured a film roughness of \mbox{$\sim$100 pm}, and observed no gaps or island-like features.

\subsection{Sample mounting and alignment}

We affix the sample under study to a kinematic tip and tilt mount and attach the stage to the Sample Positioning System (SPS) located in path $b$ of the interferometer immediately after the PBS. The SPS features motorized horizontal translation with 25~mm of travel, a typical positioning accuracy of \mbox{$\pm$2.2 \si{\micro\meter}}, and a maximum speed of 5~mm per second (manufacturer specifications). Vertical and longitudinal translations are achieved via manual micrometers with 25~mm of travel. The horizontal and vertical axes of the SPS are approximately orthogonal to the probe propagation direction. We then align the sample such that the sample surface is approximately normal to the probe by adjusting the tip and tilt of the sample mount while monitoring the signal from the Sample Targeting System. Once the sample is mounted and aligned, an automated protocol optimizes the interferometer phase for maximum measurement sensitivity. The Supplementary Materials details the protocols for the sample alignment, interferometer phase optimization, and sample measurement.

\begin{acknowledgments}
	We thank Kathy A. Walsh and Julio A. N. T. Soares for their assistance with sample characterization; Brian Williams, Joseph Lukens, Ryan Bennink, and Warren Grice at Oak Ridge National Laboratory for the initial development of the entanglement source; and James N. Eckstein, Tim Stelzer, Elizabeth A. Goldschmidt, Virginia O. Lorenz, Sanjukta Kundu, and Stanisław Kurdziałek for helpful discussions. This work was carried out in part in the Materials Research Laboratory Central Research Facilities, University of Illinois.
	
	This material is based upon work supported by the U.S. Air Force under Grant No. FA9550-21-1-0059 (P.G.K., C.P.L., S.J.J., M.V., K.A.M.), by the U.S. Department of Energy, Office of Science, Office of Biological and Environmental Research under Award Number DE-SC0023167 (P.G.K., C.P.L., S.J.J., S.I.B., S.S.), and by the National Science Foundation Graduate Research Fellowship Program under Grant No. DGE 21-46756 (C.P.L.). Portions of this work have been funded by the U.S. Government. Any opinions, findings, and conclusions or recommendations expressed in this material are those of the authors and do not necessarily reflect the views of the U.S. Air Force, the U.S. Department of Energy, the National Science Foundation, or the U.S. Government.                      
\end{acknowledgments}

\section*{AUTHOR CONTRIBUTIONS}
Conceptualization: P.G.K., C.P.L., S.J.J., and K.A.M.; Data curation: C.P.L.; Formal analysis: C.P.L., S.J.J., and S.S.; Funding acquisition: P.G.K., C.P.L., S.J.J., and K.A.M.; Investigation: C.P.L., S.J.J., S.S., M.V., and K.A.M.; Methodology: C.P.L. and S.J.J.; Project administration: C.P.L.; Resources: S.S.; Software: C.P.L., S.J.J., and M.V.; Supervision: P.G.K. and S.I.B.; Validation: C.P.L. and S.J.J.; Visualization: C.P.L. and S.J.J.; Writing–original draft: C.P.L. and S.J.J.; Writing–review and editing: C.P.L., S.J.J., S.S., M.V., P.G.K., and S.I.B.
\newline

\section*{COMPETING INTERESTS}
S.J.J., P.G.K., C.P.L., and K.A.M. are inventors on U.S. patent application no. 17/948,308 submitted by the Board of Trustees of the University of Illinois that covers precision quantum-interference-based non-local contactless measurement. All other authors declare they have no competing interests. 

\section*{DATA AVAILABILITY}
All data not already available in the manuscript or the Supplementary Materials may be accessed on Dryad (\url{https://doi.org/10.5061/dryad.1rn8pk154}).

\onecolumngrid
\clearpage

\setcounter{section}{0}
\renewcommand{\thesection}{S\arabic{section}}

\setcounter{equation}{0}
\renewcommand{\theequation}{S\arabic{equation}}

\setcounter{figure}{0}
\renewcommand{\thefigure}{S\arabic{figure}}

\setcounter{table}{0}
\renewcommand{\thetable}{S\arabic{table}}

\newcommand{\toclva}[2]{\item \textbf{#2}\leaders\hbox to 0.5em{\hss\textbf{.}\hss}\hfill \textbf{\pageref{#1}}}
\newcommand{\toclvb}[2]{\item \textit{#2}\leaders\hbox to 0.5em{\hss.\hss}\hfill \textbf{\pageref{#1}}}
\newcommand{\toclvc}[2]{\item #2\leaders\hbox to 0.5em{\hss.\hss}\hfill \textbf{\pageref{#1}}}

\begin{center}
	{\Large  \textbf{Supplementary Materials for}}
	\vspace{0.25 in}
\end{center}
\begin{center}
	{\large Fast quantum interferometry at the nanometer and attosecond scales\\with energy-entangled photons}
	\vspace{0.25 in}
\end{center}

\noindent \textbf{\large Table of contents}

\begin{list}{}{\leftmargin=0in \labelwidth=0em \itemsep=0.5em}
	\toclva{sec:sup_text}{Supplementary text}
	
	\begin{list}{}{\leftmargin=0.25in \labelwidth=0em \itemsep=0.5em}
		
		\toclvb{sec:theory}{Energy-entangled two-photon interference theory}
		\begin{list}{}{\leftmargin=0.25in \labelwidth=0em \itemsep=0em}
			\toclvc{sec:qtheory}{Quantum theory}
			\toclvc{sec:qfi}{Quantum Fisher information}
			\toclvc{sec:cfi}{Classical Fisher information}
			\toclvc{sec:saturation}{Experimental saturation of the Cram\'er–Rao bound}
			\toclvc{sec:modelloss}{Modeling the effect of loss}
			\toclvc{sec:modelbg}{Modeling the effect of background}
		\end{list}
		
		\toclvb{sec:mm}{Materials and methods}
		\begin{list}{}{\leftmargin=0.25in \labelwidth=0em \itemsep=0em}
			\toclvc{sec:smd}{Source module: Design}
			\toclvc{sec:smc}{Source module: Characterization}
			\toclvc{sec:img}{Interferometer module: General details}
			\toclvc{sec:imv}{Interferometer module: Interference visibility}
			\toclvc{sec:dm}{Detection module}
			\toclvc{sec:smo}{Spatial mode overlap}
			\toclvc{sec:eescp}{End-to-end system calibration protocol}
			\toclvc{sec:sp}{System performance}
			\toclvc{sec:ep}{Note on error propagation}
		\end{list}
		
		\toclvb{sec:samplem}{Sample measurements}
		\begin{list}{}{\leftmargin=0.25in \labelwidth=0em \itemsep=0em}
			\toclvc{sec:smp}{Sample measurement protocol}
			\toclvc{sec:sf}{Thin-film sample fabrication}
			\toclvc{sec:msm}{Model for thin-film sample measurements}
			\toclvc{sec:ncs}{Refractive index: Calibration sample}
			\toclvc{sec:nqp}{Refractive index: Quantum probe}
			\toclvc{sec:ncp}{Refractive index: Classical probe}
			\toclvc{sec:imf}{Independent measurements of test sample film thickness}
		\end{list}
		
		\toclvb{sec:aim}{Alternative interference modes}
		\begin{list}{}{\leftmargin=0.25in \labelwidth=0em \itemsep=0em}
			\toclvc{sec:cfb}{Classical frequency beating}
			\toclvc{sec:qfb}{Quantum sum-frequency beating}
		\end{list}
		
	\end{list}
	
	\toclva{sec:figures}{Figures}
	
	\begin{list}{}{\leftmargin=0.5in \labelwidth=0em \itemsep=0em}
		\toclvc{fig:figs1}{Fig. S1. Simulated optical loss and background}
		\toclvc{fig:figs2}{Fig. S2. Schematic of non-degenerate polarization entanglement source}
		\toclvc{fig:figs3}{Fig. S3. Simplified schematic of non-degenerate polarization entanglement source}
		\toclvc{fig:figs4}{Fig. S4. Entanglement source pump and signal spectra}
		\toclvc{fig:figs5}{Fig. S5. Polarization-entangled state density matrix}
		\toclvc{fig:figs6}{Fig. S6. Interferometer drift}
		\toclvc{fig:figs7}{Fig. S7. Effects of system imperfections on fringe visibility}
		\toclvc{fig:figs8}{Fig. S8. Spatial overlap of the 1550-nm and 810-nm modes}
		\toclvc{fig:figs9}{Fig. S9. Interference fringe envelope}
		\toclvc{fig:figs10}{Fig. S10. System performance as a function of entanglement source pumping power}
		\toclvc{fig:figs11}{Fig. S11. Optical image of nickel thin film sample (5-nm thickness)}
		\toclvc{fig:figs12}{Fig. S12. Illustration of model for thin-film measurements}
		\toclvc{fig:figs13}{Fig. S13. Probe beam characterization}
		\toclvc{fig:figs14}{Fig. S14. Quantum interferometer calibration measurement}
		\toclvc{fig:figs15}{Fig. S15. Test sample thickness measurements via scanning-stylus profilometry}
		\toclvc{fig:figs16}{Fig. S16. Test sample thickness measurements via 3D optical profilometry}
		\toclvc{fig:figs17}{Fig. S17. Classical frequency beating}
		\toclvc{fig:figs18}{Fig. S18. Quantum sum-frequency beating}
	\end{list}
	
	\toclva{sec:tables}{Tables}
	\begin{list}{}{\leftmargin=0.5in \labelwidth=0em \itemsep=0em}
		\toclvc{tab:tabs1}{Table S1. Detector specifications}
		\toclvc{tab:tabs2}{Table S2. Interferometer module free-space transmission}
		\toclvc{tab:tabs3}{Table S3. Detection module fiber-coupling efficiencies}
		\toclvc{tab:tabs4}{Table S4. Coincident detection jitter}
		\toclvc{tab:tabs5}{Table S5. State tomography angles}
		\toclvc{tab:tabs6}{Table S6. Key sapphire wafer specifications}
	\end{list}
	
\end{list}

\clearpage

\counterwithin{equation}{section}
\renewcommand{\thesection}{S}
\renewcommand{\theequation}{\thesection\arabic{equation}}

\phantomsection
\label{sec:sup_text}
\begin{center}
	{\large \textbf{Supplementary text}}
	\vspace{0.125 in}
\end{center}

\section*{Energy-entangled two-photon interference theory}
\label{sec:theory}

\subsection*{Quantum theory}
\label{sec:qtheory}

Treatments similar to the one provided in this section can be found in Refs. \cite{chen2019} and \cite{branczyk2017}. In general, a photon which has some finite spectral bandwidth can be described as 
\begin{equation}
	\ket{\psi} = \int d\omega_1f(\omega_1)a^\dagger(\omega_1)\ket{0},
\end{equation}
where $f(\omega_1)$ describes the frequency spread of the photon. For a Gaussian spread,
\begin{equation}
	f(\omega_1) = \frac{1}{(2\pi\sigma^2)^{1/4}}e^{-\frac{(\omega_1-\omega_1^0)^2}{4\sigma^2}}.
\end{equation} 
Here, $\omega^0_1$ is the central frequency of the photon, and the prefactor assures normalization of $\ket{\psi}$ (recalling that $\sigma$ is the Gaussian half-bandwidth). A general pure two-photon state can be written as
\begin{equation}
	\ket{\psi} =  \int d\omega_1d\omega_2f(\omega_1,\omega_2)a^\dagger(\omega_1)a^\dagger(\omega_2)\ket{0},
\end{equation}
where $f(\omega_1,\omega_2)$ now describes the spectral information of both photons, including any potential frequency entanglement. To see how this state propagates under two-photon interference, we begin with single photons of frequencies $\omega_1$ and $\omega_2$, respectively, in spatial modes 1 and 2 and follow the two creation operators $a^\dagger(\omega_1)$, $a^\dagger(\omega_2)$. The photons are incident on a 50:50 beamsplitter, such that 
\begin{equation}
	\begin{split}
		a_1^\dagger(\omega_1) &\rightarrow \frac{1}{\sqrt{2}}\left(a_1^\dagger(\omega_1)+ia_2^\dagger(\omega_1)\right)\\
		a_2^\dagger\left(\omega_2\right) &\rightarrow \frac{1}{\sqrt{2}}\left(a_2^\dagger(\omega_2)+ia_1^\dagger(\omega_2)\right), 
	\end{split}
	\label{eq:s4}
\end{equation}
where we have utilized a convention in which the spatial modes after the beamsplitter correspond to the transmitted mode of the input state. Then, 
\begin{equation}
	\begin{split}
		a_1^\dagger(\omega_1)a_2^\dagger(\omega_2) \rightarrow &\frac{1}{2}\left(a_1^\dagger(\omega_1)+ia_2^\dagger(\omega_1)\right)\left(a_2^\dagger(\omega_2)+ia_1^\dagger(\omega_2)\right)\\
		= &\frac{1}{2}\left[a_1^\dagger(\omega_1)a_2^\dagger(\omega_2) + i\left(a_1^\dagger(\omega_1)a_1^\dagger(\omega_2)+a_2^\dagger(\omega_1)a_2^\dagger(\omega_2)\right)- a_2^\dagger\left(\omega_1\right)a_1^\dagger\left(\omega_2\right)\right]\\
		= &\frac{1}{2}\left[a_1^\dagger(\omega_1)a_2^\dagger(\omega_2) - a_1^\dagger(\omega_2)a_2^\dagger(\omega_1) \right],
	\end{split}
\end{equation}
where in the last line we have removed terms which will not contribute coincidences, and we have reordered the terms such that $a_1^\dagger$ is always first. Suppose we have a delay in spatial mode 1, 
such that
\begin{equation}
	\ket{\psi}\rightarrow\ket{\psi(\tau)} = \int d\omega_1d\omega_2f(\omega_1,\omega_2)e^{i\omega_1\tau}a_1^\dagger(\omega_1)a_2^\dagger(\omega_2)\ket{0}. 
\end{equation}
After the beamsplitter we are left with the state 
\begin{equation}
	\ket{\psi(\tau)} = \frac{1}{2}\int d\omega_1d\omega_2f(\omega_1,\omega_2)e^{i\omega_1\tau}\left[a_1^\dagger(\omega_1)a_2^\dagger(\omega_2) - a_1^\dagger(\omega_2)a_2^\dagger(\omega_1)\right]\ket{0}. 
\end{equation}
The photons are assumed to be detected via a pair of detectors with flat frequency response:
\begin{equation}
	\hat{S} = \int d\omega d\omega'a_1^\dagger(\omega)a_2^\dagger(\omega')\ket{0}\bra{0}a_1(\omega)a_2(\omega').
\end{equation}
This projects the photon in spatial mode 1 (2) onto the frequency $\omega$ ($\omega'$). We note that this operator could include a frequency-dependent detector efficiency or a spectral filter by including some weighting function $\eta(\omega)$. Because of the properties of creation and annihilation operators, 
\begin{equation}
	\bra{0}a_1(\omega)a_2(\omega')a_1^\dagger(\omega_1)a_2^\dagger(\omega_2)\ket{0} = \delta(\omega-\omega_1)\delta(\omega'-\omega_2),
\end{equation}
and the probability of coincidence is then 
\begin{equation}
	\begin{split}
		P_C &= \bra{\psi(\tau)}\hat{S}\ket{\psi(\tau)}\\
		&= \frac{1}{4} \int d\omega_1d\omega_2d\omega_1^@d\omega_2^@d\omega d\omega'f(\omega_1,\omega_2)f^*\left(\omega_1^@,\omega_2^@\right)e^{i\left(\omega_1-\omega_1^@\right)\tau}\\
		&\phantom{=} \times\left[\delta\left(\omega-\omega_1^@\right)\delta\left(\omega'-\omega_2^@\right) - \delta\left(\omega-\omega_2^@\right)\delta\left(\omega'-\omega_1^@\right) \right] \\
		&\phantom{=} \times\left[\delta(\omega-\omega_1)\delta(\omega'-\omega_2) - \delta(\omega-\omega_2)\delta(\omega'-\omega_1) \right],
	\end{split}
\end{equation}
where we have used $\omega^@$ to distinguish the frequencies of $\bra{\psi(\tau)}$ and $\ket{\psi(\tau)}$. Distributing the two delta function products and computing the $\omega$, $\omega'$ integrals, we are left with
\begin{equation}
	\begin{split}
		P_C &= \frac{1}{2}\int d\omega_1d\omega_2d\omega_1^@d\omega_2^@f(\omega_1,\omega_2)f^*\left(\omega_1^@,\omega_2^@\right)e^{i\left(\omega_1-\omega_1^@\right)\tau} \\
		&\phantom{=} \times\left[\delta\left(\omega_1-\omega_1^@\right)\delta\left(\omega_2-\omega_2^@\right) - \delta\left(\omega_1-\omega_2^@\right)\delta\left(\omega_2-\omega_1^@\right) \right].
	\end{split}
\end{equation}
Doing the $\omega^@$ integrals, 
\begin{equation}
	P_C = \frac{1}{2}\int d\omega_1d\omega_2\left|f(\omega_1,\omega_2)\right|^2-\frac{1}{2}\int d\omega_1d\omega_2f(\omega_1,\omega_2)f^*(\omega_2,\omega_1)e^{i(\omega_1-\omega_2)\tau}.
\end{equation}
Since for a normalized state we have 
\begin{equation}
	\int d\omega_1d\omega_2\left|f(\omega_1,\omega_2)\right|^2 = \braket{\psi|\psi} = 1,
\end{equation}
it then follows that
\begin{equation}
	P_C = \frac{1}{2}-\frac{1}{2}\int d\omega_1d\omega_2f(\omega_1,\omega_2)f^*(\omega_2,\omega_1)e^{i(\omega_1-\omega_2)\tau}.
	\label{eq:s14}
\end{equation}
For a pair of photons produced by continuous-wave SPDC with pump frequency $\omega_p$, the perfect spectral correlations $\omega_1 + \omega_2 = \omega_p$ allow us to write the state $\ket{\psi_{SPDC}}$ as
\begin{equation}
	\begin{split}
		\ket{\psi_{SPDC}} &= \frac{1}{(2\pi\sigma^2)^{1/4}}\int d\Omega e^{-\frac{\Omega^2}{4\sigma^2}}a_1^\dagger\left(\omega_1^0+\Omega\right)a_2^\dagger\left(\omega_2^0-\Omega\right)\ket{0}\\
		&= \frac{1}{(2\pi\sigma^2)^{1/4}}\int d\omega_1d\omega_2 e^{-\frac{\left(\omega_1-\omega_1^0\right)^2}{4\sigma^2}}a_1^\dagger(\omega_1)a_2^\dagger(\omega_2)\delta\left(\left(\omega_1-\omega_1^0\right)+\left(\omega_2-\omega_2^0\right)\right)\ket{0},
	\end{split}
	\label{eq:s15}
\end{equation}
where we have defined
\begin{equation}
	\begin{split}
		\Omega &\equiv \omega_1-\omega_1^0 \\
		&= -\left(\omega_2-\omega_2^0\right),
	\end{split}
	\label{eq:s16}
\end{equation}
and the second form is used so that we may write the frequency spectrum as  
\begin{equation}
	f_{SPDC}(\omega_1,\omega_2) = \frac{1}{(2\pi\sigma^2)^{1/4}}e^{-\frac{\left(\omega_1-\omega_1^0\right)^2}{4\sigma^2}}\delta\left(\left(\omega_1-\omega_1^0\right)+\left(\omega_2-\omega_2^0\right)\right).
	\label{eq:s17}
\end{equation}
Plugging Eq.~\ref{eq:s17} into Eq.~\ref{eq:s14}, we focus on the integral in the second term: 
\begin{equation}
	\begin{split}
		\int d\omega_1d\omega_2f_{SPDC}(\omega_1,\omega_2)f_{SPDC}^*(\omega_2,\omega_1)e^{i(\omega_1-\omega_2)\tau} &= \frac{1}{(2\pi\sigma^2)^{1/2}}\int d\omega_1d\omega_2e^{-\frac{\left(\omega_1-\omega_1^0\right)^2}{4\sigma^2}}e^{-\frac{\left(\omega_2-\omega_1^0\right)^2}{4\sigma^2}}e^{i(\omega_1-\omega_2)\tau}\\
		&\phantom{=} \times \delta\left(\left(\omega_1-\omega_1^0\right)+\left(\omega_2-\omega_2^0\right)\right).
		\label{eq:s18}
	\end{split}
\end{equation}
Utilizing the delta function to complete the $\omega_2$ integral, we have
\begin{equation}
	\frac{1}{(2\pi\sigma^2)^{1/2}}\int d\omega_1e^{-\frac{\left(\omega_1-\omega_1^0\right)^2}{4\sigma^2}}e^{-\frac{\left(\omega_1-\omega_2^0\right)^2}{4\sigma^2}}e^{i\left(2\omega_1-\omega_1^0-\omega_2^0\right)\tau}.
\end{equation}
Focusing for a moment only on the exponents, 
\begin{equation}
	\begin{split}
		&-\frac{1}{4\sigma^2}\left[\left(\omega_1-\omega_1^0\right)^2+\left(\omega_1-\omega_2^0\right)^2\right]+i\left(2\omega_1-\omega_1^0-\omega^2_0\right)\tau\\
		&\phantom{=}=-\frac{1}{4\sigma^2}\left[2(\omega_1)^2-2\omega_1\omega_1^0+\left(\omega_1^0\right)^2-2\omega_1\omega_2^0+\left(\omega_2^0\right)^2\right]+i\left(2\omega_1-\omega_1^0-\omega^0_2\right)\tau\\
		&\phantom{=}=-\frac{1}{2\sigma^2}\left[(\omega_1)^2-\omega_1\left(\omega_1^0+\omega_2^0+4i\sigma^2\tau\right)\right]-\frac{1}{4\sigma^2}\left[\left(\omega_1^0\right)^2+\left(\omega_2^0\right)^2\right]-i\left(\omega_1^0+\omega^0_2\right)\tau\\
		&\phantom{=}=-\frac{1}{2\sigma^2}\left[\omega_1-\frac{1}{2}\left(\omega_1^0+\omega_2^0+4i\sigma^2\tau\right)\right]^2-\frac{1}{4\sigma^2}\left[\left(\omega_1^0\right)^2+\left(\omega_2^0\right)^2\right]+\frac{1}{8\sigma^2}\left(\omega_1^0+\omega_2^0+4i\sigma^2\tau\right)^2-i\left(\omega_1^0+\omega^0_2\right)\tau\\
		&\phantom{=}=-\frac{1}{2\sigma^2}\left[\omega_1-\frac{1}{2}\left(\omega_1^0+\omega_2^0+4i\sigma^2\tau\right)\right]^2-\frac{1}{8\sigma^2}\left[\omega_1^0-\omega_2^0\right]^2-2\sigma^2\tau^2.
	\end{split}
\end{equation}
Defining $\Delta\omega \equiv \omega_1^0 - \omega_2^0$ , we finally have
\begin{equation}
	\begin{split}
		\frac{1}{(2\pi\sigma^2)^{1/2}}\int d\omega_1e^{-\frac{\left(\omega_1-\omega_1^0\right)^2}{4\sigma^2}}e^{-\frac{\left(\omega_1-\omega_2^0\right)^2}{4\sigma^2}}e^{i\left(2\omega_1-\omega_1^0-\omega_2^0\right)\tau} &= \frac{e^{-\frac{(\Delta\omega)^2}{8\sigma^2}}e^{-2\sigma^2\tau^2}}{(2\pi\sigma^2)^{1/2}}\int d\omega_1 e^{-\frac{1}{2\sigma^2}\left[\omega_1-\frac{1}{2}\left(\omega_1^0+\omega_2^0+4i\sigma^2\tau\right)\right]^2}\\
		&= e^{-\frac{(\Delta\omega)^2}{8\sigma^2}}e^{-2\sigma^2\tau^2},
	\end{split}
\end{equation}
where we have used the normalization of the spectral spread to replace the integral with $(2\pi\sigma^2)^{1/2}$. Thus,
\begin{equation}
	P_C = \frac{1}{2}\left[1-\beta e^{-2\sigma^2\tau^2} \right],
\end{equation}
where we have defined
\begin{equation}
	\beta \equiv e^{-\frac{(\Delta\omega)^2}{8\sigma^2}}.
\end{equation}
If the photons are energy entangled, we have, in terms of frequency,
\begin{equation}
	\begin{split}
		f_{ent}(\omega_1,\omega_2) &= \frac{1}{\sqrt{1+\beta}}\frac{1}{\sqrt{2}}\left(f_{SPDC}(\omega_1,\omega_2)+f_{SPDC}(\omega_2,\omega_1)\right)\\
		&= \frac{1}{\sqrt{1+\beta}}\frac{1}{\sqrt{2}}\frac{1}{(2\pi\sigma^2)^{1/4}}\left[e^{-\frac{\left(\omega_1-\omega_1^0\right)^2}{4\sigma^2}}+e^{-\frac{\left(\omega_2-\omega_1^0\right)^2}{4\sigma^2}}\right]\delta\left(\left(\omega_1-\omega_1^0\right)+\left(\omega_2-\omega_2^0\right)\right).
	\end{split}
\end{equation}
Going forward we will ignore the normalization factor $(1+\beta)^{-1}$, since $\beta \approx 0$ for experimental values of $\Delta\omega$, $\sigma$. Utilizing the facts that $f_{ent}(\omega_2,\omega_1)=f_{ent}(\omega_1,\omega_2)$ and $f_{ent}^*(\omega_1,\omega_2)=f_{ent}(\omega_1,\omega_2)$, we plug this into Eq.~\ref{eq:s14}, leaving us with the second-term integral
\begin{equation}
	\begin{split}
		\int d\omega_1d\omega_2f_{ent}(\omega_1,\omega_2)f_{ent}^*(\omega_2,\omega_1)e^{i(\omega_1-\omega_2)\tau} &= \frac{1}{2}\int d\omega_1d\omega_2\left(f_{SPDC}(\omega_1,\omega_2)+f_{SPDC}(\omega_2,\omega_1)\right)^2e^{i(\omega_1-\omega_2)\tau}\\
		&= \frac{1}{2}\int d\omega_1d\omega_2\Big(\left|f_{SPDC}(\omega_1,\omega_2)\right|^2+\left|f_{SPDC}(\omega_2,\omega_1)\right|^2\\
		&\phantom{=} +2f_{SPDC}(\omega_1,\omega_2)f_{SPDC}(\omega_2,\omega_1)\Big)e^{i(\omega_1-\omega_2)\tau}.
	\end{split}
	\label{eq:s25}
\end{equation} 
We can identify the integral $\int d\omega_1d\omega_2f_{SPDC}(\omega_1,\omega_2)f_{SPDC}(\omega_2,\omega_1)e^{i(\omega_1-\omega_2)\tau}$ as equivalent to the non-degenerate SPDC seen above in Eq.~\ref{eq:s18}, and so that term is equal to
\begin{equation}
	\begin{split}
		&\beta e^{-2\sigma^2\tau^2},\\
		&\beta \equiv e^{-\frac{(\Delta\omega)^2}{8\sigma^2}}.
	\end{split}
\end{equation}
Looking at the first two terms of Eq.~\ref{eq:s25}, if we relabel $\omega_1$ and $\omega_2$ in the second term we see
\begin{equation}
	\frac{1}{2}\int d\omega_1d\omega_2\left|f_{SPDC}(\omega_2,\omega_1)\right|^2e^{i(\omega_1-\omega_2)\tau} \rightarrow \frac{1}{2}\int d\omega_1d\omega_2\left|f_{SPDC}(\omega_1,\omega_2)\right|^2e^{-i(\omega_1-\omega_2)\tau},
\end{equation}
such that we only need to compute 
\begin{equation}
	\begin{split}
		\frac{1}{2}\int d\omega_1d\omega_2\left|f_{SPDC}(\omega_1,\omega_2)\right|^2e^{\pm i(\omega_1-\omega_2)\tau} &= \frac{1}{2}\frac{1}{(2\pi\sigma^2)^{1/2}}\int d\omega_1d\omega_2e^{-\frac{\left(\omega_1-\omega_1^0\right)^2}{4\sigma^2}}e^{-\frac{\left(\omega_2-\omega_2^0\right)^2}{4\sigma^2}}e^{\pm i(\omega_1-\omega_2)\tau}\\
		&\phantom{=} \times \delta\left(\left(\omega_1-\omega_1^0\right)+\left(\omega_2-\omega_2^0\right)\right)\\
		&= \frac{1}{2}\frac{1}{(2\pi\sigma^2)^{1/2}}\int d\omega_1e^{-\frac{\left(\omega_1-\omega_1^0\right)^2}{2\sigma^2}}e^{\pm i\left(2\omega_1-\omega_1^0-\omega_2^0\right)\tau}.
	\end{split}
\end{equation}
Focusing on exponents as before, we have
\begin{equation}
	\begin{split}
		-\frac{\left(\omega_1-\omega_1^0\right)^2}{2\sigma^2} \pm i\left(2\omega_1-\omega_1^0-\omega_2^0\right)\tau &= -\frac{\left(\omega_1^2-2\omega_1\omega_1^0 \mp 4i\omega_1\sigma^2\tau \right)}{2\sigma^2} - \frac{\left(\omega_1^0\right)^2}{2\sigma^2} \mp i\left(\omega_1^0 + \omega_2^0\right)\tau \\
		&= -\frac{\left(\omega_1-\left(\omega_1^0 \pm 2i\sigma^2\tau \right)\right)^2}{2\sigma^2} +\frac{1}{2\sigma^2}\left(\omega_1^0 \pm 2i\sigma^2\tau\right)^2 - \frac{\left(\omega_1^0\right)^2}{2\sigma^2} \mp i\left(\omega_1^0 + \omega_2^0\right)\tau  \\
		&= -\frac{\left(\omega_1-\left(\omega_1^0 \pm 2i\sigma^2\tau \right)\right)^2}{2\sigma^2} - 2\sigma^2\tau^2 \pm i(\Delta\omega)\tau,
	\end{split}
\end{equation}
and so
\begin{equation}
	\begin{split}
		\frac{1}{2}\frac{1}{(2\pi\sigma^2)^{1/2}}\int d\omega_1e^{-\frac{\left(\omega_1-\omega_1^0\right)^2}{2\sigma^2}}e^{\pm i\left(2\omega_1-\omega_1^0-\omega_2^0\right)\tau} &= \frac{1}{2}\frac{1}{(2\pi\sigma^2)^{1/2}}e^{-2\sigma^2\tau^2}e^{\pm i (\Delta\omega)\tau}\int d\omega_1e^{-\frac{\left(\omega_1-\left(\omega_1^0 \pm 2i\sigma^2\tau \right)\right)^2}{2\sigma^2}}\\
		&= \frac{1}{2}e^{-2\sigma^2\tau^2}e^{\pm i (\Delta\omega)\tau}.
	\end{split}
\end{equation}
We then have,
\begin{equation}
	\begin{split}
		\frac{1}{2}\int d\omega_1d\omega_2\left(\left|f_{SPDC}(\omega_1,\omega_2)\right|^2+\left|f_{SPDC}(\omega_2,\omega_1)\right|^2\right)e^{i(\omega_1-\omega_2)\tau} &= \frac{e^{-2\sigma^2\tau^2}}{2}\left(e^{i(\Delta\omega)\tau}+e^{-i(\Delta\omega)\tau}\right)\\
		&= \cos((\Delta\omega)\tau)e^{-2\sigma^2\tau^2}.
	\end{split}
\end{equation}
Combining everything at the end,
\begin{equation}
	P_C = \frac{1}{2}\left[1-\cos((\Delta\omega)\tau)e^{-2\sigma^2\tau^2} - \beta e^{-2\sigma^2\tau^2}\right],
\end{equation}
where $\beta = e^{-\frac{(\Delta\omega)^2}{8\sigma^2}}$ as before. Finally, taking the limit $\beta \rightarrow 0$, we are left with Eq.~\ref{eq:2} in the main text:
\begin{equation}
	P_C = \frac{1}{2}\left[1-\cos((\Delta\omega)\tau)e^{-2\sigma^2\tau^2}\right].
	\label{eq:s33}
\end{equation}

\subsection*{Quantum Fisher information}
\label{sec:qfi}

In evaluating the quantum Fisher information (Eq.~\ref{eq:6}) for the case of energy-entangled two-photon interference, the probe state $\ket{\psi(\tau)}$ (the state after acquiring the interferometric phase shift $\omega\tau$ but before reaching the final interfering beamsplitter) is (using the second form of Eq.~\ref{eq:s15})
\begin{equation}
	\ket{\psi} = \frac{1}{\sqrt{2(1+\beta)}}\int d\Omega f(\Omega)\left[e^{-i\left(\omega_1^0+\Omega\right)\tau}a_1^\dagger\left(\omega_1^0+\Omega\right)a_2^\dagger\left(\omega_2^0-\Omega\right)+e^{-i\left(\omega_2^0+\Omega\right)\tau}a_1^\dagger\left(\omega_2^0+\Omega\right)a_2^\dagger\left(\omega_1^0-\Omega\right)\right]\ket{0},
\end{equation}
where the two photons have center frequencies $\omega_1^0$, $\omega_2^0$, $\beta = e^{-\frac{\left(\omega_1^0-\omega_2^0\right)^2}{8\sigma^2}}$ fixes the normalization of the entangled state, $\Omega$ is defined as in Eq.~\ref{eq:s16}, and
\begin{equation}
	f(\Omega) = e^{-\frac{\Omega^2}{4\sigma^2}}
\end{equation}
is the normalized SPDC spectral amplitude. Note that we have used $f(-\Omega) = f(\Omega)$ to change 
the variable of integration $\Omega \rightarrow -\Omega$ in the second term. The first term in Eq.~\ref{eq:6} requires
\begin{equation}
	\begin{split}
		\left|\frac{\partial\psi(\tau)}{\partial\tau}\right\rangle &= \frac{-i}{\sqrt{2(1+\beta)}}\int d\Omega 	f(\Omega)\left[\left(\omega_1^0+\Omega\right)e^{-i\left(\omega_1^0+\Omega\right)\tau}a_1^\dagger\left(\omega_1^0+\Omega\right)a_2^\dagger\left(\omega_2^0-\Omega\right)\right.\\
		&\phantom{=}\left.+\left(\omega_2^0+\Omega\right)e^{-i\left(\omega_2^0+\Omega\right)\tau}a_1^\dagger\left(\omega_2^0+\Omega\right)a_2^\dagger\left(\omega_1^0-\Omega\right)\right],
	\end{split}
\end{equation}
and gives
\begin{equation}
	\begin{split}
		\left\langle\frac{\partial\psi(\tau)}{\partial\tau}\middle|\frac{\partial\psi(\tau)}{\partial\tau}\right\rangle &= \frac{1}{2(1+\beta)}\int d\Omega d\Omega'f(\Omega)f(\Omega')[\kappa_1+\kappa_2+\kappa_3+\kappa_4]\\
		&=\frac{1}{2}\left(\left(\omega_1^0\right)^2+\left(\omega_2^0\right)^2 + 2\sigma^2\right),
	\end{split}
\end{equation}
where
\begin{equation}
	\begin{split}
		\kappa_1 &\equiv \left(\omega_1^0+\Omega'\right)\left(\omega_1^0+\Omega\right)e^{i\left(\omega_1^0+\Omega'\right)\tau}e^{-i\left(\omega_1^0+\Omega\right)\tau}a_1\left(\omega_1^0+\Omega'\right)a_2\left(\omega_2^0-\Omega'\right)a_1^\dagger\left(\omega_1^0+\Omega\right)a_2^\dagger\left(\omega_2^0-\Omega\right)\\
		\kappa_2 &\equiv \left(\omega_1^0+\Omega'\right)\left(\omega_2^0+\Omega\right)e^{i\left(\omega_1^0+\Omega'\right)\tau}e^{-i\left(\omega_2^0+\Omega\right)\tau}a_1\left(\omega_1^0+\Omega'\right)a_2\left(\omega_2^0-\Omega'\right)a_1^\dagger\left(\omega_2^0+\Omega\right)a_2^\dagger\left(\omega_1^0-\Omega\right)\\
		\kappa_3 &\equiv \left(\omega_2^0+\Omega'\right)\left(\omega_1^0+\Omega\right)e^{i\left(\omega_2^0+\Omega'\right)\tau}e^{-i\left(\omega_1^0+\Omega\right)\tau}a_1\left(\omega_2^0+\Omega'\right)a_2\left(\omega_1^0-\Omega'\right)a_1^\dagger\left(\omega_1^0+\Omega\right)a_2^\dagger\left(\omega_2^0-\Omega\right)\\
		\kappa_4 &\equiv \left(\omega_2^0+\Omega'\right)\left(\omega_2^0+\Omega\right)e^{i\left(\omega_2^0+\Omega'\right)\tau}e^{-i\left(\omega_2^0+\Omega\right)\tau}a_1\left(\omega_2^0+\Omega'\right)a_2\left(\omega_1^0-\Omega'\right)a_1^\dagger\left(\omega_2^0+\Omega\right)a_2^\dagger\left(\omega_1^0-\Omega\right),
	\end{split}
\end{equation}
and we have used the facts that 
\begin{equation}
	\begin{split}
		&\int d\Omega|f(\Omega)|^2\Omega = 0\\
		&\int d\Omega|f(\Omega)|^2\Omega^2 = \sigma^2.
	\end{split}
\end{equation}
Looking at the second term in Eq.~\ref{eq:6}, we have
\begin{equation}
	\begin{split}
		\left\langle\psi(\tau)\middle|\frac{\partial\psi(\tau)}{\partial\tau}\right\rangle &= \frac{1}{2(1+\beta)}\int d\Omega d\Omega' f(\Omega)f(\Omega')\\
		&\phantom{=}\left[\left(\omega_1^0+\Omega\right)e^{i\left(\omega_1^0+\Omega'\right)\tau}e^{-i\left(\omega_1^0+\Omega\right)\tau}\delta(\Omega-\Omega')\right.\\
		&\phantom{=}+\left(\omega_2^0+\Omega\right)e^{i\left(\omega_1^0+\Omega'\right)\tau}e^{-i\left(\omega_2^0+\Omega\right)\tau}\delta\left(\left(\omega_1^0+\Omega'\right)-\left(\omega_2^0+\Omega\right)\right)\\
		&\phantom{=}+\left(\omega_1^0+\Omega\right)e^{i\left(\omega_2^0+\Omega'\right)\tau}e^{-i\left(\omega_1^0+\Omega\right)\tau}\delta\left(\left(\omega_2^0+\Omega'\right)-\left(\omega_1^0+\Omega\right)\right)\\
		&\phantom{=}+\left.\left(\omega_2^0+\Omega\right)e^{i\left(\omega_2^0+\Omega'\right)\tau}e^{-i\left(\omega_2^0+\Omega\right)\tau}\delta(\Omega-\Omega')\right]\\
		&=\frac{1}{2}\left(\omega_1^0+\omega_2^0\right),
	\end{split}
\end{equation}
and it therefore follows that 
\begin{equation}
	\left|\left\langle\psi(\tau)\middle|\frac{\partial\psi(\tau)}{\partial\tau}\right\rangle\right|^2 = \frac{1}{4}\left(\left(\omega_1^0\right)^2+2\omega_1^0\omega_2^0+\left(\omega_2^0\right)^2\right).
\end{equation}
The quantum Fisher information $Q$ can then be calculated as 
\begin{equation}
	\begin{split}
		\frac{Q}{4} &= \left\langle\frac{\partial\psi(\tau)}{\partial\tau}\middle|\frac{\partial\psi(\tau)}{\partial\tau}\right\rangle-\left|\left\langle\psi(\tau)\middle|\frac{\partial\psi(\tau)}{\partial\tau}\right\rangle\right|^2\\
		&=\frac{1}{2}\left(\left(\omega_1^0\right)^2+\left(\omega_2^0\right)^2 + 2\sigma^2\right)-\frac{1}{4}\left(\left(\omega_1^0\right)^2+\left(\omega_2^0\right)^2+2\omega_1^0\omega_2^0\right)\\
		&=\frac{1}{4}\left[\left(\omega_1^0\right)^2-2\omega_1^0\omega_2^0+\left(\omega_2^0\right)^2\right]+\sigma^2\\
		&=\frac{\left(\Delta\omega\right)^2}{4} + \sigma^2.
	\end{split}
\end{equation}

\subsection*{Classical Fisher information}
\label{sec:cfi}

For ideal energy-entangled two-photon interference, Eq.~\ref{eq:8} becomes
\begin{equation}
	\begin{split}
		\mathcal{I} &= \frac{\left(\frac{\partial P_C}{\partial\tau}\right)^2}{P_C}+\frac{\left(\frac{\partial P_A}{\partial\tau}\right)^2}{P_A}\\
		&= \frac{\left(\frac{\partial P_C}{\partial\tau}\right)^2}{P_C}+\frac{\left(\frac{\partial (1-P_C)}{\partial\tau}\right)^2}{1-P_C}\\
		&= \frac{\left(\frac{\partial P_C}{\partial\tau}\right)^2}{P_C(1-P_C)}.
	\end{split}
\end{equation}
We then calculate the single-event Fisher information $\mathcal{I}$: 
\begin{equation}
	\begin{split}
		\frac{\partial P_C}{\partial\tau} &= \frac{1}{2}e^{-2\sigma^2\tau^2}\left((\Delta\omega)\sin((\Delta\omega)\tau)+4\sigma^2\tau\cos((\Delta\omega)\tau)\right)\\
		P_C(1-P_C) &= \frac{1}{4}\left(1-\cos^2((\Delta\omega)\tau)e^{-4\sigma^2\tau^2}\right),
	\end{split}
\end{equation}
and so
\begin{equation}
	\begin{split}
		\mathcal{I} &= \frac{\left(e^{-2\sigma^2\tau^2}\left((\Delta\omega)\sin((\Delta\omega)\tau)+4\sigma^2\tau\cos((\Delta\omega)\tau)\right)\right)^2}{1-\cos^2((\Delta\omega)\tau)e^{-4\sigma^2\tau^2}}\\
		&=\frac{\left((\Delta\omega)\sin((\Delta\omega)\tau)+4\sigma^2\tau\cos((\Delta\omega)\tau)\right)^2}{e^{4\sigma^2\tau^2}-\cos^2((\Delta\omega)\tau)}.
	\end{split}
\end{equation}

\subsection*{Experimental saturation of the Cram\'er–Rao bound}
\label{sec:saturation}

A mixed energy-entangled (EE) state
\begin{equation}
	\begin{split}
		\rho_{EE,mixed}(\epsilon) &= \frac{1+\epsilon}{2}\ket{\Psi^+}\bra{\Psi^+}+\frac{1-\epsilon}{2}\ket{\Psi^-}\bra{\Psi^-}\\
		\ket{\Psi^+} &\equiv \frac{1}{\sqrt{2}}\left(\ket{\omega_1}_a\ket{\omega_2}_b+\ket{\omega_2}_a\ket{\omega_1}_b\right)\\
		\ket{\Psi^-} &\equiv \frac{1}{\sqrt{2}}\left(\ket{\omega_1}_a\ket{\omega_2}_b-\ket{\omega_2}_a\ket{\omega_1}_b\right)
	\end{split}
\end{equation}
has purity $(1+\epsilon^2)/2$ and produces interference fringes
\begin{equation}
	\begin{split}
		P_C(\tau) &= \frac{1+\epsilon}{2} \times \frac{1}{2}\left(1-\cos((\Delta\omega)\tau)e^{-2\sigma^2\tau^2}\right) + \frac{1-\epsilon}{2} \times \frac{1}{2}\left(1+\cos((\Delta\omega)\tau)e^{-2\sigma^2\tau^2}\right)\\
		&=\frac{1}{2}\left(1-\epsilon\cos((\Delta\omega)\tau)e^{-2\sigma^2\tau^2}\right).
	\end{split}
\end{equation}
Using error propagation, the variance of $P_C$ is equal to
\begin{equation}
	\text{Var}[P_C] = \left(\frac{\partial P_C}{\partial  \tau}\right)^2\sigma_\tau^2,
\end{equation}
and so the single-measurement error $\sigma_\tau$ is given by
\begin{equation}
	\begin{split}
		\sigma_\tau &= \frac{\sqrt{\text{Var}[P_C]}}{\left|\frac{\partial P_C}{\partial\tau}\right|}\\
		&= \frac{\sqrt{E[(P_C)^2]-E[P_C]^2}}{\left|\frac{\partial P_C}{\partial\tau}\right|}\\
		&= \frac{\sqrt{E[P_C]-E[P_C]^2}}{\left|\frac{\partial P_C}{\partial\tau}\right|},
	\end{split}
\end{equation}
where $E[P_C]$ is the expectation value of $P_C$, and in the last line we have used the fact that 
$E[P_C] = E[(P_C)^2]$ since $P_C$ is a projective measurement. We note that this form is similar to the classical Fisher information described above. In this case, 
\begin{equation}
	\begin{split}
		\text{Var}[P_C] &= \frac{1}{4}\left(1-\epsilon^2\cos^2((\Delta\omega)\tau)e^{-4\sigma^2\tau^2}\right)\\
		\left|\frac{\partial P_C}{\partial\tau}\right| &= \frac{\epsilon}{2}e^{-2\sigma^2\tau^2}\left((\Delta\omega)\sin((\Delta\omega)\tau)+4\sigma^2\tau\cos((\Delta\omega)\tau)\right)
	\end{split}
\end{equation}
and so
\begin{equation}
	\begin{split}
		\sigma_\tau &= \frac{\sqrt{\text{Var}[P_C]}}{\left|\frac{\partial P_C}{\partial\tau}\right|}\\
		&= \frac{\frac{1}{2}\sqrt{1-\epsilon^2\cos^2((\Delta\omega)\tau)e^{-4\sigma^2\tau^2}}}{\frac{\epsilon}{2}e^{-2\sigma^2\tau^2}\left((\Delta\omega)\sin((\Delta\omega)\tau)+4\sigma^2\tau\cos((\Delta\omega)\tau)\right)}\\
		&= \frac{1}{\epsilon}\frac{\sqrt{e^{4\sigma^2\tau^2}-\epsilon^2\cos^2((\Delta\omega)\tau)}}{(\Delta\omega)\sin((\Delta\omega)\tau)+4\sigma^2\tau\cos((\Delta\omega)\tau)}.
	\end{split}
\end{equation}

\subsection*{Modeling the effect of loss}
\label{sec:modelloss}

Figure~\hyperref[fig:figs1]{\ref*{fig:figs1}A} shows the induced system transmission $\eta$ tested in our experiment as a function of half-wave plate angle, indicating a minimum observed transmission of $\eta=5(1)\times10^{-4}$. The experimental transmission is obtained for each waveplate angle $\theta$ by calculating the relative coincidence rate $\eta(\theta) \equiv N_c(\theta)/N_c(0)$, where the coincidence rate $N_c(\theta)$ has been corrected to account for accidental coincident detections and erroneous events from leakage.

As noted in the main text, in the case of the two-photon interferometer, loss reduces the coincident detection rate. In some circumstances, a reduced detection rate may be compensated for by increasing the integration time. For example, when measuring the visibilities shown in Fig.~\hyperref[fig:fig4]{\ref*{fig:fig4}A}, we progressively increased the per-point integration time for the fringe scans above the default 1 s (up to 10 s) as loss increased to maintain sufficient counts. However, we still see a reduction in fit quality for the interference fringes, leading to unreliable visibility extraction. The quantum visibilities shown in Fig.~\hyperref[fig:fig4]{\ref*{fig:fig4}A} are therefore instead calculated according to
\begin{equation}
	V_\eta \equiv \frac{\text{max}_{\delta x}(P_C) - \text{min}_{\delta x}(P_C)}{\text{max}_{\delta x}(P_C) + \text{min}_{\delta x}(P_C)},
\end{equation}
where the extrema are taken from the $P_C$ fringe scan performed at a given transmission $\eta$. Error 
bars are estimated via error propagation assuming Poissonian photon statistics.

\subsection*{Modeling the effect of background}
\label{sec:modelbg}

The total number of accidental coincident detections $A_i$ may be written in terms of $R_i$, the ratio of mean total single-detector events in the presence of no background, $\left\langle S_0 \right\rangle$, and the mean total single-detector events coming from the incident background, $\left\langle S_{BG,i} \right\rangle$:
\begin{equation}
	\begin{split}
		A_i &\approx \left\langle S_i \right\rangle^2\Delta T \\
		&= (\left\langle S_0 \right\rangle + \left\langle S_{BG,i} \right\rangle)^2 \Delta T \\
		&= A_0 + (2\left\langle S_0 \right\rangle\left\langle S_{BG,i} \right\rangle + \left\langle S_{BG,i} \right\rangle^2) \Delta T \\
		&= A_0(1+2R_i+(R_i)^2),\\
		R_i &\equiv \frac{\left\langle S_{BG,i} \right\rangle}{\left\langle S_{0} \right\rangle}.
	\end{split}
\end{equation}
For quantum interference measurements, the single-detector events are summed across four 
detectors, while for classical interference measurements they are summed across two detectors. 
Eq.~\ref{eq:27} can then be rewritten as 
\begin{equation}
	V_i = \frac{1}{1+(2R_i+(R_i)^2)\frac{A_0}{C_0}}V_0.
	\label{eq:s54}
\end{equation}
The background is quantified as the fraction of total singles, defined as 
\begin{equation}
	\begin{split}
		B_i &= \frac{\left\langle S_i \right\rangle - \left\langle S_0 \right\rangle}{\left\langle S_i \right\rangle} = \frac{\left\langle S_{BG,i} \right\rangle}{\left\langle S_i \right\rangle} \\
		&\rightarrow R_i = \frac{\left\langle S_i \right\rangle - \left\langle S_0 \right\rangle}{\left\langle S_0 \right\rangle} = \frac{B_i}{B_i - 1},
	\end{split}
\end{equation}
where $B_i$ is background corresponding to the background light source being set to the $i^\text{th}$ 
brightness setting. Substituting this into Eq.~\ref{eq:s54}, 
\begin{equation}
	V_i = \frac{1}{1+\frac{(3B_i - 2)B_i}{(B_i-1)^2}\frac{A_0}{C_0}}V_0.
\end{equation}
The corresponding result for classical interference is  
\begin{equation}
	V_i = (1-B_i)V_0.
\end{equation}
The effect of the introduced background on single and coincident detection rates are shown in 
Fig.~\hyperref[fig:figs1]{\ref*{fig:figs1}B}. 

\section*{Materials and methods}
\label{sec:mm}

\subsection*{Source module: Design}
\label{sec:smd}

Figure~\ref{fig:figs2} shows a schematic of the entanglement source. The half-wave plate and quarter-wave plate immediately after the input coupler are calibrated to maximize transmission through a \mbox{532-nm} polarizing beamsplitter such that the pump polarization is $\ket{H}_{532}$. Next, a second half-wave plate in a motorized rotation mount is used to rotate the pump polarization to approximately $\ket{D}_{532} = (\ket{H)_{532}+\ket{V}_{532}})/\sqrt{2}$, though the exact pump polarization is adjusted to balance the resulting entangled state (see the \textbf{Polarization-entangled state balancing step} in the \textbf{End-to-end system calibration protocol} section).  

The underlying beam-displacer interferometer is detailed in Fig.~\hyperref[fig:figs3]{\ref*{fig:figs3}A}. The calcite beam displacers transmit vertically polarized light, and laterally displace horizontally polarized light by $\sim$2.4~mm. The beam displacers are cut to different lengths for each wavelength to achieve the same lateral displacement: 21.6 mm for 532 nm, 22.5 mm for 810 nm, and 23.9 mm for 1550 nm.

Figure~\hyperref[fig:figs3]{\ref*{fig:figs3}B} shows the custom housing in which the MgO:PPLN SPDC crystals are mounted. The crystals are configured to produce the down-conversion 
\begin{equation}
	\begin{split}
		\ket{H}_{532} &\rightarrow \ket{H}_{810}\ket{H}_{1550}, \\
		\ket{V}_{532} &\rightarrow \ket{V}_{810}\ket{V}_{1550}.
	\end{split}
\end{equation}

\subsection*{Source module: Characterization}
\label{sec:smc}

Figure~\ref{fig:figs4} shows the measured pump (532~nm) and signal (810~nm) spectra from our source. Gaussian fits yield center wavelengths of 531.9120(5)~nm and 810.504(1)~nm for the pump and signal spectra, respectively, as well as bandwidths of 0.167(1)~nm and 0.495(2)~nm. Energy conservation implies an idler wavelength of 1547.484(5)~nm, which is close to the nominal wavelength of 1550~nm.

We observed a mean brightness of 91,649(1,429) and 85,980(1,298) detected pairs per second per mW for the $\ket{H}_{1550}\ket{H}_{810}$ and $\ket{V}_{1550}\ket{V}_{810}$ processes, respectively (the numbers in parentheses are the standard deviations). For this measurement, the overall pumping power was set to 1 mW and the polarization of the pump beam incident on the first beam displacer was set to $\ket{D}$. The pump beam for each process was blocked one at a time to observe the contribution of the other process to the overall detected pair rate. The fiber compensation half-wave and quarter-wave plates were set to arbitrary known angles (the zero angles of their rotation mounts) and the tomography quarter-wave plates were set to their calibrated zero angles (where the waveplate axes are aligned with the $HV$ basis). Since the detectors have polarization-dependent efficiency and the two processes involve orthogonal polarizations, the pair detection rate was manually maximized for each process by adjusting the polarization control paddles attached to the fibers leading to the 1550 and \mbox{810-nm} detectors. The same detectors from port $A$ of the interferometer were used for this measurement, and were connected directly to the source output fibers, bypassing the interferometer module. We recorded 50 consecutive detector measurements with a \mbox{1-second} integration time per measurement. While the precise brightness varies depending on the source configuration (i.e., the pumping power split between each process, the angle settings of the output waveplates, etc.), the sum of the per-process brightnesses indicates a detected source brightness of $>$$10^5$ detected pairs per second per mW.

During the same measurement, we also obtained the mean net Klyshko heralding efficiencies \cite{klyshko1980} for both wavelengths for both processes. For $\ket{H}_{1550}\ket{H}_{810}$, we observed $8.7(1)\%$ and $19.9(2)\%$ for 1550 and 810 nm, respectively. For $\ket{V}_{1550}\ket{V}_{810}$, we observed $9.2(1)\%$ and $20.8(1)\%$ for 1550 and 810 nm, respectively. The given uncertainties are the standard deviations. These net efficiencies do not include corrections for fiber-coupling losses, detector efficiency and background, or accidentals (detections of uncorrelated photon pairs). The detector efficiencies are summarized in Table \ref{tab:tabs1}, and the background and accidentals were negligible relative to the observed detection rates. Correcting for the fiber link transmissions and detector efficiencies (but not the fiber coupling loss), we obtain estimated lower bounds of $\sim$$10\%$ and $\sim$$22\%$ for 1550 and 810~nm, respectively, for $\ket{H}_{1550}\ket{H}_{810}$. The corresponding values for $\ket{V}_{1550}\ket{V}_{810}$ are $\sim$$10\%$ and $\sim$$23\%$. 

The \textbf{End-to-end system calibration protocol} section describes the procedure for calibrating our source to optimize the generated entangled state, and the \textbf{State tomography} subsection details our procedure for characterizing the entanglement. The resulting state density matrix is shown in Fig.~\ref{fig:figs5}. We observed a purity, concurrence, and singlet fraction of $90.3(2)\%$, $89.6(2)\%$, and $94.8(1)\%$, respectively. 

\subsection*{Interferometer module: General details}
\label{sec:img}

Table~\ref{tab:tabs2} summarizes the end-to-end free-space transmission of the interferometer 
module. 

Figure~\ref{fig:figs6} shows a typical drift in $P_C$ over 100~seconds. Also shown is the normalized noise, defined as the variance in $P_C$ over Poissonian variance $\lambda$, with a \mbox{10-s} rolling window. Over 100~seconds our interferometer exhibits a mean \mbox{10-second} normalized noise of 1.3(6) with passive stabilization and no thermal isolation. This result illustrates how our basic passive measures are largely sufficient, given our ability to perform measurements with nanometer-scale resolution on a timescale much faster than the interferometer drift.

For details on how we reconfigured our interferometer to obtain the classical interference data discussed in this work, please refer to the \textbf{Classical frequency beating} section.

\subsection*{Interferometer module: Interference visibility}
\label{sec:imv}

The interference visibility achievable with our interferometer is largely determined by the purity of the polarization entangled state, the extinction ratio of both output ports of the polarizing beamsplitter (PBS) at the interferometer input, and the splitting ratio of the non-polarizing beamsplitter (NPBS) at the interferometer output. 

The state purity is discussed in the main text. We characterized the PBS extinction ratio in situ with classical alignment lasers nominally at 1550 and 810~nm and power meters. A calibrated polarizer was used to set the incident polarization to either horizontal or vertical, and the transmission for both output ports was measured for both input polarizations (correcting for background on the power meters). All measurements were performed in free space. From these we obtain $T_p$:$T_s$ and $R_s$:$R_p$ of $\sim$7,100 and $\sim$500 for 1550 nm, respectively. The corresponding values for 810 nm are $\sim$5,600 and $\sim$500.

Similarly, we characterized the transmission and reflectance of the NPBS with classical alignment lasers, for both the transmitted (i.e., through the PBS) and reflected (i.e., off the PBS) paths of the interferometer (modes $a$ and $b$, respectively). The polarization of the light in the reflected path was rotated to match that of the transmitted path by an achromatic half-wave plate (775-1550~nm). All measurements were performed in free space. For the transmitted path, we observed $\sim$$62\%$ and $\sim$$36\%$ transmission and reflectance, respectively, for 1550~nm. The corresponding values for 810~nm are $\sim$$50\%$ and $\sim$$50\%$. For the reflected path, we observed $\sim$$61\%$ and $\sim$$36\%$ for 1550~nm, and $\sim$$50\%$ and $\sim$$49\%$ for 810~nm. 

In Fig.~\ref{fig:figs7} we plot the effects of imperfect PBS extinction ratio and NPBS splitting efficiency on the final fringe visibility using a simplified numerical model. The effect of an imperfect PBS, shown in Fig.~\hyperref[fig:figs7]{\ref*{fig:figs7}A}, is modeled by taking the PBS as a polarization-dependent beamsplitter with transmission and reflection coefficients given by 
\begin{equation}
	\begin{split}
		T_p = \frac{ER_T(1-ER_R)}{1-ER_TER_R}~~&;~~R_p = \frac{1-ER_T}{1-ER_TER_R} \\
		T_s = \frac{1-ER_R}{1-ER_TER_R}~~&;~~R_s = \frac{ER_R(1-ER_R)}{1-ER_TER_R}, 
	\end{split}
\end{equation}
where $ER_T$ ($ER_R)$ is the extinction ratio (ER) of the transmitted (reflected) port of the PBS. For a perfect PBS, $ER_T$, $ER_R \rightarrow \infty$, such that $T_p$, $R_s\rightarrow 1$, $T_s$, $R_p\rightarrow 0$, as expected. In contrast, for a completely non-polarizing beamsplitter,  $ER_T$, $ER_R \rightarrow 1$, such that $T_s$, $T_p$, $R_s$, $R_p \rightarrow 0.5$. The effect of an imperfect PBS is to probabilistically route the polarization-entangled state into the incorrect interferometer mode, leading to ``leakage'' events in which both photons are routed into the same path. This can be seen by observing the state of the $\ket{H}_{1550}\ket{V}_{810}$ process after the (first) PBS and half-wave plate in the interferometer,
\begin{equation}
	\begin{split}
		\ket{H}_{1550}\ket{V}_{810} &\rightarrow \sqrt{T_{p,1550}T_{s,810}}\ket{H,a}_{1550}\ket{V,a}_{810}\\
			&\phantom{=}+ e^{-i\omega_{810}\tau}\sqrt{T_{p,1550}R_{s,810}}\ket{H,a}_{1550}\ket{H,b}_{810}\\
			&\phantom{=}+ e^{-i\omega_{1550}\tau}\sqrt{R_{p,1550}T_{s,810}}\ket{V,b}_{1550}\ket{V,a}_{810}\\
			&\phantom{=}+ e^{-i(\omega_{1550}+\omega_{810})\tau}\sqrt{R_{p,1550}R_{s,810}}\ket{V,b}_{1550}\ket{H,b}_{810},
	\end{split}
	\label{eq:s60}
\end{equation}
with a similar equation holding for the $\ket{V}_{1550}\ket{H}_{810}$ process; here relative phase is introduced in mode $b$. We end up with four distinguishable interference processes corresponding to the four combinations of photon polarization. The $\ket{H}\ket{H}$ and $\ket{V}\ket{V}$ terms (terms 2 and 3 in Eq.~\ref{eq:s60}, respectively) will lead to interference at the beat-note frequency $\Delta\omega$, with their visibilities dependent on the PBS extinction ratio. Similarly, the $\ket{H}\ket{V}$ and $\ket{V}\ket{H}$ terms (terms 1 and 4 in Eq.~\ref{eq:s60}, respectively) will lead to interference at the sum frequency $\omega_1 + \omega_2$. 

The net fringe can then be used to calculate the resulting interference visibility, shown in Fig.~\hyperref[fig:figs7]{\ref*{fig:figs7}A}. We have fixed the transmission port extinction ratio (ER) $T_p$:$T_s$ at 10,000, and varied the reflected port ER $R_s$:$R_p$, for both the case where a single PBS is used, or, as in our experiment, where a second PBS is used in the reflected port to provide additional filtering. In both cases, we see that for ERs $>$ 100, the visibility loss is negligible. In our experiment, where we both have a second filtering PBS and a reflected port ER of $\sim$500, this effect can be safely ignored. 

Fig.~\hyperref[fig:figs7]{\ref*{fig:figs7}B} plots the fringe visibility as a function of NPBS splitting ratio for a single wavelength, assuming the other is fixed at 50:50. To model this effect, we replace the two beamsplitters in Eq.~\ref{eq:s4} with unbalanced transmission and reflection probabilities 
\begin{equation}
	\begin{split}
		a_1^\dagger(\omega_1) &\rightarrow \frac{1}{\sqrt{T_{\omega_1}+R_{\omega_1}}}\left(\sqrt{T_{\omega_1}}a_1^\dagger(\omega_1)+i\sqrt{R_{\omega_1}}a_2^\dagger(\omega_1)\right)\\
		a_2^\dagger\left(\omega_2\right) &\rightarrow \frac{1}{\sqrt{T_{\omega_2}+R_{\omega_2}}}\left(\sqrt{T_{\omega_2}}a_2^\dagger(\omega_2)+i\sqrt{R_{\omega_2}}a_1^\dagger(\omega_2)\right). 
	\end{split}
\end{equation}
This changes the coincidence fringes to  
\begin{equation}
	P_C = \frac{1}{(T_{\omega_1}+R_{\omega_1})(T_{\omega_2}+R_{\omega_2})}\left[T_{\omega_1}T_{\omega_2} + R_{\omega_1}R_{\omega_2} - 2\sqrt{T_{\omega_1}T_{\omega_2}R_{\omega_1}R_{\omega_2}}\cos((\Delta\omega)\tau)e^{-2\sigma^2\tau^2}\right].
\end{equation}
We note that when $T_{\omega_1} = R_{\omega_1} = T_{\omega_2} = R_{\omega_2} \rightarrow 0.5$, $P_C \rightarrow$ Eq.~\ref{eq:s33}, as expected. The visibility can then be calculated from the resulting fringes, substituting experimental values for the beamsplitter ratios. With a splitting ratio of $\sim$63:37 --- corresponding to the $\sim$$61\%$ transmission and $\sim$$36\%$ reflection probabilities associated with the \mbox{1550-nm} NPBS --- we predict a fringe visibility of $\sim$$97\%$.
 
Relatedly, the four coincident detection fringes have fitted visibilities that vary slightly with respect to each other: $88.2(4)\%$, $89.2(5)\%$, $87.7(4)\%$, and $89.2(5)\%$ for 1550$A$-810$B$, 1550$B$-810$A$, 1550$A$-810$A$, and 1550$B$-810$B$, respectively. We suspect this variation rises from small differences in optical alignment between the four output modes of the interferometer. The four single-detection fringes all have fitted visibilities less than $1\%$: $0.9(2)\%$, $0.6(2)\%$, $0.8(2)\%$, and $0.9(2)\%$ for 1550$A$, 1550$B$, 810$A$, and 810$B$, respectively. 

\subsection*{Detection module}
\label{sec:dm}

The four fiber couplers in the detection module couple photons exiting the interferometer into single-mode fibers (\mbox{SMF-28} for 1550~nm and 780HP for 810~nm) with anti-reflection-coated tips. The coupling efficiencies are summarized in Table \ref{tab:tabs3}.

Each \mbox{2-m} collection fiber is mated to a \mbox{20-m} transfer fiber that connects to the fiber input of a superconducting nanowire single-photon detector (SNSPD). Each transfer fiber is equipped with a three-paddle polarization controller to optimize the polarization of light incident on the detectors, which have polarization-dependent efficiencies. Key detector specifications are summarized in Table \ref{tab:tabs1}.

Table \ref{tab:tabs4} summarizes the combined timing jitter for coincident detections. The observed jitter led us to select 50~ps as the radius for the coincident detection window. Unless otherwise noted, the default integration time for all measurements is 1~second.

\subsection*{Spatial mode overlap}
\label{sec:smo}

At the interferometer input, the \mbox{1550-nm} and \mbox{810-nm} light are launched from fiber into free space and multiplexed into a common spatial mode via a dichroic mirror. We verify the spatial mode overlap by performing knife-edge scans with classical light from nominally \mbox{1550-nm} and \mbox{810-nm} alignment lasers. A dichroic mirror downstream of the knife edge directs each wavelength into its own power sensor for power measurement. Knife-edge scans were performed transversely to the common mode in both the $x$ (horizontal) and $y$ (vertical) directions at two longitudinal points ($z=0$~mm and $z=45$~mm). We assume Gaussian beams and fit the resulting normalized power versus knife-edge displacement curves to the function
\begin{equation}
	P(\delta) = \frac{P_0}{2}\left[1 \pm \text{Erf}\left(\frac{\sqrt{2}(\delta-\delta_0)}{w}\right)\right],
	\label{eq:s63}
\end{equation} 
where $P$ is the measured normalized power, $\delta$ is the position displacement of the knife edge in the scan direction, and $P_0$ is the normalized power incident on the knife edge. $\delta_0$ and $w$ are the centroid and $1/e^2$ radius of the beam’s transverse intensity profile, respectively. The sign of the error function is determined by the scan direction relative to the beam.    

From the fit parameters we obtain $\Delta \equiv \delta_0^{810} - \delta_0^{1550}$. As shown in Fig. \ref{fig:figs8}, the observed $\Delta$ values all are zero within fitting error, indicating that the \mbox{1550-nm} and \mbox{810-nm} beams are well-overlapped and parallel to each other. Averaging the $w$ values for the four scans per wavelength gives a mean $w$ of 1.47(1)~mm for 1550~nm and 1.59(1)~mm for 810~nm, indicating that the beams are $x$-$y$ symmetric and well-collimated. The values in parentheses are the standard deviations.  

\subsection*{End-to-end system calibration protocol}
\label{sec:eescp}

To prepare our experiment for nanometer-scale measurements, we perform an end-to-end system calibration to achieve optimal performance. The six-step procedure is as follows:

\begin{enumerate}
	\item \textbf{Transfer fiber compensation:} Stress-induced birefringence in the two single-mode fibers linking the source and interferometer modules apply an arbitrary two-qubit polarization rotation on the state generated by the source module (Eq.~\ref{eq:4}). To facilitate subsequent operations, we use compensation half-wave and quarter-wave plates to correct for the fiber transformation such that the waveplates-fiber system for each wavelength acts as the identity. Our procedure is as follows:
	\begin{enumerate}[label=\alph*.]
		\item The pump driving the $\ket{H}_{1550}\ket{H}_{810}$ process in the source (“Path $A$”) is blocked such that the source produces only photon pairs in the state $\ket{V}_{1550}\ket{V}_{810}$.
		\item The reflected path of the interferometer (mode $b$) is blocked such that only horizontally polarized photons that transmit through the first polarizing beamsplitter are detected in the detection module.
		\item The half-wave and quarter-wave plates inserted in front of both the \mbox{1550-nm} and \mbox{810-nm} output fiber couplers in the source module are rotated to minimize the number of photon detection events on all four detectors. A search algorithm based on the principles of gradient descent is utilized, and each wavelength is optimized separately. Once the counts are minimized, the combined waveplates-fiber system for each wavelength acts as the identity, transforming the input state $\ket{V}$ to the output state $\ket{V}$ , which reflect off the polarizing beamsplitter into the blocked reflected path of the interferometer, and are therefore not detected.
	\end{enumerate}
	
	\hspace{3mm}Finally, to verify the compensation, both half-wave plates are rotated by $45^\circ$ to maximize detector counts. The ratio of the observed maximum and minimum counts is then taken to obtain the extinction ratio. Then, all blocked paths are unblocked while the waveplates are left at their optimized angles, all of which are redefined as $0^\circ$ to simplify subsequent calibration steps.
	
	\hspace{3mm}The described process is fully automated via motorized rotation mounts. The runtime depends on the initial conditions. A typical calibration ran for 14~minutes and yielded background-corrected extinction ratios of 1,845(102) for 1550~nm and 900(13) for 810~nm. The given uncertainties assume Poissonian statistics for detector counts. 
	
	\item \textbf{Polarization-entangled state balancing:} We realize the optimal polarization-entangled state when the $\ket{H}_{1550}\ket{H}_{810}$ and $\ket{V}_{1550}\ket{V}_{810}$ terms in Eq.~\ref{eq:4} have equal amplitudes upon incidence on the polarizing beamsplitter at the interferometer input. The amplitudes may be unequal because of unbalanced pumping powers or heralding efficiencies for each of the two SPDC processes in the source. To equalize the amplitudes, we perform the following procedure:
	\begin{enumerate}[label=\alph*.]
		\item The reflected path of the interferometer (mode $b$) is blocked such that only horizontally polarized photons that transmit through the input polarizing beamsplitter are detected in the detection module.
		\item The half-wave plates for each transfer fiber are rotated by $45^\circ$ to project the entangled state onto $\ket{V}_{1550}\ket{V}_{810}$ at the first polarizing beamsplitter in the interferometer.
		\item The half-wave plates are then rotated back to $0^\circ$ to project the state onto $\ket{H}_{1550}\ket{H}_{810}$.
		\item We compute the difference in the observed photon pair detection rates from the two projections in steps $b$ and $c$. If the difference is less than $2\%$ of the rate for $\ket{H}_{1550}\ket{H}_{810}$, the two terms are considered balanced, and the optimization ends.
		\item If the difference is equal to or greater than $2\%$, the optimization continues. The half-wave plate in the source that controls the polarization of the \mbox{532-nm} pump incident on the first beam displacer is rotated by $0.2^\circ$, with the rotation direction dependent on the sign of the difference. Rotating this half-wave plate changes the relative pair production rate for the $\ket{H}_{1550}\ket{H}_{810}$ and $\ket{V}_{1550}\ket{V}_{810}$ processes in the source, and thereby their relative amplitudes in the entangled state.
		\item Steps $b$, $c$, and $d$ are then repeated. If the difference falls below the $2\%$ threshold, the optimization ends.
		\item Steps $e$ and $f$ are repeated however many times are necessary to achieve the $2\%$ threshold. 
	\end{enumerate}
	
	\hspace{3mm}The described process is fully automated via motorized rotation mounts. The runtime depends on the initial state of the source. A typical calibration ran for 7~minutes and returned $20.5^\circ$ as the optimized angle, with 89,773 counts for $\ket{H}_{1550}\ket{H}_{810}$ and 90,017 counts for $\ket{V}_{1550}\ket{V}_{810}$ (\mbox{5-second} integration time). The optimized angle for the pump half-wave plate is typically a few degrees away from the nominal $22.5^\circ$ angle (where both processes are pumped equally). For example, when the half-wave plate is at $20.5^\circ$, the pumping power is split 57:43 between the $\ket{H}_{1550}\ket{H}_{810}$ and $\ket{V}_{1550}\ket{V}_{810}$ processes, respectively. Pumping one process harder than the other compensates for asymmetric loss between the processes such that after all losses are factored in the state amplitudes become balanced.
	
	\item \textbf{State tomography:} The next step is to verify the entangled state via a two-qubit state tomography. With the transfer fibers calibrated to act as the identity (Step 1), we form the state projectors for the tomography by rotating the compensation half-wave plate and a second quarter-wave plate immediately preceding the compensation waveplates. The compensation quarter-wave plate remains fixed at its optimized angle from Step 1. After traveling through the transfer fibers, photons from the source are projected onto $\ket{H}$ by transmitting through the first polarizing beamsplitter at the interferometer input (with the reflected port blocked). These photons then continue to the detection module for detection.
	
	\hspace{3mm}To form the state projectors a second quarter-wave plate is necessary since the two state rotations realized with a half-wave and a quarter-wave plate (one equatorial and one azimuthal on the Poincar\'e sphere) can only transform an arbitrary polarization state to a linear state or vice versa. To transform an arbitrary polarization state to another arbitrary polarization state, it is necessary to add a second azimuthal rotation with a second quarter-wave plate. We perform three rotations (one azimuthal, one equatorial, and another azimuthal) using a quarter-wave plate, a half-wave plate, and a quarter-wave plate, in that order. With the transfer fibers between the projecting beamsplitter and the source, the beamsplitter may be interpreted (from the source side) as projecting the source photons into an arbitrary polarization basis state. With three waveplates on the source side, we can therefore transform this arbitrary projection to another arbitrary projection, namely, onto each of the six ``canonical'' polarization basis states required for a complete state tomography:
	\begin{enumerate}[label=\arabic*.]
		\item Horizontal: $\ket{H}$
		\item Vertical: $\ket{V}$
		\item Diagonal: $\ket{D}$
		\item Anti-diagonal: $\ket{A}$
		\item Left circular: $\ket{L}$
		\item Right circular: $\ket{R}$
	\end{enumerate}
	
	\hspace{3mm}Prior to installation, the first quarter-wave plate is calibrated against the polarizing beamsplitter (without intervening transfer fibers) to identify the angle where its fast axis is parallel to $\ket{H}$ as defined by the beamsplitter. This angle and the optimized angles for the half-wave plate and second quarter-wave plate from Step 1 are taken as the new ``zero'' angles for the tomography.
	
	\hspace{3mm}Table \ref{tab:tabs5} summarizes the angles corresponding to projection onto each basis state for each waveplate, relative to their zero angles. These angles transform the projector to $\ket{H}$ since our physical projector is a polarizing beamsplitter in the $HV$ basis, e.g., $\ket{D}$ at the source is transformed into $\ket{H}$ at the polarizing beamsplitter.
	
	\hspace{3mm}A tomography is taken by projecting the entangled state produced by the source onto all 36 possible two-qubit permutations of the six polarization basis states above. For each projection, the total number of coincident detections across all four coincident detection channels and the corresponding accidental counts are recorded. The coincident counts are corrected by subtracting the accidentals counts, with an enforced lower bound of zero counts. The resulting counts are then processed by a maximum likelihood estimation analysis \cite{altepeter2005} to recover the entangled state density matrix $\rho$ most likely to give rise to these measurement outcomes (Fig.~\ref{fig:figs5}). From this matrix we compute the state purity
	\begin{equation}
		\gamma \equiv \text{tr}(\rho^2),
	\end{equation}
	and the concurrence
	\begin{equation}
		\mathcal{C} \equiv \text{max}(0, \lambda_1 - \lambda_2 - \lambda_3 - \lambda_4),
	\end{equation}
	where $\lambda_{(1,2,3,4)}$ are the eigenvalues of the matrix
	\begin{equation}
		\sqrt{\sqrt{\rho}\tilde{\rho}\sqrt{\rho}},
	\end{equation}
	where $\tilde{\rho}$ is defined as
	\begin{equation}
		\tilde{\rho} \equiv (\sigma_y \otimes \sigma_y)\rho^*(\sigma_y \otimes \sigma_y),
	\end{equation}
	where $\sigma_y$ is the Pauli-Y matrix. We also compute the singlet fraction
	\begin{equation}
		SF \equiv \underset{U}{\text{max}}\bra{\Psi^-}U^\dagger\rho U\ket{\Psi^-}
	\end{equation}
	for the unitary transformation $U$. The singlet fraction can be interpreted either as the fidelity between $\rho$ and the nearest maximally entangled state, or as the maximal fidelity between $\rho$ and the polarization Bell state $\ket{\Psi^-}=(\ket{HV}-\ket{VH})/\sqrt{2}$, assuming we have local unitary transformation $U$ to correct $\rho$. Fig.~\ref{fig:figs5} shows the density matrix $\rho_{SF}$ produced by this second interpretation, though optimized to achieve the equivalent state $\ket{\Psi^+}=(\ket{HV}+\ket{VH})/\sqrt{2}$ for visual clarity.
	
	\hspace{3mm}The tomography-measurement process is fully automated via motorized rotation mounts and takes 6~minutes to perform. The process is performed 10 times and the mean purity, concurrence, and singlet fraction are calculated, with the standard deviation taken as the uncertainty. The density matrices (Figs.~\hyperref[fig:fig1]{\ref*{fig:fig1}B} and \ref{fig:figs5}) are generated by taking the mean of the real and imaginary elements of the 10 individual density matrices obtained from the tomographies.
	
	\item \textbf{Bit flip:} To convert the polarization entangled state generated by our source, $\ket{\phi_{SPDC}}$ (Eq.~\ref{eq:4}), to the desired energy entangled state (Eq.~\ref{eq:1}), the state incident on the first polarizing beamsplitter in the interferometer must be of the form
	\begin{equation}
		I_{1550}X_{810}\ket{\phi_{SPDC}} = \frac{1}{\sqrt{2}}\big(\ket{H}_{1550}\ket{V}_{810}+e^{i\phi}\ket{V}_{1550}\ket{H}_{810}\big),
		\label{eq:s69}
	\end{equation} 
	where $I$ and $X$ are the identity and Pauli-X gates, respectively. We perform a bit flip on the \mbox{810-nm} photon by rotating the compensation half-wave plate in the \mbox{810-nm} side of the source by $45^\circ$ (relative to the calibrated angle from Step 1).
	
	\item \textbf{Interferometer path-length balancing:} To maximize interference visibility, the effective lengths of the two optical paths of the interferometer must be equal (assuming other degrees of freedom are indistinguishable). During initial interferometer construction, the path lengths were made as equal as possible with the use of a ruler. This was followed by sending 810-nm laser pulses with \mbox{$O(10^1)$-ps} FWHM pulse duration through the interferometer (with roughly equal powers in both paths) and monitoring the optical power in free space at one of the interferometer outputs while scanning the relative path lengths with the optical trombone. With the laser pulses having a short coherence length, high-visibility classical interference fringes are observed only when the path lengths are close to being equalized. 
	
	\hspace{3mm}Fine optimization takes place during the system end-to-end calibration. The trombone scans $\pm$1~mm around the initial position in 0.2~mm steps, using the servo motor. At each step, the sinusoidal quantum interference fringe is measured by scanning the nano-positioning stage over 1.9~\si{\micro\meter} in \mbox{100-nm} steps and recording the resulting coincident detection fringes. After extracting the fitted visibility from each fringe scan, the visibility versus trombone servo motor position dataset is then fitted to
	\begin{equation}
		V(x) = V_0e^{-2\sigma^2\left(\frac{x-x_0}{c}\right)^2},
	\end{equation}
	where $V_0$ is the maximum quantum interference visibility, $\sigma$ is the photons’ half bandwidth (angular frequency), $x_0$ is the trombone servo motor position corresponding to $V_0$, and $c$ is the speed of light. The trombone is then moved to position $x_0$ and the visibility re-measured to verify consistency with $V_0$. At this point, the interferometer is balanced up to the nano-positioner position; the trombone servo motor has a guaranteed 
	bi-directional repeatability of $\pm$1~\si{\micro\meter}, well below the \mbox{30-\si{\micro\meter}} travel of the nano-positioner. The fully automated procedure takes 7~minutes. A typical optimization returned a balanced visibility of $89.7(6)\%$, in agreement with the $90(1)\%$ predicted from the fit. 
	
	\hspace{3mm}This balancing procedure was used to obtain the data shown in Fig.~\hyperref[fig:fig1]{\ref*{fig:fig1}D}, but with a different step size for the servo motor, and the position data redefined from the trombone motor position to the relative change in optical path length. The given uncertainty is from the fitting error. A selection of the interference fringes from this measurement is shown in Fig.~\ref{fig:figs9}. We observe interference fringes over millimeters of relative delay with the fringe visibility decreasing when moving away from the zero-delay position, consistent with Eq.~\ref{eq:2}. 
	
	\item \textbf{Detector balancing:} The four superconducting nanowire single-photon detectors in the detection module have polarization-dependent detection efficiencies, in addition to small variations in their intrinsic efficiencies as well as fiber coupling and transmission losses.
	
	\hspace{3mm}While imbalanced detector efficiencies will not change the visibility of any given fringe (assuming the counts remain well above the background), the maximum-likelihood estimation phase extraction can be affected if the four interference fringes $N_{AA}$, $N_{AB}$, $N_{BA}$, and $N_{BB}$ have different efficiencies that are not properly accounted for. This can be done by measuring the relative efficiency of each fringe for incorporation into the interference model, or by equalizing the system detection efficiency.
	
	\hspace{3mm}We equalize the four detectors by leveraging their polarization-dependent efficiencies. First, the trombone position is moved a few millimeters away from the optimized position (from Step 5) such that the count rates for all detectors are unaffected by interference. Then, the count rates of the two \mbox{1550-nm} detectors are equalized and maximized by manually adjusting the three-paddle fiber polarization controllers attached to the detectors’ input fibers. The same procedure is performed for the two \mbox{810-nm} detectors. Upon completion, the trombone position is returned to the optimized position. 
\end{enumerate}
With the conclusion of Step 6, the system is now fully calibrated and ready for measurements.

\subsection*{System performance}
\label{sec:sp}

Figure \ref{fig:figs10} illustrates the performance of the calibrated system as a function of the source pumping power. While the individual detector count rates scale linearly with the pump power, the total coincident detection rate saturates slightly above 3~mW of pumping power. We attribute this to our tight coincidence window ($\pm$50~ps). Accordingly, for most measurements we set 3~mW as the upper bound for the pumping power. We observe a coincidence to accidental ratio (CAR) of 1044(153) at 1~mW of pumping power, which falls to 310(16) at 3~mW. We define the CAR as
\begin{equation}
	CAR = \frac{C-A}{A},
\end{equation}
where $C$ is the number of coincident detections and $A$ is the number of accidentals. We measure accidentals by recording the number of coincident detections at an arbitrary time delay between individual channels and leaving unchanged the coincident detection window.   

Accidentals are uncorrelated photon pairs. They can arise from the fact that entangled photon sources based on spontaneous parametric down-conversion have some probability of generating multiple entangled pairs in the same temporal mode. If the opposite members of each pair are lost, a coincident detection may still occur with the two remaining photons; however, these photons are not entangled with each other and contribute to unwanted background noise. In our case, the CAR is an important consideration as coincidences scale linearly with pumping power while accidentals scale quadratically. Consequently, for certain measurements where the probe experiences high loss, increasing the pumping power to compensate for a reduced coincident detection rate may not be the optimal strategy, as the accidentals may start to dominate and degrade the interference visibility; increasing the integration time instead may be more effective. 

Lastly, we observe that the interference visibility is largely independent of the pumping power, indicating that increasing the pumping power does not significantly affect the entangled state produced by the source for up to at least 4~mW. We also observe that the measurement resolution (estimated from the interference fringes) scales as the inverse square root of the pumping power, as expected given that the number of coincident detections $N$ scales linearly with the pumping power (at least through 3~mW) and Poissonian uncertainty scales as $1/\sqrt{N}$.

\subsection*{Note on error propagation}
\label{sec:ep}

In this work, the uncertainties for some derived quantities are estimated via error propagation and are indicated as such. For a derived quantity $f(x,y,\cdots)$ based on the values of $x$, $y$, $\cdots$ and their corresponding errors $\sigma_x$, $\sigma_y$,$\cdots$, we obtain $\sigma_f$ (the resulting error for $f$) via the following formula:
\begin{equation}
	\sigma_f = \sqrt{\left(\frac{\partial f}{\partial x}\right)^2(\sigma_x)^2 + \left(\frac{\partial f}{\partial y}\right)^2(\sigma_y)^2 + \cdots}.
\end{equation}
In doing so, we assume that the variables $x$, $y$, $\cdots$ are uncorrelated.

\section*{Sample measurements}
\label{sec:samplem}

\subsection*{Sample measurement protocol}
\label{sec:smp}

After the sample is inserted into the Sample Positioning System (SPS), we activate the Sample Targeting System (STS) by flipping into place a mirror flipper at the interferometer input, which allows a collimated beam from a red Helium-Neon (HeNe) laser to enter the interferometer along approximately the same path as the energy-entangled probe photons (as marked by iris diaphragms) and illuminate the sample. After using the SPS to optimize the sample’s horizontal and vertical positions based on this visible indicator of the probe beam location, we adjust the sample tip and tilt to overlap back-reflected (off the sample) HeNe light with the forward-propagating beam. This ensures that the sample surface is approximately normal to the probe beam. Once the sample is positioned and aligned, the STS is deactivated and its mirror flipped out of the way. Note that the STS as described is not suitable for photosensitive samples; alternative sample positioning techniques may be used instead. 

We then illuminate the sample with energy-entangled probe photons. As the presence of the sample changes the relative lengths of the two interferometer paths, we re-balance these lengths to maximize the interference visibility using the same procedure described in the \textbf{End-to-end system calibration protocol} section to maximize the interference visibility. Then, reference interference fringes (see the \textbf{Experimental displacement extraction} section in the main text) are obtained at the initial sample position by keeping the sample fixed and sweeping the interferometer through 4~\si{\micro\meter} of path-length difference (2~\si{\micro\meter} trombone displacement) and recording the two coincidence and two anti-coincidence fringes in 0.1~\si{\micro\meter} steps (0.05~\si{\micro\meter} trombone steps).  

A search protocol then automatically recenters the coincident detection probability at $P_C \approx 0.5$, where $dP_C/d\tau$ is maximized. Depending on the amount of delay introduced by the sample, it is sometimes advantageous to shift the starting $P_C$ slightly off-center to remain close to the high-sensitivity region of the fringe (e.g., a sample measurement where $P_C$ ranges from 0.4 through 0.6 may return better results than one where $P_C$ ranges from 0.5 through 0.7). While our current methods limit the total measurable displacement to half of a fringe (corresponding to scanning $P_C$ from 0 to 1), larger displacements can be measured by keeping track of the total 
number of fringes passed through (``fringe counting'') as well as the position within a fringe.

Finally, we translate the sample transversely to the probe beam (horizontally) and record the counts from the four coincident detection channels as a function of the sample’s transverse position. We then compare these recorded counts against the reference fringes via the procedure described in the \textbf{Experimental displacement extraction} section (main text) to extract the interferometer displacement as a function of sample position, which may be fitted to extract useful quantities (e.g., see the \textbf{Model for thin-film sample measurements} section).

Since the thin-film sample measurements performed as a part of this work involved scanning the probe beam across a sharp boundary between the uncoated and coated regions of a substrate (in either direction), the resulting data is similar to those obtained via the knife-edge scans described in the \textbf{Spatial mode overlap} section. Accordingly, we estimate the quantum probe transmission through the thin film by fitting the total coincident detection counts versus sample position data with a modified form of Eq.~\ref{eq:s63}, and take the fitting error as the uncertainty. Examples of such an analysis are shown in Figs.~\hyperref[fig:fig5]{\ref*{fig:fig5}A} (top panel) and \ref{fig:figs14}. 

When selecting samples for study, it is important to keep in mind that transmissive measurements with energy-entangled photons function best when probing samples featuring roughly comparable transmissions for the two frequencies involved.

\subsection*{Thin-film sample fabrication}
\label{sec:sf}

Table \ref{tab:tabs6} summarizes the specifications for the wafers used as substrates for our samples. Fig. \ref{fig:figs11} illustrates how our sample fabrication process yields strips of uncoated and coated regions with well-defined edges.

\subsection*{Model for thin-film sample measurements}
\label{sec:msm}

With a probe beam diameter on the order of millimeters, data acquired during sample measurements will be a convolution of the probe and sample features. Additionally, a large probe diameter necessitates scanning over a distance on the order of millimeters to obtain a zero-thickness reference (uncoated region) and a relative thickness measurement (coated region). In scanning over this long of a distance, our measurement will be influenced by any effective curvature that may be present in the substrate (e.g., due to local variations in substrate thickness).

To extract the film thickness from our measurement given these considerations, we developed an analysis model. In our model, we treat our target feature as an infinitely sharp step in height atop a substrate with a linear wedge and quadratic curvature (Fig.~\hyperref[fig:figs12]{\ref*{fig:figs12}A}) and consider a Gaussian optical probe (Fig.~\hyperref[fig:figs12]{\ref*{fig:figs12}B}). The convolution of the two is shown in Fig.~\hyperref[fig:figs12]{\ref*{fig:figs12}C}, and has the following form (assuming an effective refractive index of 2 for the sample and substrate and 1 for the ambient atmosphere):
\begin{equation}
	\sigma_x(y) = a + b(y-y_0)+c\left((y-y_0)^2 + \sigma^2\right) - \frac{a}{2}\text{Erfc}\left(\frac{y-y_0}{\sigma\sqrt{2}}\right) + d.
	\label{eq:s73}
\end{equation}
Here, $\sigma_x$ is the displacement measured by our interferometer, $y$ is the sample position (transverse to probe beam propagation), $y_0$ is the step edge position, $a$ is the film thickness, $b$ is the linear wedge, $c$ is the quadratic curvature, $\sigma$ is the standard deviation of the Gaussian transverse intensity profile of the probe beam, Erfc is the complementary error function, and $d$ is the vertical offset. By fitting the displacement data with \ref{eq:s73}, we can obtain values for the film thickness as well as the linear and quadratic curvatures from the fit parameters $a$, $b$, and $c$. 

Converting the fitted $\sigma$ to $1/e^2$ diameter, we can check for general agreement with the physical probe beam diameter, which can be measured via knife-edge scans using a procedure largely identical to the one described in the \textbf{Spatial mode overlap} section. However, instead of using classical alignment lasers and power meters, we use single photons from our entanglement source, which are coupled into single-mode fiber at the interferometer output and detected with single-photon detectors. The resulting counts versus knife-edge displacement data (Fig.~\ref{fig:figs13}) therefore provides information regarding the spatial modes relevant to interference measurements. We also perform the knife-edge scan at only one longitudinal location. We observe a mean diameter of 1.21(4)~mm over all eight scans (horizontal and vertical for both ports $A$ and $B$ for both 810 and 1550~nm). Based on this diameter, the 2.4(2)~mm fitted $1/e^2$ diameter for the \mbox{5-nm} test sample is reasonable. We note that this comparison is limited as the fitted waist depends on the actual steepness of the step edge. However, the fitted quantity of interest --- the step height --- is independent of the beam waist in this analysis.

\subsection*{Refractive index: Calibration sample}
\label{sec:ncs}

The effective thickness $x_0$ of our thin film sample (as measured by our interferometer) is related to the physical thickness $x$ via the relation
\begin{equation}
	x = \frac{x_0}{n_{film}-1},
	\label{eq:s74}
\end{equation}
where $n_{film}$ is the refractive index of the film material. The denominator being $n-1$ rather than $n$ arises from the fact that the film is replacing the air that the optical probe would have otherwise propagated through. The net difference in path length introduced by the film when inserted in the interferometer (i.e., the effective thickness) is therefore $x_0 = xn_{film} - xn_{air}$. Taking $n_{air} = 1$ and rearranging, we recover Eq.~\ref{eq:s74}.

For best results, we utilize experimentally measured values of $n_{film}$ for both the quantum and classical sample measurements. The uncertainty in the measured $n_{film}$ can be combined with the fitting error for $x_0$ via error propagation to obtain the uncertainty in $x$. Obtaining such experimental values requires a calibration sample with a well-defined thickness. We designed a calibration sample by considering the nature of our test sample and its fabrication process.

In general, electron-beam physical vapor deposition can reliably produce films of the desired thickness for sufficiently thick films (e.g., tens of nanometers or greater). This accuracy is realized by calibrating the electron-beam machine by depositing a few hundred nanometers of film at a particular power. The deposition time is noted, and the film thickness is measured via X-ray reflectometry, which generally gives accurate results for films with \mbox{$>$10-nm} thickness. Combining the deposition time and thickness yields a deposition rate; the electron-beam machine uses this rate to time depositions at a particular power to achieve the desired film thickness. However, for the specific machine used to support this work, ramping up to the target power takes some time, during which time the deposition rate is not well-defined. For very thin films with thickness on the order of nanometers, like our test sample, the deposition occurs over the ramp-up period. The resulting thickness is therefore less accurate, as evidenced by the fact that our test sample has a nominal thickness of 5~nm but a measured thickness of $\sim$7~ nm.

Accordingly, we fabricated a calibration sample with \mbox{50-nm} thickness using identical materials and methodology as the test sample. We selected \mbox{50-nm} thickness as it is a thickness that our electron-beam machine can reliably deliver. Spot measurements with an atomic force microscope confirmed a film thickness of $\sim$50~nm. The next two sections describe how we used this calibration sample to obtain $n_{film}$ for the quantum and classical probes.  

\subsection*{Refractive index: Quantum probe}
\label{sec:nqp}

To obtain $n_{film}$ for the quantum probe, we measured the effective thickness of the calibration sample in our interferometer using the same procedure as with the \mbox{5-nm} sample (described in the main text). With a \mbox{50-nm} film the quantum probe transmission is $1.6(2)\%$, yet going from the uncoated to the coated regions the quantum interference visibility decreases only moderately, from $88.1(5)\%$ to $71(2)\%$. This moderate drop in visibility may be corrected for by measuring $V_{P_C}(y)$, the $P_C$ fringe visibility at sample position $y$, for every value of $y$ in the sample scan. Then, for each sample position, the visibilities of the four coincident detection 
reference fringes (measured at sample position $y=0$) are scaled by $V_{P_C}(y)/V_{P_C}(y=0)$. 

These visibility-corrected reference fringes are then used to extract the interferometer displacement as a function of sample position via the procedure described in the \textbf{Experimental displacement extraction} section (main text). To simplify subsequent analysis steps, the displacement was extracted in units of phase, i.e., $k\sigma_x(y)$ where $k$ is the $k$-vector of the quantum probe and $\sigma_x(y)$ is the interferometer displacement as a function of sample position. We fit the resulting phase data to a modified version of Eq.~\ref{eq:s73}:
\begin{equation}
	k\sigma_x(y) = k\left(n_fa + n_s\left(b(y-y_0)+c\left((y-y_0)^2 + \sigma^2\right)\right) - n_f\frac{a}{2}\text{Erfc}\left(\frac{y-y_0}{\sigma\sqrt{2}}\right) + n_sd\right).
\end{equation}
In this version, the equation is multiplied by the quantum $k$-vector $k$, the terms relating to the 
film are multiplied by $n_f$, and the terms relating to the substrate are multiplied by $n_s$. Here, $n_f$ and $n_s$ are defined as
\begin{equation}
	\begin{split}
		n_f &\equiv n_{film}-1\\
		n_s &\equiv n_{substrate}-1,
	\end{split}
\end{equation} 
and are used to convert physical displacement to interferometer displacement (see Eq,~\ref{eq:s74}). $n_{substrate}$ is the refractive index of sample substrate. For this fit, the parameter of interest is $n_f$, from 
which we can extract $n_{film}$. We fix the film thickness $a$ to the calibrated value of 50~nm and 
assume nominal wavelengths for $k$ such that
\begin{equation}
	k \equiv k_{810} - k_{1550} \equiv \frac{2\pi}{810~\text{nm}} - \frac{2\pi}{1550~\text{nm}}.
\end{equation} 
We assume nominal wavelengths because the measured wavelengths are close to these (Fig.~\ref{fig:figs4}). For this analysis, the substrate parameters (e.g., the linear and quadratic curvatures) are not relevant, so the exact value of $n_{substrate}$ is not critical. For informational purposes, we consider a three-term Sellmeier equation and Ref.~\cite{malitson1962} for the relevant constants for synthetic sapphire; the equation yields indices of 1.7462 and 1.7599 for 1550 and 810~nm, respectively. Since the index is similar for both wavelengths, for simplicity we treat the two indices as approximately equal and fix $n_{substrate}$ to 1.75. 

With these parameters, the fit returns an $n_{film}$ of 3.3(3) for the quantum probe. As shown in Fig.~\ref{fig:figs14}, the fit curve closely tracks the experimental results. As an ``effective'' index of refraction, $n_{film}$ may be used in analysis tasks as if our interferometer utilizes a single ``wavelength'' (the fringe period).

\subsection*{Refractive index: Classical probe}
\label{sec:ncp}

To obtain $n_{film}$ for the classical \mbox{1550-nm} probe, we attempted to utilize the same procedure we used for the quantum probe (described above). However, the lossy nature of the film yielded poor measurement results, including a severe drop in the interference visibility from $96.0(2)\%$ for the uncoated region to $16.6(1)\%$ for the coated region; these measurement results proved to be too large for effective correction during analysis. 

Instead, we utilized ellipsometry, a standard technique for characterizing the refractive index of thin films. The coated region of the $\mbox{50-nm}$ calibration sample was measured directly with an ellipsometer; data was taken at six angles of incidence ($15^\circ$, $25^\circ$, $35^\circ$, $45^\circ$, $55^\circ$, and $65^\circ$) for 710~nm through 1650~nm in \mbox{10-nm} steps, except for 1350 through 1430~nm (because of an absorption band for a fiber within the instrument). Fitting the resulting data and assuming a 50-nm film thickness yielded an $n_{film}$ of 3.07(5) at 1550 nm; the given error is the fitting error from the ellipsometer data analysis. Our value is close to values reported in the literature, e.g., Ref.~\cite{johnson1974} reports a refractive index of 3.14 at 0.77~eV (1610~nm) for a vacuum-evaporated nickel thin film.

\subsection*{Independent measurements of test sample film thickness}
\label{sec:imf}

We performed three independent thickness measurements for the test sample (\mbox{5-nm} nominal thickness) using three established nano-characterization techniques: atomic force microscopy, scanning-stylus profilometry, and 3D optical profilometry. The details of these measurements are discussed below. 

\begin{enumerate}
	\item \textbf{Atomic force microscopy:} We utilized an Asylum Research MFP-3D atomic force microscope (AFM). A portion of the test sample containing both coated and uncoated regions was cleaved from the main wafer to accommodate sample insertion into the microscope.
	
	\hspace{3mm}Scans over the uncoated-coated boundary resulted in unreliable thickness measurements as film non-uniformity near the boundary complicated the identification of the reference (uncoated region) and measured (coated region) heights. Film non-uniformity was prominent in these measurements because the AFM features high lateral and vertical resolutions; a relatively small field of view (90~\si{\micro\meter} by 90~\si{\micro\meter} maximum) prevented us from performing large scans that incorporate regions away from the boundary that exhibit better uniformity.
	
	\hspace{3mm}We instead targeted a coated region away from the edge, where the film was observed to be highly uniform. The AFM tip was then brought into contact with the film to mechanically scrape away the film and expose the substrate, creating a sharp boundary between highly uniform coated and uncoated regions (as shown in Fig.~\hyperref[fig:fig5]{\ref*{fig:fig5}B}). The final image is 1.01~\si{\micro\meter} wide and 0.59~\si{\micro\meter} tall. Tilt correction was applied to the image by defining the substrate (uncoated) portion of the image as flat.
	
	\hspace{3mm}Given the high uniformity of the coated and uncoated regions, we obtained the thickness and its uncertainty by extracting a cross-sectional profile and averaging the first and last 340~nm of the coated and uncoated regions, respectively. The mean height values were subtracted to obtain the film thickness, and the standard deviations of the heights were combined via error propagation to obtain the uncertainty value.
	
	\item \textbf{Scanning-stylus profilometry:} We utilized a Hysitron TI-950 TriboIndenter nanoindenter in scanning-probe mode: the probe was scanned over the boundary at a constant contact force while tracking the probe height. The instrument accommodated the test sample wafer in its entirety; no cleaving was required. An uncoated-coated boundary region was selected for measurement. A \mbox{40-\si{\micro\meter}} by \mbox{19-\si{\micro\meter}} image of the sample surface was generated. 
	
	\hspace{3mm}Tilt correction was applied to the image by defining the substrate (uncoated) portion of the image as flat. Five adjacent, parallel, and non-overlapping cross-sectional profiles were generated from the corrected image (Fig.~\ref{fig:figs15}). The profiles are approximately orthogonal to the uncoated-coated boundary (see Fig.~\hyperref[fig:figs16]{\ref*{fig:figs16}B} for an illustration of similar profile placement and orientation with respect to the boundary, but from a different measurement). Each profile is the average of 20 scan lines (20~pixels, 3.125~\si{\micro\meter} total width). As these profiles are similar in form to those obtained with our interferometer, we quantitatively extract the film thickness from each profile by using Eq.~\ref{eq:s73} as a fitting function. The five fitted thickness values have a mean fit error of $1\%$, indicating excellent fits to the data. The mean of these five thicknesses is taken as the final thickness, with the standard deviation as the measurement uncertainty.
	
	\item \textbf{3D optical profilometry:} We utilized a Keyence VK-X1000 3D laser scanning confocal microscope. The instrument accommodated the test sample wafer in its entirety; no cleaving was required. An uncoated-coated boundary region was selected for measurement. A \mbox{282-\si{\micro\meter}} by \mbox{212-\si{\micro\meter}} image of the sample was taken. 
	
	\hspace{3mm}Six adjacent, parallel, and non-overlapping cross-sectional profiles were generated from this image (Fig.~\hyperref[fig:figs16]{\ref*{fig:figs16}A}). The profiles are approximately orthogonal to the uncoated-coated boundary, as can be seen in Fig.~\hyperref[fig:figs16]{\ref*{fig:figs16}B}. Each profile is the average of 101 scan lines (\mbox{0.276-\si{\micro\meter}} line spacing). We note that this measurement is close to the limit of the vertical resolution of confocal optical profilometry, which resulted in significantly noisier data compared to the other classical techniques tested. Since fitting the noisy data to Eq.~\ref{eq:s73} yielded poor fits, we instead performed a semi-qualitative analysis to determine the height difference across the uncoated-coated boundary. For each profile, the heights of the uncoated and coated regions were estimated by identifying the maximum and minimum vertical points of the step in the sample. The film thickness is the difference of these two points. The mean of the six resulting heights is taken as the final thickness, with the standard deviation as the measurement uncertainty.
\end{enumerate}

\section*{Alternative interference modes}
\label{sec:aim}

\subsection*{Classical frequency beating}
\label{sec:cfb}

Instead of starting with a non-degenerate polarization-entangled state (Eq.~\ref{eq:4}) and converting it to energy entanglement via propagation through a polarizing beamsplitter, we can start with a non-entangled, non-degenerate two-photon state of the form
\begin{equation}
	\ket{\phi_{SPDC}}=\ket{\Theta}_{1550}\ket{\Theta}_{810},
\end{equation}
where $\Theta$ refers to a polarization basis state involving an equal superposition of $\ket{H}$ and $\ket{V}$, e.g., those from the diagonal or circular bases. Such a state may be generated with our entanglement source by pumping only the crystal that contributes to the $\ket{H}_{1550}\ket{H}_{810}$ term of Eq.~\ref{eq:4} and rotating the compensation and tomography waveplates to transform $\ket{H}_{1550}\ket{H}_{810}$ into $\ket{\Theta}_{1550}\ket{\Theta}_{810}$.

With the polarization state being an equal superposition of $\ket{H}$ and $\ket{V}$, the polarizing beamsplitter is effectively a 50:50 beamsplitter such that the entire interferometer functions as an ordinary dual-frequency Mach-Zehnder interferometer. Indeed, the classical interference measurements reported in this work were performed by reconfiguring our experiment in the manner described in the paragraph above and monitoring the counts on the \mbox{1550-nm detectors} (ignoring the \mbox{810-nm} detectors). When analyzing the classical data (e.g., extracting the interference visibility), we utilize the normalized probability of photon detection in output port $A$ of the interferometer (obtained by dividing the counts from the $1550A$ detector by the sum of the counts from the $1550A$ and $1550B$ detectors).

However, if we monitor both wavelengths, we can observe classical frequency beating. To model this behavior, we start with the fact that for a photon with a frequency of $\omega$, the probability that the photon exits the interferometer in a particular output port is given by
\begin{equation}
	\begin{split}
		P_A &= \sin^2\left(\frac{\phi}{2}\right) \\
		P_B &= \cos^2\left(\frac{\phi}{2}\right),
	\end{split}
	\label{eq:s79}
\end{equation}
where the $A$ and $B$ subscripts refer to interferometer output ports $A$ and $B$, and $\phi = \omega\tau$ is the relative phase between the two interferometer paths resulting from a relative temporal delay of $\tau$.

In the dual-frequency case, two photons of frequencies $\omega_1$ and $\omega_2$, respectively, propagate through the same interferometer. We are interested in four possible outcomes: both photons exiting in (1) port $A$ or (2) port $B$, (3) the $\omega_1$ photon in port $A$ and the $\omega_2$ photon in port $B$, and (4) the $\omega_1$ photon in port $B$ and the $\omega_2$ photon in port $A$. Using Eq.~\ref{eq:s79} and keeping track of the frequency-dependent relative phase, we conclude that the probabilities for these four outcomes are given by
\begin{equation}
	\begin{split}
		P_{AA} &= \sin^2\left(\frac{\phi_1}{2}\right)\sin^2\left(\frac{\phi_2}{2}\right)\\
		P_{BB} &= \cos^2\left(\frac{\phi_1}{2}\right)\cos^2\left(\frac{\phi_2}{2}\right)\\
		P_{AB} &= \sin^2\left(\frac{\phi_1}{2}\right)\cos^2\left(\frac{\phi_2}{2}\right)\\
		P_{BA} &= \cos^2\left(\frac{\phi_1}{2}\right)\sin^2\left(\frac{\phi_2}{2}\right),
	\end{split}
	\label{eq:s80}
\end{equation} 
where the $AA$ subscript indicates that the $\omega_1$ and $\omega_2$ photons, respectively, both exit in port $A$, 
and so on. It then follows that the coincidence probability is given by
\begin{equation}
	\begin{split}
		P_C &= \frac{P_{AB}+P_{BA}}{P_{AA}+P_{AB}+P_{BA}+P_{BB}} \\
		&= \frac{1}{2}\left(1-\cos(2\phi_1)\cos(2\phi_2)\right),
	\end{split}
	\label{eq:s81}
\end{equation}
via substitution of the probabilities from Eq.~\ref{eq:s80} and trigonometric simplification.

We experimentally observed classical frequency beating (Fig.~\ref{fig:figs17}). The coincident detection fringes were fitted to fitting functions based on the form of the probabilities given in Eq.~\ref{eq:s80}. The coincident probability fringes were fitted to a fitting function based on the form of Eq.~\ref{eq:s81}. The single-detection fringes were fitted to a sinusoidal function, being standard classical single-photon interference fringes. We observe excellent agreement between the experimental data and the fits.

\subsection*{Quantum sum-frequency beating}
\label{sec:qfb}

While we are primarily interested in the two-photon interference effect arising from the energy-entangled state 
\begin{equation}
	\ket{\psi} = \frac{1}{\sqrt{2}}\left(\ket{\omega_1}_a\ket{\omega_2}_b+\ket{\omega_2}_a\ket{\omega_1}_b\right),
\end{equation}
a similar effect can be explored for the state in which both colors always travel the same path in the interferometer,
\begin{equation}
	\ket{\psi_{SF}} = \frac{1}{\sqrt{2}}\left(\ket{\omega_1}_a\ket{\omega_2}_a+\ket{\omega_2}_b\ket{\omega_1}_b\right),
	\label{eq:s83}
\end{equation}
Experimentally, in our system this would correspond to replacing the polarization-entangled input state with  
\begin{equation}
	\frac{1}{\sqrt{2}}\big(\ket{H}_{\omega_1}\ket{H}_{\omega_2}+\ket{V}_{\omega_1}\ket{V}_{\omega_2}\big).
\end{equation} 
When the state $\ket{\psi_{SF}}$ travels through the interferometer, both photons will acquire a frequency-dependent-relative phase,
\begin{equation}
	\ket{\psi_{SF}} \rightarrow \frac{1}{\sqrt{2}}\left(\ket{\omega_1}_a\ket{\omega_2}_a+e^{-i\left(\omega_1+\omega_2\right)\tau}\ket{\omega_2}_b\ket{\omega_1}_b\right).
\end{equation}
After the combining 50:50 beamsplitter,  
\begin{equation}
	\begin{split}
		\ket{\psi_{SF}} &\rightarrow \frac{1}{\sqrt{8}}\left(\left(\ket{\omega_1}_a+i\ket{\omega_1}_b\right)\left(\ket{\omega_2}_a+i\ket{\omega_2}_b\right)+e^{-i\left(\omega_1+\omega_2\right)\tau}\left(\ket{\omega_1}_b+i\ket{\omega_1}_a\right)\left(\ket{\omega_2}_b+i\ket{\omega_2}_a\right)\right)\\
		&= \frac{1}{\sqrt{8}}\left(\left(1-e^{-i\left(\omega_1+\omega_2\right)\tau}\right)\left(\ket{\omega_1}_a\ket{\omega_2}_a-\ket{\omega_1}_b\ket{\omega_2}_b\right) + i\left(1+e^{-i\left(\omega_1+\omega_2\right)\tau}\right)\left(\ket{\omega_1}_a\ket{\omega_2}_b + \ket{\omega_1}_b\ket{\omega_2}_a \right) \right),
	\end{split}
\end{equation}
leading to coincidence probability  
\begin{equation}
	P_C = \frac{1}{2}\left(1+\cos((\omega_1+\omega_2)\tau)\right).
\end{equation}
This fringe beats at the sum-frequency of the two photons, or, in the case of a photon pair generated via spontaneous parametric down-conversion, the pump frequency $\omega_p = \omega_1+\omega_2$. 

Metrological measurements utilizing this sum-frequency beating effect can provide higher resolution per photon pair, with a Cram\'er-Rao bound of 
\begin{equation}
	\sigma_{\tau,CR,SF} \geq \frac{1}{\sqrt{\mathstrut N}}\frac{1}{\sqrt{\mathstrut (\omega_1+\omega_2)^2 + 4\sigma^2}}.
\end{equation}
However, since this interference requires both photons to travel the same path in the interferometer, it is susceptible to loss similar to classical interference. In Fig.~\ref{fig:figs18}, we demonstrate sum-frequency beating in our experimental system, measuring the interference pattern and resulting Fisher information. The coincidence probability $P_C$ oscillates sinusoidally with a fitted period of 535.85(3)~nm, close to the fitted \mbox{531.9120(5)-nm} center wavelength of the entanglement source pump (Fig.~\ref{fig:figs4}).

\counterwithin{figure}{section}
\renewcommand{\thesection}{S}
\renewcommand{\thefigure}{\thesection\arabic{figure}}

\clearpage

\phantomsection
\label{sec:figures}
\begin{center}
	{\large \textbf{Figures}}
	\vspace{0.125 in}
\end{center}

\begin{figure}[h]
	\begin{center}
		\includegraphics{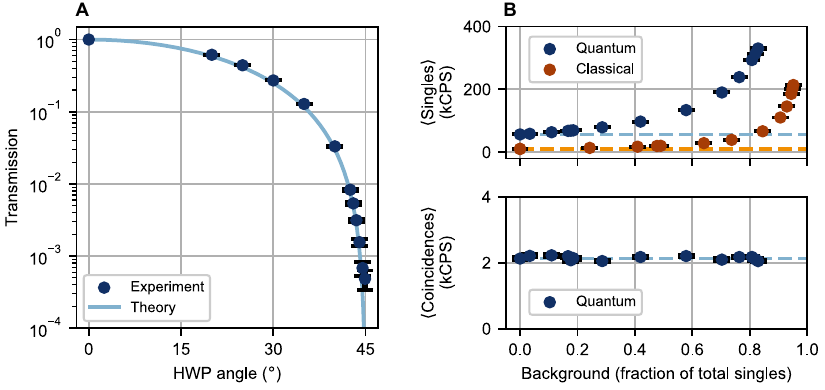}
	\end{center}
	\caption{\textbf{Simulated optical loss and background.} \textbf{(A)} We implement a tunable optical loss in mode $b$ of the interferometer with a half-wave plate (HWP) followed by a polarizing beamsplitter (PBS). The observed PBS transmission for the quantum probe as a function of the HWP angle is in excellent agreement with the theoretical curve for an ideal HWP-PBS system obeying Malus’ law (solid curve); the measured transmission does not reach zero when the angle approaches $45^\circ$ because of a finite extinction ratio arising from HWP and PBS imperfections. \textbf{(B)} The mean total count rates for single detections (top) and coincident detections (bottom) as a	function of simulated optical background, quantified as the fraction of total singles. The dashed lines indicate the zero-background baseline. Unlike the single-detection rate, which increases as background increases, the coincident detection rate is effectively independent of the background because of the tight coincident detection window used.}
	\label{fig:figs1}
\end{figure}

\clearpage

\begin{figure}[h]
	\begin{center}
		\includegraphics[width=\textwidth]{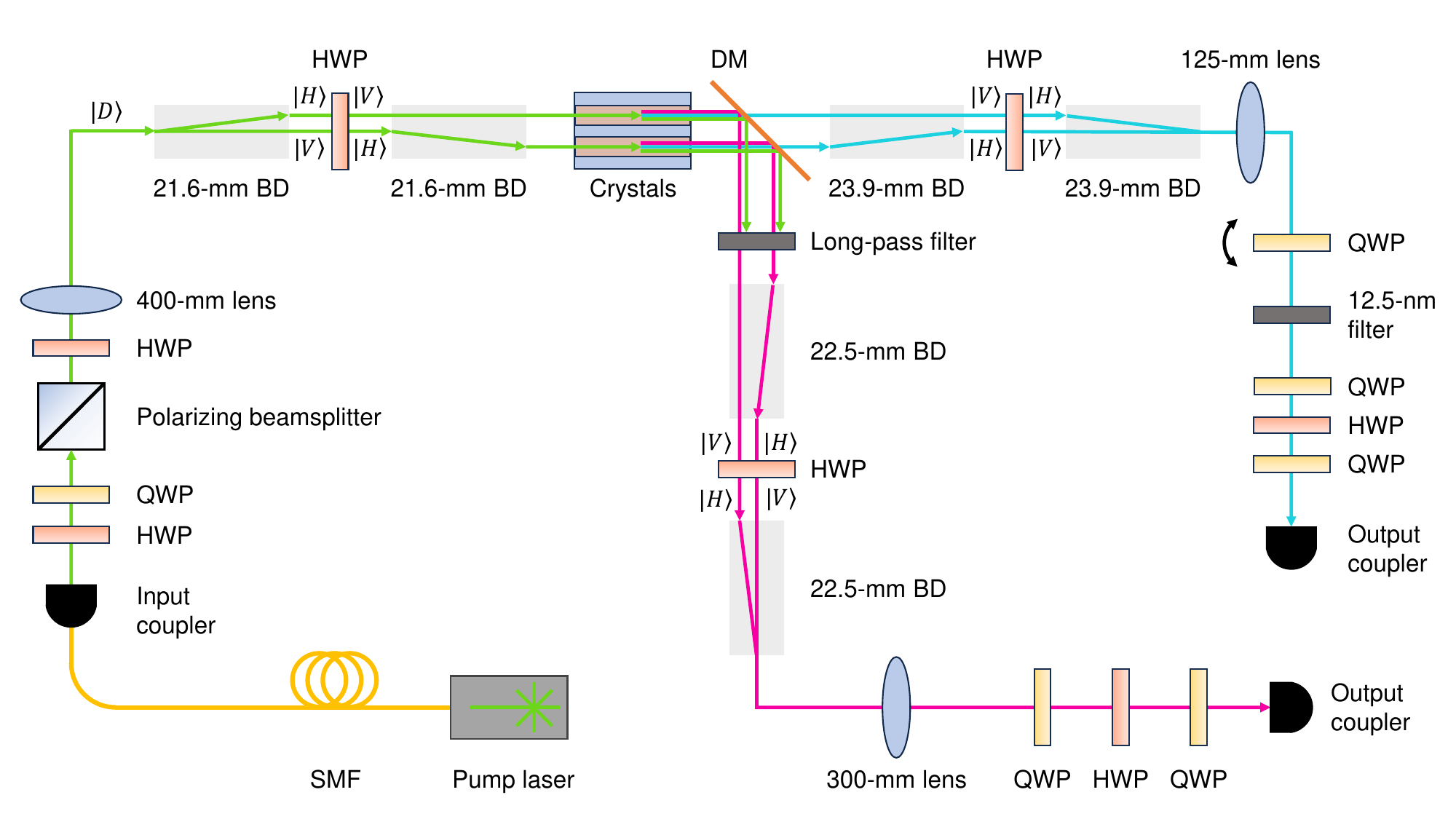}
	\end{center}
	\caption{\textbf{Schematic of non-degenerate polarization entanglement source.} A \mbox{532-nm} (green) continuous-wave (CW) laser is polarization-filtered and focused into a pair of beam displacers (BDs), creating a polarization superposition of paths. Each beam is used to pump a MgO:PPLN down-conversion crystal, creating photon pairs at 810~nm (magenta) and 1550~nm (cyan) via \mbox{Type-0} spontaneous parametric down-conversion (SPDC). A dichroic mirror (DM) separates the two wavelengths, after which the processes are recombined using a second pair of beam displacers in each arm, leading to a non-degenerate polarization-entangled state. Residual pump light is removed by interference filters, and the photons are coupled into single-mode fibers (SMFs). A trio of waveplates --- two quarter-wave plates (QWP) and a half-wave plate (HWP) --- built into the source is used to pre-compensate for any polarization transformations accrued in the fibers.}
	\label{fig:figs2}
\end{figure}

\clearpage

\begin{figure}[h]
	\begin{center}
		\includegraphics[width=\textwidth]{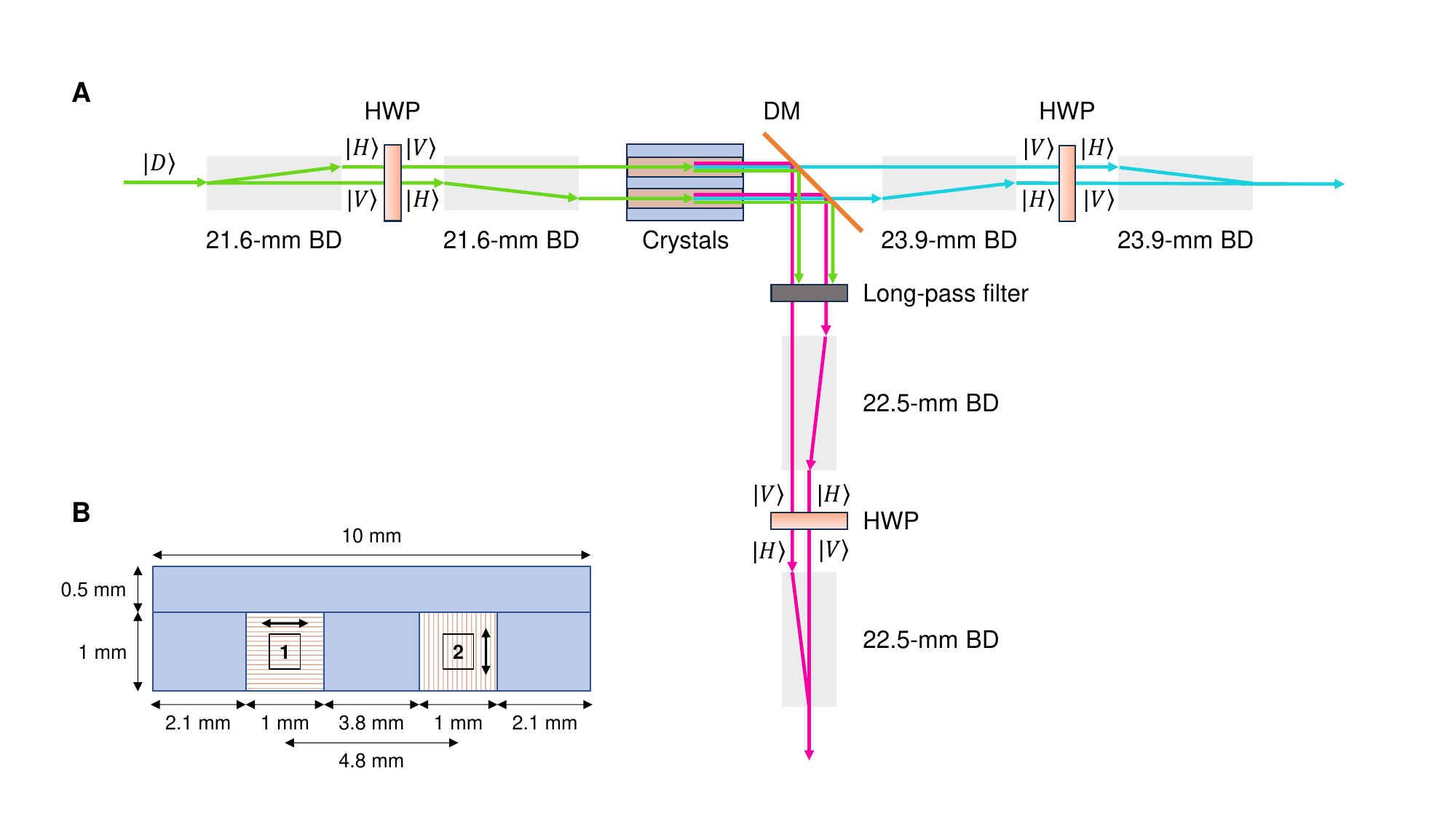}
	\end{center}
	\caption{\textbf{Simplified schematic of non-degenerate polarization entanglement source.} \textbf{(A)} Beam displacer interferometer, in which a diagonally polarized photon at 532~nm (green) is split via two beam displacers into a superposition of paths separated laterally by 4.8~millimeters. After down-conversion, the paths are recombined with a second set of beam displacers. \textbf{(B)} SPDC crystal mount, with two \mbox{1-mm $\times$ 1-mm $\times$ 20-mm} MgO:PPLN crystals separated by a \mbox{3.8-mm} LN spacer and rotated $90^\circ$ with respect to one another about the optical axis. This ensures that the center of each crystal aligns with one of the source interferometer beams.}
	\label{fig:figs3}
\end{figure}

\clearpage

\begin{figure}[h]
	\begin{center}
		\includegraphics{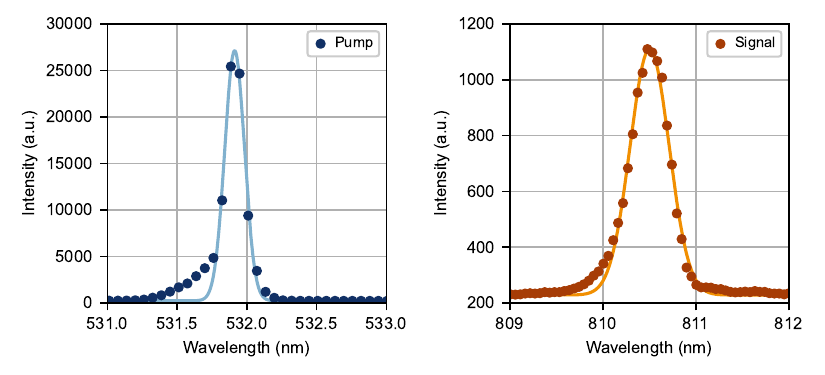}
	\end{center}
	\caption{\textbf{Entanglement source pump and signal spectra.} Gaussian fits to the pump (left) and signal (right) spectra yield center wavelengths of 531.9120(5)~nm and 810.504(1)~nm, close to the nominal wavelengths of 532 and 810~nm, respectively. Energy conservation implies an idler wavelength of 1547.484(5)~nm, which is close to the nominal wavelength of 1550~nm. The given uncertainties are based on fit errors.}
	\label{fig:figs4}
\end{figure}

\clearpage

\begin{figure}[h]
	\begin{center}
		\includegraphics{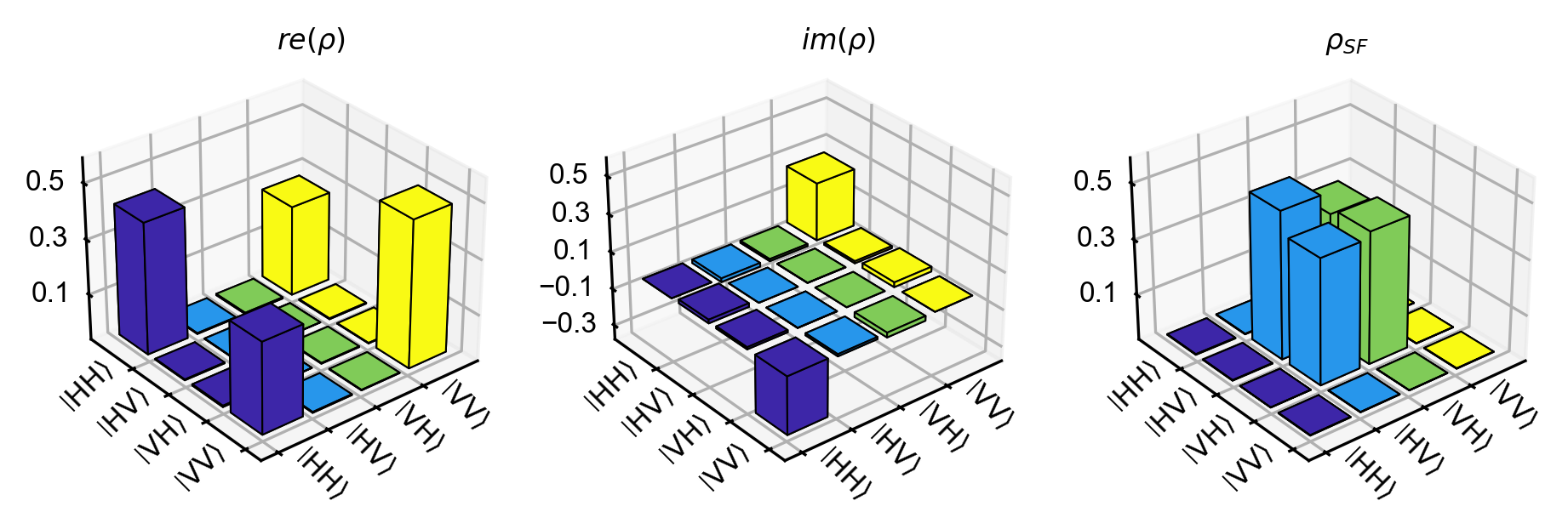}
	\end{center}
	\caption{\textbf{Polarization-entangled state density matrix.} The mean real (left) and imaginary (center) components of the density matrix $\rho$ recovered during the state tomography step of the end-to-end system calibration protocol (Step 3). Identical to Fig.~\hyperref[fig:fig1]{\ref*{fig:fig1}B}, $\rho_{SF}$ (right) illustrates how the unitary transformations realized via wave plates may be used to transform the state generated by the entanglement source (Eq.~\ref{eq:4}) and directly measured via state tomography into a given maximally entangled state, here chosen to be the nominal state required to generate energy entanglement (Eq.~\ref{eq:s69}). The bit-flip step in the calibration protocol (Step 4) performs the unitary transformation to recover the singlet fraction matrix, up to a relative phase.}
	\label{fig:figs5}
\end{figure}

\clearpage

\begin{figure}[h]
	\begin{center}
		\includegraphics{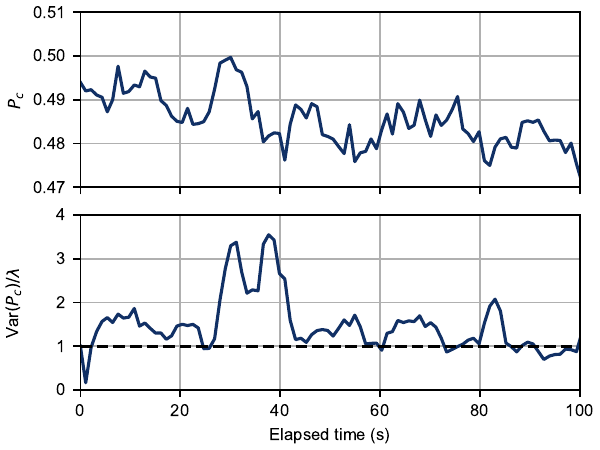}
	\end{center}
	\caption{\textbf{Interferometer drift.} Top panel: Time trace showing a typical drift in the interferometer phase, leading to a change in $P_C$, over 100 seconds. Bottom panel: The normalized noise (defined as the ratio of variance in $P_C$ over Poissonian variance $\lambda$) for a \mbox{10-s} rolling window. The black dashed line shows the expected trace $(\text{Var}(P_C)/\lambda = 1)$ absent interferometer drift and other noise sources.}
	\label{fig:figs6}
\end{figure}

\clearpage

\begin{figure}[h]
	\begin{center}
		\includegraphics{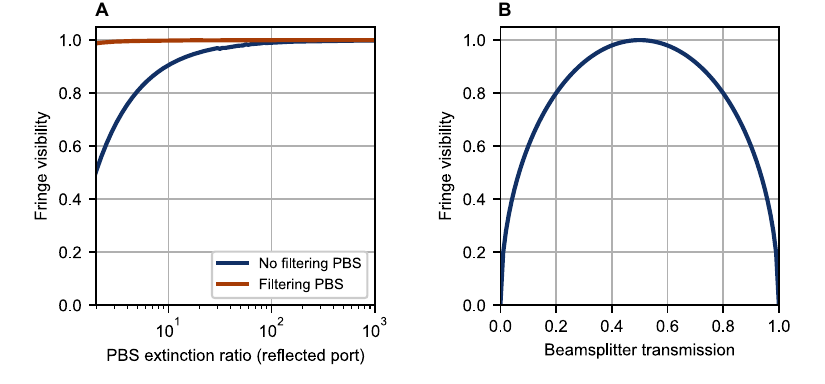}
	\end{center}
	\caption{\textbf{Effects of system imperfections on fringe visibility.} \textbf{(A)} The reduction in interference visibility resulting from imperfect polarizing beamsplitter (PBS) extinction ratio (ER). The transmission-port ER is fixed at 10,000, while the reflected-port ER is varied. Two scenarios are considered, one in which only a single PBS is used (blue), and one in which a second PBS is placed in the reflected arm of the interferometer to achieve additional polarization filtering (red). \textbf{(B)} The reduction in interference visibility resulting from imperfect non-polarizing beamsplitter (NPBS) splitting ratio. Here the transmission coefficient $T$ of the NPBS is varied for one wavelength in the system, with the reflection coefficient $R = 1 - T$ adjusted correspondingly. For the other wavelength, $T = R = 0.5$ is assumed.}
	\label{fig:figs7}
\end{figure}

\clearpage

\begin{figure}[h]
	\begin{center}
		\includegraphics{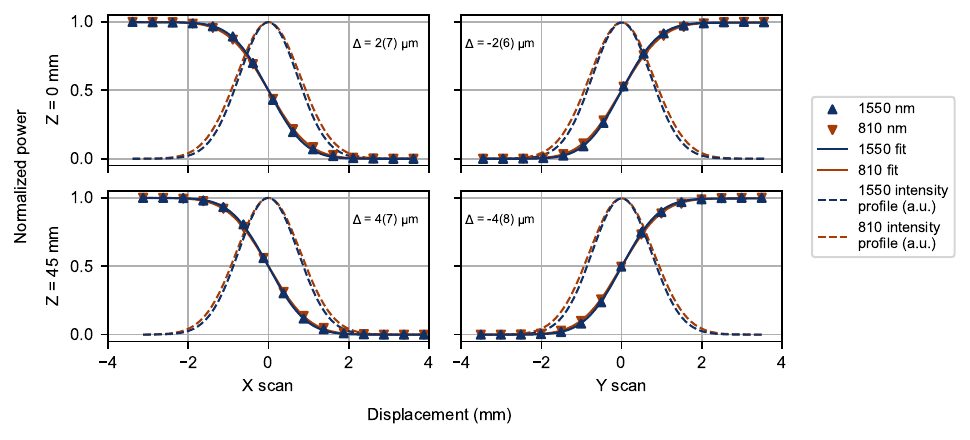}
	\end{center}
	\caption{\textbf{Spatial overlap of the \mbox{1550-nm} and \mbox{810-nm} modes.} We measure the spatial overlap of the \mbox{1550-nm} and \mbox{810-nm} modes at the input of our interferometer by performing knife-edge scans and fitting the curves to Eq.~\ref{eq:s63} to obtain the beam centroid $\delta_0$. The relative offsets of the \mbox{1550-nm} and \mbox{810-nm} centroids $\Delta \equiv \delta_0^{810} - \delta_0^{1550}$ are close to zero at two points along the multiplexed beam (separated longitudinally by 45~mm), indicating excellent spatial mode overlap and beam parallelism. The uncertainty in $\Delta$ is estimated via error propagation. To illustrate the overlap in the intensity profiles for both beams, we also plot the normalized magnitude (in arbitrary units) of the first derivative of the fitted curves.}
	\label{fig:figs8}
\end{figure}

\clearpage

\begin{figure}[h]
	\begin{center}
		\includegraphics{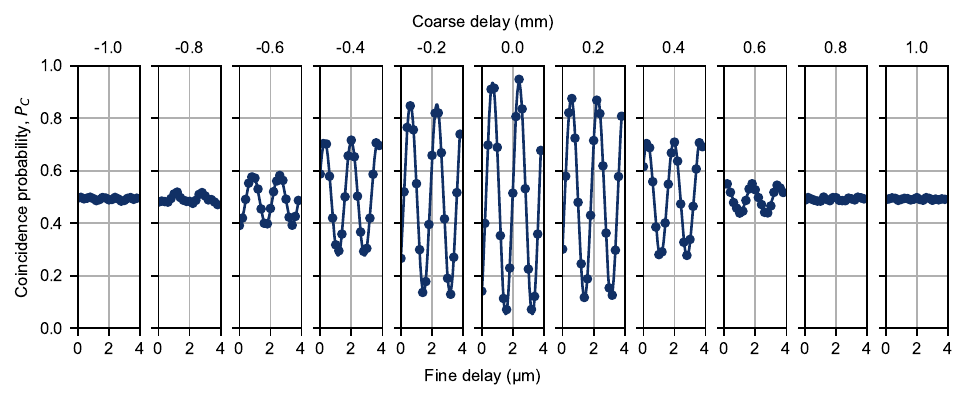}
	\end{center}
	\caption{\textbf{Interference fringe envelope.} A selection of individual fringe scans used to calculate the visibilities shown in Fig.~\hyperref[fig:fig1]{\ref*{fig:fig1}D}. The lower $x$-axis labels indicate the fine relative delay (piezoelectric nano-positioning stage) and the upper labels indicate the coarse relative delay (DC servo actuator) introduced by the optical trombone. A fine fringe scan was performed at each coarse delay position. The delays are in terms of optical path length. The solid curves are fits to the data.}
	\label{fig:figs9}
\end{figure}

\clearpage

\begin{figure}[h]
	\begin{center}
		\includegraphics{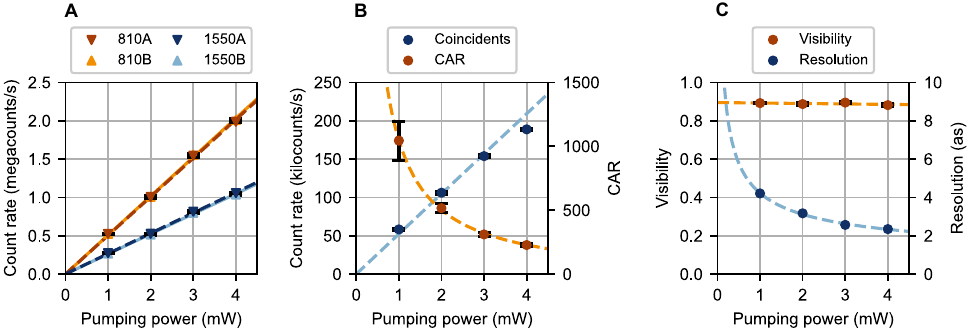}
	\end{center}
	\caption{\textbf{System performance as a function of entanglement source pumping power.} \textbf{(A)} The count rates on individual detectors scale linearly with the pump power, at least for count rates up to $\sim$2 megacounts per second. This behavior is expected as the detector deadtime $\tau_{dt}$ is $\leq$20~ns for the \mbox{810-nm} detectors and $\leq$30~ns for the \mbox{1550-nm} detectors (vendor-supplied values); the nominal time tagger deadtime is $\sim$2~ns. The corresponding detector saturation rate $1/\tau_{dt}$ is therefore at least $\sim$50 and $\sim$33 megacounts per second for the \mbox{810-nm} and \mbox{1550-nm} detectors, respectively. Since the $A$ and $B$ detectors are balanced for each wavelength as a part of the end-to-end system calibration protocol, their count rates appear almost identical. Solid and dashed lines are linear fits with the $y$-intercept fixed at zero. The fitted slopes for the $810A$ and $810B$ detectors are 503,753(5,031) and 508,167(4,450) counts per second per mW, respectively. The corresponding values for the \mbox{1550-nm} detectors are 266,211(2,099) and 261,650(2,059) counts per second per mW, respectively. \textbf{(B)} The total coincident detection rate also scales linearly with the pump power but begins to saturate above 3 mW of pumping power. The dashed line is a linear fit with the $y$-intercept fixed at zero; the fitted slope is 52,270(1,297) counts per second per mW. The coincidence to accidental ratio (CAR) scales inversely with the pumping power and exhibits excellent agreement with the fit (dashed line). \textbf{(C)} The interference visibility is largely independent of the pumping power; a linear fit yields a slope of $-0.2(3)\%$ per mW and a $y$-intercept of $89.5(8)\%$ (dashed line). The resolution scales as square root of the pump power and closely tracks the fit (dashed line).}
	\label{fig:figs10}
\end{figure}

\clearpage

\begin{figure}[h]
	\begin{center}
		\includegraphics[width=0.75\textwidth]{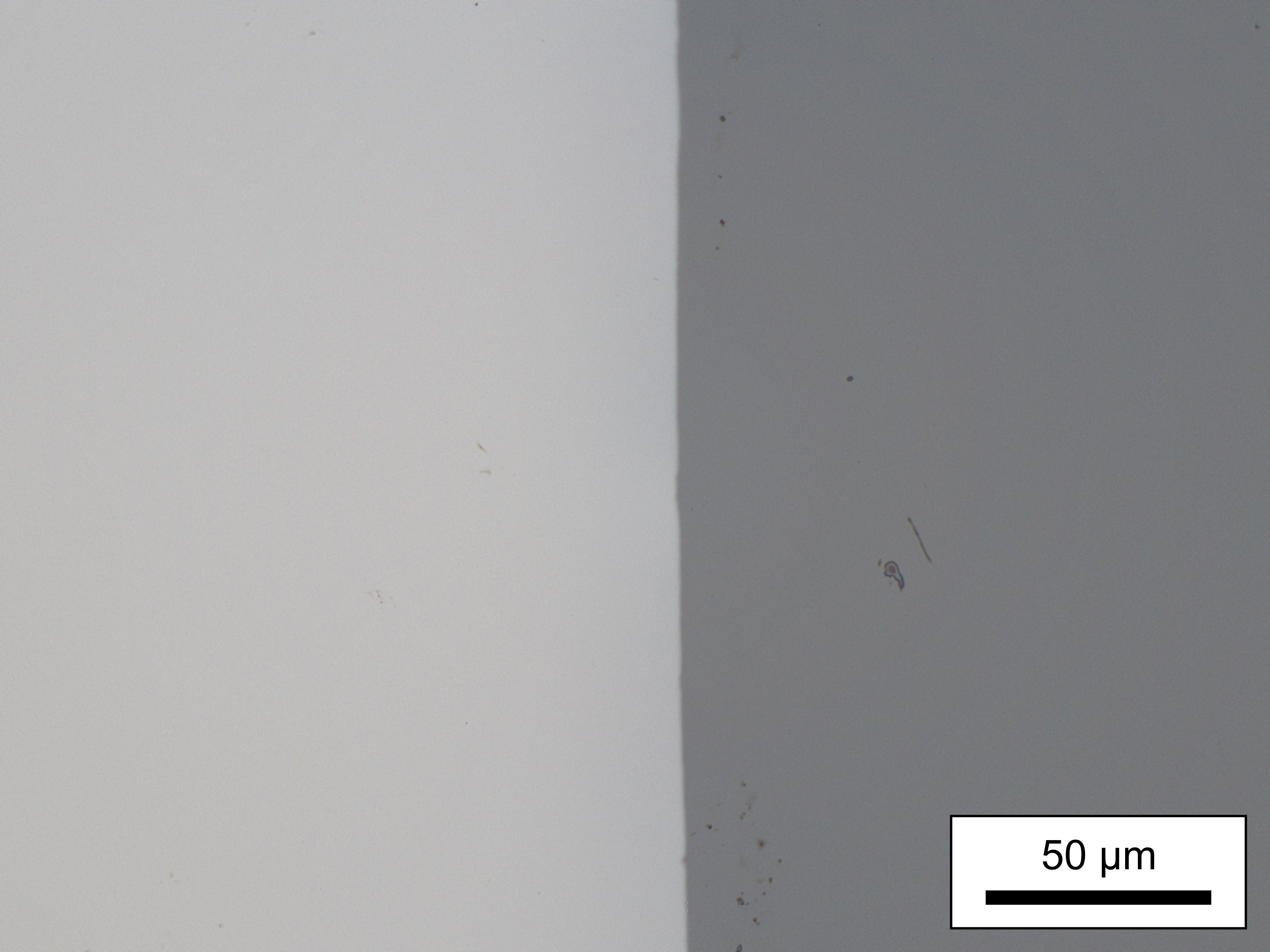}
	\end{center}
	\caption{\textbf{Optical image of nickel thin film sample (5-nm thickness).} The left side is the 
		coated region, and the right is uncoated (bare substrate). Image taken with a Keyence VK-X1000 
		3D laser scanning confocal microscope.}
	\label{fig:figs11}
\end{figure}

\clearpage

\begin{figure}[h]
	\begin{center}
		\includegraphics{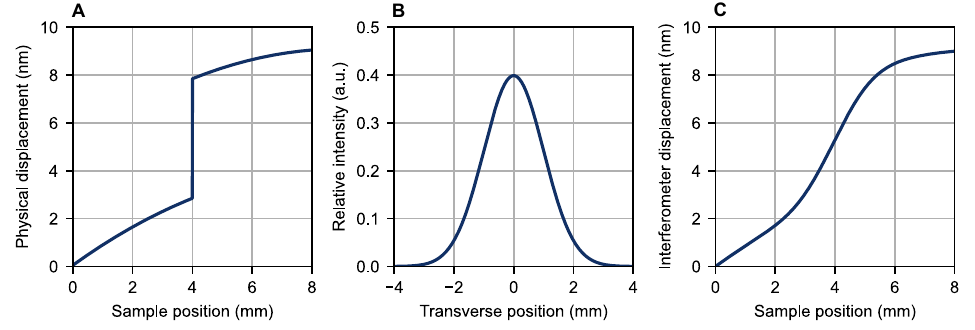}
	\end{center}
	\caption{\textbf{Illustration of model for thin-film measurements.} \textbf{(A)} Hypothetical sample consisting of a \mbox{5-nm} thick film atop a substrate with a linear curvature of 0.5~nm/mm and a quadratic curvature of $-0.05$ nm/mm$^2$. The edge of the uncoated and coated regions is located at sample position 4~mm. \textbf{(B)} Hypothetical Gaussian transverse intensity distribution for the probe beam, with a full width at half maximum of $2\sqrt{2\ln 2}$~mm. \textbf{(C)} Hypothetical measurement of the sample shown in (A) using the probe beam geometry shown in (B). The resulting displacement versus position data is a convolution of (A) and (B). For this illustration, an effective refractive index of 2 is assumed for the film and substrate of the sample, and 1 for the ambient atmosphere.}
	\label{fig:figs12}
\end{figure}

\clearpage

\begin{figure}[h]
	\begin{center}
		\includegraphics{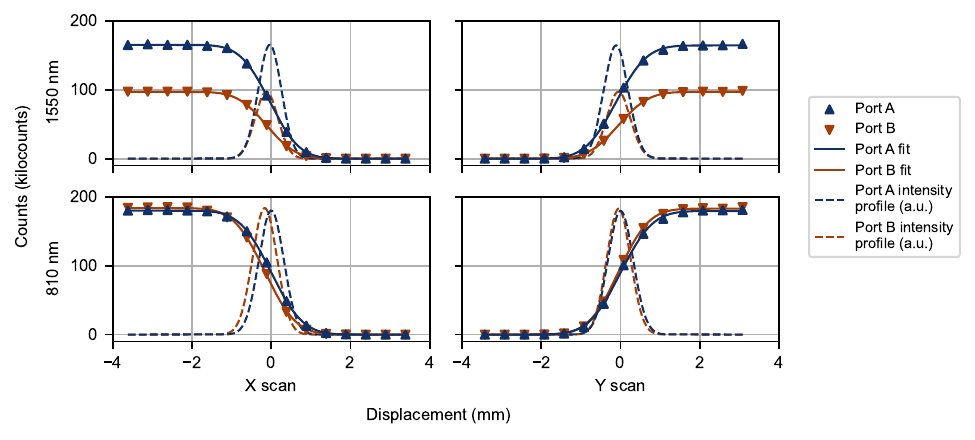}
	\end{center}
	\caption{\textbf{Probe beam characterization.} We characterize the spatial mode of the probe beam for both wavelengths by performing knife-edge scans and fitting the curves to Eq.~\ref{eq:s63}. To illustrate the intensity profiles for both beams, we also plot the normalized magnitude (in arbitrary units) of the first derivative of the fitted curves.}
	\label{fig:figs13}
\end{figure}

\clearpage

\begin{figure}[h]
	\begin{center}
		\includegraphics{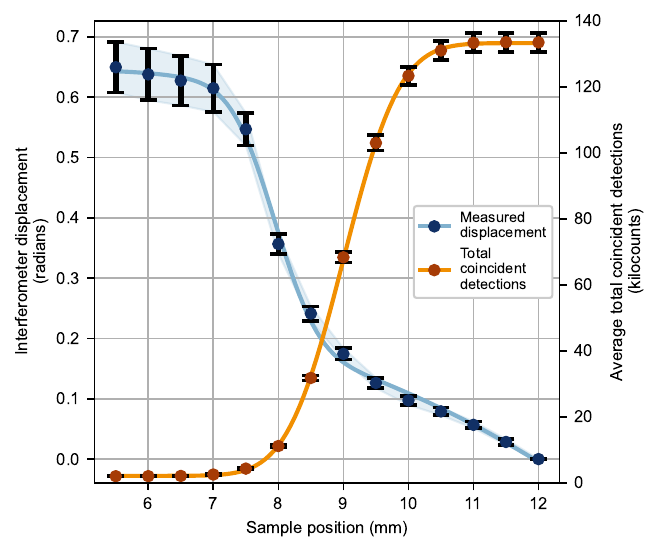}
	\end{center}
	\caption{\textbf{Quantum interferometer calibration measurement.} The interferometer displacement (in units of radians) and the average total coincident detections as a function of sample position for the calibration sample (\mbox{50-nm} thickness). The error bars show the standard deviation of 100 trials. For each trial, the scan started at position 12~mm, with an integration time of 1~second per point. The solid curves are fits to the data. A quantum probe transmission of $1.6(2)\%$ is observed for the coated region (relative to the uncoated region), and fixing the film thickness at 50 nm yields an $n_{film}$ of 3.3(3).}
	\label{fig:figs14}
\end{figure}

\clearpage

\begin{figure}[h]
	\begin{center}
		\includegraphics{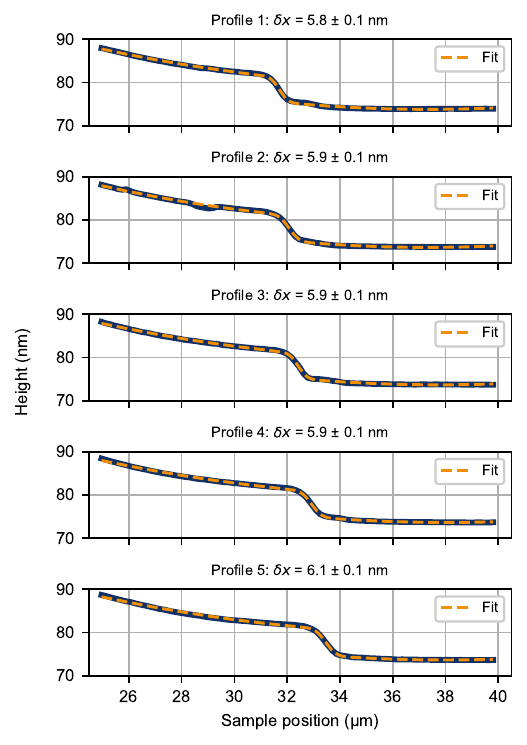}
	\end{center}
	\caption{\textbf{Test sample thickness measurements via scanning-stylus profilometry.} Measured height as a function of sample position for five adjacent, parallel, and non-overlapping cross-sectional profiles (solid curves). A change in height across the uncoated-coated boundary is clearly visible. The film thickness $\delta x$ is extracted from fits to each profile (dashed curves).}
	\label{fig:figs15}
\end{figure}

\clearpage

\begin{figure}[h]
	\begin{center}
		\includegraphics{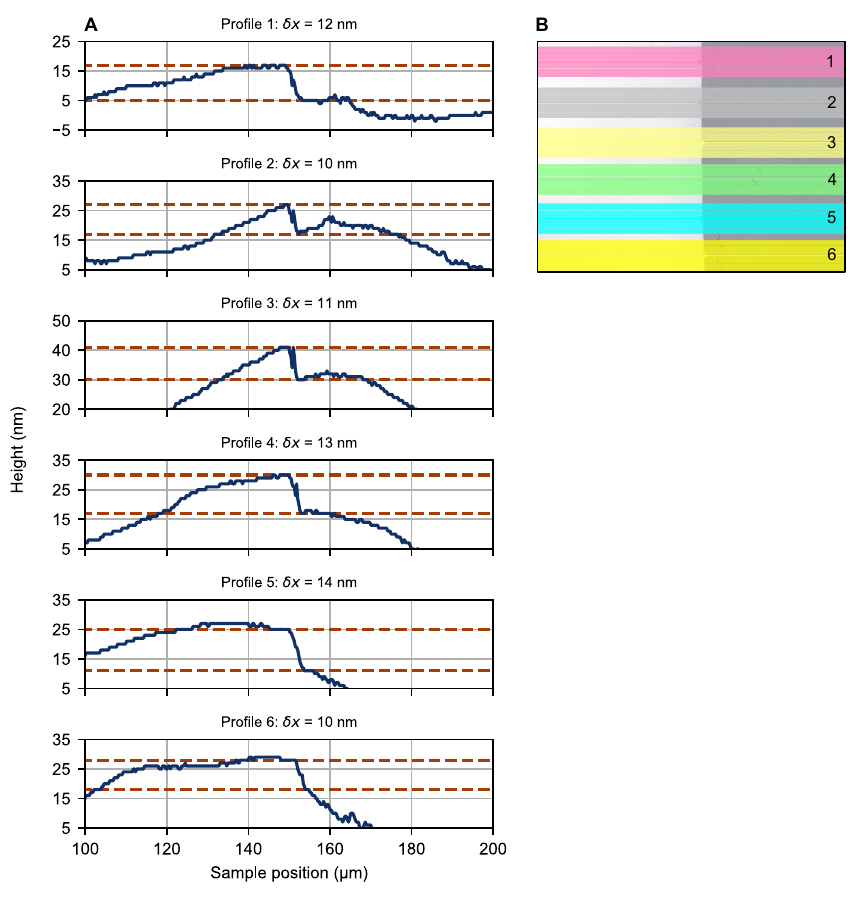}
	\end{center}
	\caption{\textbf{Test sample thickness measurements via 3D optical profilometry.} \textbf{(A)} Measured height as a function of sample position for six adjacent, parallel, and non-overlapping cross-sectional profiles (solid blue). A change in height across the uncoated-coated boundary is clearly visible. The dashed red lines indicate the heights used to calculate the film thickness $\delta x$. \textbf{(B)} Illustration of the regions from which the cross-sectional profiles were generated (superimposed on Fig. \ref{fig:figs11}). The vertical width of each color band indicates the region where horizontal scan lines were averaged in generating a profile. The region numbering follows the profile numbering in (A).}
	\label{fig:figs16}
\end{figure}

\clearpage

\begin{figure}[h]
	\begin{center}
		\includegraphics{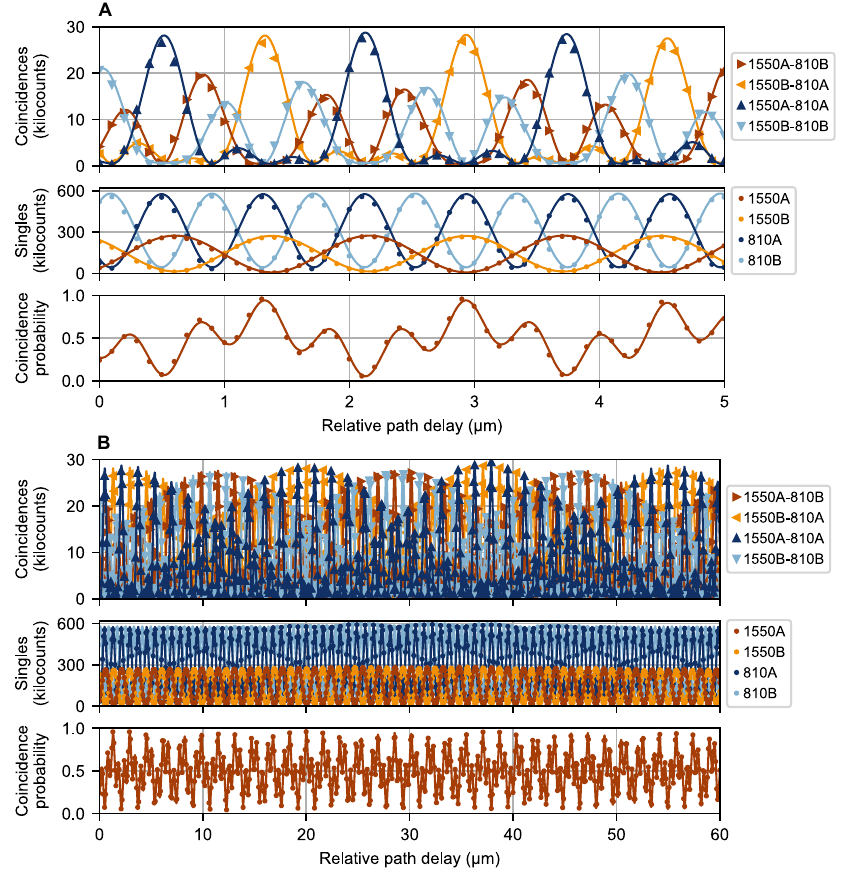}
	\end{center}
	\caption{\textbf{Classical frequency beating.} A zoomed-in section of a \mbox{60-\si{\micro\meter}} scan is shown in \textbf{(A)} while the full scan is shown in \textbf{(B)}. Solid curves are fits; see text. Note the envelope beating with the coincidences fringes in (B). The four coincident detection fits have fitted photon wavelengths of 1554.32(4)~nm and 812.96(4)~nm for 1550$A$-810$B$, 1554.22(4)~nm and 812.95(4)~nm for 1550$B$-810$A$, 1554.27(4)~nm and 812.98(4)~nm for 1550$A$-810$A$, and 1554.25(4)~nm and 812.95(4)~nm for 1550$B$-810$B$. Similarly, the four single-detection fits return wavelengths of 1554.76(4)~nm, 1554.75(4)~nm, 812.98(1)~nm, and 812.99(1)~nm for 1550$A$, 1550$B$, 810$A$, and 810$B$, respectively. The coincidence probability fit returns photon wavelengths of 1554.67(5) and 812.90(2)~nm. These fitted wavelengths are close to the actual photon wavelengths (Fig. \ref{fig:figs4}) of 810.504(1)~nm (measured) and 1547.484(5) nm (inferred via energy conservation). The fitted visibilities for the 1550$A$, 1550$B$, 810$A$, and 810$B$ fringes are $95.0(2)\%$, $90.4(2)\%$, $85.9(2)\%$, and $86.0(2)\%$, respectively.}
	\label{fig:figs17}
\end{figure}

\clearpage

\begin{figure}[h]
	\begin{center}
		\includegraphics{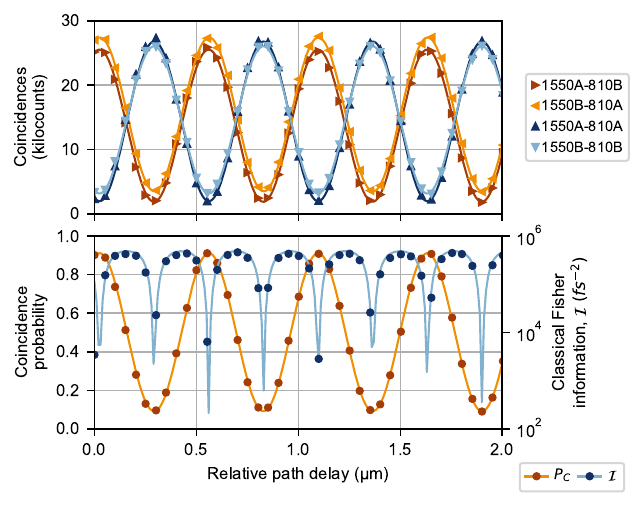}
	\end{center}
	\caption{\textbf{Quantum sum-frequency beating.} A \mbox{2-\si{\micro\meter}} snippet of a \mbox{10-\si{\micro\meter}} scan is shown; the fits (solid curves) utilize the full scan. As the photons’ relative time of arrival $\tau$ is scanned by adjusting the relative path delay between modes $a$ and $b$, $P_C$ oscillates sinusoidally with a fitted period of 535.85(3)~nm, close to the 531.9120(5)~nm center wavelength of the entanglement source pump (Fig. \ref{fig:figs4}). The fringe visibility is $81.6(2)\%$. Fits to the four coincident detection fringes yield visibilities of $86.3(3)\%$, $77.3(2)\%$, $86.3(3)\%$, and $78.3(2)\%$ for 1550$A$-810$B$, 1550$B$-810$A$, 1550$A$-810$A$, and 1550$B$-810$B$, respectively. In contrast, fits to the four individual detector fringes yield visibilities of $0.8(1)\%$, $0.5(1)\%$, $0.9(1)\%$, and $1.3(1)\%$ for 1550$A$, 1550$B$, 810$A$, and 810$B$, respectively. With 1~mW of pumping power for the entanglement source and an integration time of 1~second, we observe a mean of 58,200(600) total coincident detections per measurement. The corresponding resolution extracted from the Fisher information is 0.43~nm (1.43~as), which represents an $82\%$ saturation of the Cram\'er–Rao bound, assuming we have Eq.~\ref{eq:s83} as our probe state. For 10,000 coincident detections, the expected experimental resolution is 3.5~as, which contrasts with 10.2~as for the difference-frequency beating (the main effect considered in this work), despite the latter achieving an $88\%$ saturation of the Cram\'er–Rao bound. This enhancement occurs because of the shorter fringe period.}
	\label{fig:figs18}
\end{figure}

\clearpage

\phantomsection
\label{sec:tables}
\begin{center}
	{\large \textbf{Tables}}
	\vspace{0.125 in}
\end{center}

\begin{table}[h]
\raggedright
\begin{tblr}{
		colspec = {Q[c,0.45in]|Q[c,0.65in]|Q[c,1.1in]|Q[c,1.1in]|Q[c,0.6in]|Q[c,0.75in]}, 
		rows = {m, rowsep = 3pt},
		hline{1,2,6} = {solid},
	}
	\textit{Detector channel} & \textit{Design wavelength} & \textit{Detector absolute efficiency, Vendor measured} & \textit{Fiber link  transmission, measured} & \textit{Detector FWHM timing jitter, Vendor measured} & \textit{Time-tagger channel FWHM timing jitter, nominal} \\
	$1550A$ & 1590~nm & $89(2)\%$ (1570 nm) & $\sim$$100\%$ (1550 nm) & 51 ps  & 14 ps \\
	$1550B$ & 1590~nm & $87(2)\%$ (1570 nm) & $\sim$$100\%$ (1550 nm)  & 48 ps & 14 ps  \\
	$810A$ & 777~nm & $101(4)\%$ (778 nm) & $\sim$$91\%$ (810 nm)  & 41 ps & 14 ps  \\
	$810B$ & 777~nm & $100(4)\%$ (778 nm)  & $\sim$$97\%$ (810 nm)  & 54 ps & 14 ps  \\
\end{tblr}
\caption{\textbf{Detector specifications.} The detector absolute efficiency is measured from the optical input at the detector front panel and includes front-panel insertion loss. Optimal incident polarization is assumed. The given approximate fiber link transmission values are intended to illustrate the absence of relatively large sources of loss between the fiber couplers and the detectors; a precise characterization is beyond the scope of this work. These values do not include the fiber-coupling losses in the detection module (see Table~\ref{tab:tabs3}) but include fiber mating loss at the input to the fiber link. Also not included are small additional losses (insertion and transmission) from a short patch fiber connecting the fiber link to the detector front panel; their precise values are unknown but are estimated to be $<$$10\%$.}
\label{tab:tabs1}
\end{table}

\clearpage

\begin{table}[h]
	\raggedright
	\begin{tblr}{
			colspec = {Q[c]|Q[c]},
			hline{1,2,3,7,8,12,13} = {solid},
			vline{3,4} = {solid},
			column{1,3} = {c,0.5in},
			column{2} = {c, 0.75in},
			column{4} = {c},
			rows = {m, rowsep = 3pt}
		}
		\SetCell[c=3]{c} \textit{Optical path} & & & \SetCell[r=2]{c} \textit{Transmission} \\
		\textit{Fiber launch} & \textit{Free-space path} & \textit{Fiber coupler} &\\
		1550 & Transmitted & $1550A$ & $\sim$$16\%$\\
		1550 & Reflected & $1550A$ & $\sim$$20\%$\\
		1550 & Transmitted & $1550B$ & $\sim$$9\%$\\
		1550 & Reflected & $1550B$ & $\sim$$35\%$\\
		\SetCell[c=3]{r} \textit{Total} & & & $\sim$$80\%$\\
		810 & Transmitted & $810A$ & $\sim$$19\%$\\
		810 & Reflected & $810A$ & $\sim$$21\%$\\
		810 & Transmitted & $810B$ & $\sim$$19\%$\\
		810 & Reflected & $810B$ & $\sim$$21\%$\\
		\SetCell[c=3]{r} \textit{Total} & & & $\sim$$79\%$\\
	\end{tblr}
	\caption{\textbf{Interferometer module free-space transmission.} Transmission is defined as the quotient of the power immediately exiting the input fiber launch and the power incident on the output fiber coupler (see the interferometer and detection modules in Fig.~\hyperref[fig:fig1]{\ref*{fig:fig1}A}). Transmission through the transmitted and reflected paths of the interferometer is measured separately by blocking the transmitted and reflected paths as appropriate. Because of system geometry, the $1550A$, $1550B$, and $810A$ measurements include loss from a single reflection off a mirror. However, since the dielectric mirrors used have high reflectivity, their contribution to loss can be assumed to be negligible. The power split between the transmitted and reflected paths depends on the polarization of the incident light. The approximate values given are intended to illustrate the relatively high transmission of the interferometer; a precise characterization is beyond the scope of this work.}
	\label{tab:tabs2}
\end{table}

\clearpage

\begin{table}[h]
	\raggedright
	\begin{tblr}{
			colspec = {Q[c]|Q[c]},
			hline{1,2,3,7} = {solid},
			vline{2,3,4} = {solid},
			column{1,3,4} = {c,0.5in},
			column{2} = {c,0.65in},
			rows = {m, rowsep = 3pt}
		}
		\SetCell[r=2]{c} \textit{Detection fiber} & \SetCell[c=3]{c} \textit{Coupling efficiency} \\
	     & \textit{Transmitted path} & \textit{Reflected path} & \textit{Average} \\
		$1550A$ & $\sim$$87\%$ & $\sim$$90\%$ & $\sim$$88.5\%$ \\
		$1550B$ & $\sim$$89\%$ & $\sim$$89\%$ & $\sim$$89\%$ \\
		$810A$ & $\sim$$92\%$ & $\sim$$92\%$ & $\sim$$92\%$ \\
		$810B$ & $\sim$$97\%$ & $\sim$$95\%$ & $\sim$$96\%$ \\
	\end{tblr}
	\caption{\textbf{Detection module fiber-coupling efficiencies.} Coupling efficiency is defined as the quotient of the power incident on the fiber coupler and the power exiting the other end of the fiber. Note that the exiting end has no anti-reflection coating, which contributes $\sim$$4\%$ of loss to the measured efficiency. All measurements were made using classical alignment lasers and power meters at the suitable wavelength. Because of system geometry, the $1550A$, $1550B$, and $810A$ measurements include loss from a single reflection off a mirror. However, since these dielectric mirrors have high reflectivity, their contributions to loss can be assumed to be negligible. The approximate values given are intended to illustrate the relatively high and balanced coupling efficiencies realized; a precise characterization is beyond the scope of this work.}
	\label{tab:tabs3}
\end{table}

\newpage

\begin{table}[h]
	\raggedright
	\begin{tblr}{
			colspec = {Q[c,1.1in]|Q[c,1.1in]|Q[c,1.1in]},
			hline{1,2,6} = {solid}, 
			rows = {m, rowsep = 3pt}
		}
		\textit{Coincident detection channel} & \textit{Estimated FWHM, timing jitter} & \textit{Observed FWHM, detector correlation}\\
		$1550A$-$810A$ & 68 ps & 70.4(0.2) ps \\
		$1550A$-$810B$ & 77 ps & 80.2(0.4) ps \\
		$1550B$-$810A$ & 66 ps & 69.4(0.2) ps \\
		$1550B$-$810B$ & 75 ps & 79.5(0.4) ps \\
	\end{tblr}
	\caption{\textbf{Coincident detection jitter.} By summing (in quadrature) the vendor-supplied full width at half maximum (FWHM) jitter for the two detector and time-tagger channels for each coincident detection channel, we obtain an estimate for the combined jitter for each channel. These values are comparable to the FWHM of the correlation peak for each detector pair, which was obtained by fitting the correlation peak with a Gaussian function. The given FWHM error is the fitting error.}
	\label{tab:tabs4}
\end{table}

\newpage

\begin{table}[h]
	\raggedright
	\begin{tblr}{
			colspec = {Q[c,0.5in]|Q[c,0.8in]|Q[c,0.8in]|Q[c,0.8in]},
			hline{1,2,8} = {solid}, 
			rows = {m, rowsep = 3pt}
		}
		\textit{Basis state} & \textit{Quarter-wave plate \#1} & \textit{Half-wave plate} & \textit{Quarter-wave plate \#2} \\
		$\ket{H}$ & $0^\circ$ & $0^\circ$ & $0^\circ$ \\
		$\ket{V}$ & $0^\circ$ & $45^\circ$ & $0^\circ$ \\
		$\ket{D}$ & $45^\circ$ & $22.5^\circ$ & $0^\circ$ \\
		$\ket{A}$ & $45^\circ$ & $-22.5^\circ$ & $0^\circ$ \\
		$\ket{L}$ & $45^\circ$ & $0^\circ$ & $0^\circ$ \\
		$\ket{R}$ & $45^\circ$ & $45^\circ$ & $0^\circ$ \\
	\end{tblr}
	\caption{\textbf{State tomography angles.} Each state projector corresponds to a set of angles for the three waveplates (for each wavelength) in the source module. The shown angles are relative to the waveplates’ ``zero'' angles (i.e., corresponding to the waveplates’ fast or slow axes) obtained during calibration.}
	\label{tab:tabs5}
\end{table}

\newpage

\begin{table}[h]
	\raggedright
	\begin{tblr}{
			colspec = {Q[c,0.8in]|Q[c,0.8in]},
			hline{1,2,7} = {solid}, 
			rows = {m, rowsep = 3pt}
		}
		\textit{Metric} & \textit{Specification} \\
		Roughness & $<$0.3 nm \\
		Warp & $\leq$15 \si{\micro\meter} \\
		Bow & $\leq$10 \si{\micro\meter} \\
		TTV & $<$3 \si{\micro\meter} \\
		LTV & $\leq$1.5 \si{\micro\meter} \\
	\end{tblr}
	\caption{\textbf{Key sapphire wafer specifications.} Vendor supplied values. TTV: Total thickness 
		variation. LTV: Local thickness variation (5 $\times$ 5 mm).}
	\label{tab:tabs6}
\end{table}

\end{document}